\documentclass[draftclsnofoot, onecolumn, comsoc, 12pt]{IEEEtran} 
\usepackage{cite}
\usepackage[normalem]{ulem}
\usepackage{amsmath,amssymb,amsfonts}
\usepackage{graphicx}
\usepackage{caption}
\usepackage{textcomp}
\usepackage{xcolor}
\usepackage{graphicx}
\usepackage{amssymb}
\usepackage{amsmath}
\usepackage{cite}
\usepackage{stfloats}
\usepackage{epstopdf}
\usepackage{mdwlist}
\usepackage[section]{algorithm}
\usepackage[noend]{algpseudocode}

\usepackage[all,2cell]{xy} \UseAllTwocells
\usepackage{booktabs}

\usepackage{tikz}
\usetikzlibrary{arrows.meta}
\usetikzlibrary{decorations.pathreplacing}
\usepackage{xcolor}
\usepackage{enumitem}
\usepackage{nicefrac}

\usepackage{subcaption}
\usepackage{dsfont}
\usepackage{mathrsfs}

\usepackage{relsize}
\usepackage{float}
\usepackage{xr}
\newtheorem{theorem}{Theorem}
\newtheorem{proposition}{Proposition}
\newtheorem{lemma}{Lemma}

\newtheorem{proof}{Proof}
\newtheorem{example}{Example}
\newtheorem{definition}{Definition}

\AtBeginDocument{}

\newcommand{\beq}{\begin{equation}}
	\newcommand{\eeq}{\end{equation}}
\newcommand{\beqa}{\begin{eqnarray}}
	\newcommand{\eeqa}{\end{eqnarray}}

\newcommand{\comment}[1]{}

\DeclareMathOperator*{\argmin}{arg\,min}

\usepackage{ulem}

\newif\ifbulletlist
\newif\iftext
\texttrue 

\usepackage[affil-it]{authblk} 
\usepackage{etoolbox}
\usepackage{lmodern}

\makeatletter
\patchcmd{\@maketitle}{\LARGE \@title}{\fontsize{16}{19.2}\selectfont\@title}{}{}
\makeatother

\begin{document}
	
	\title{
		 Optimal Task Offloading with Firm Deadlines\\
		for Mobile Edge Computing Systems
	}

    \author{
    	Khai Doan,
    	Wesley Araujo,
    	Evangelos Kranakis,
    	Ioannis Lambadaris,
    	Yannis Viniotis,
	and Wonjae Shin
    	\thanks{This work was supported in part by a grant from the Natural Sciences and Engineering Research Council (NSERC) of Canada and Ericsson Canada, and in part by the National Research Foundation of Korea (NRF) under Grant No. RS-2025-00562095.}
    	\thanks{
    		Khai Doan, Wesley Araujo, and Ioannis Lambadaris are with
    		Carleton University, Department of Systems and Computer Engineering, Ottawa, ON, Canada.
    		(e-mail: \{khaidoan@sce, wesleyaraujo@cmail, ioannis@sce\}.carleton.ca). 
    	}
    	\thanks{
    	  Evangelos Kranakis is with
    		Carleton University, School of Computer Science, Ottawa, ON, Canada. (e-mail: kranakis@scs.carleton.ca).
    	}
    	\thanks{
    		Yannis Viniotis is with North Carolina State University, Department of Electrical and Computer Engineering, Raleigh, NC, USA. (e-mail: candice@ncsu.edu).
    	}
	\thanks{W. Shin is with Korea University, School of Electrical Engineering, Seoul, South Korea (e-mail: wjshin@korea.ac.kr).
	}
    }

\providecommand{\keywords}[1]{\textbf{\textit{Index terms---}} #1}

	
	\maketitle
	
	
	\begin{abstract}
		Under a dramatic increase in mobile data traffic, a promising solution for edge computing systems to maintain their local service is the task migration that may be implemented by means of {Autonomous mobile agents (AMA)}. In designing an optimal scheme for task offloading to AMA, we define a system cost as a minimization objective function that comprises two parts. First, an offloading cost which can be interpreted as the cost of using computational resources from the AMA. Second, a penalty cost due to potential task expiration. To minimize the expected (time-average) cost over a given time horizon, we formulate a {Dynamic programming (DP)}. However, the DP Equation suffers from the well-known {\it curse of dimensionality}, which makes computations intractable, especially for infinite system state space. To reduce the computational burden, we identify three important properties of the optimal policy and show that it suffices to evaluate the DP Equation on a finite subset of the state space only. We then prove that the optimal task offloading decision at a state can be inferred from that at its \emph{adjacent} states, further reducing the computational load. We present simulations to verify the theoretical results and to provide insights into the considered system.
	\end{abstract}

	%
	%
	

\section{Introduction}

	\iftext {
	{Mobile edge computing (MEC)
	and Mobile cloud computing (MCC)} are important paradigms in addressing the limited computational capability of mobile devices. In particular, MEC is an architecture that brings computational power and data storage closer to the network edge, where end-users or devices generate data. Instead of routing data to a distant centralized data center, MEC enables data processing at local edge servers, which are often integrated with base stations or located near user equipment. This proximity to data sources significantly reduces latency, enhances real-time processing, and increases the responsiveness of applications, making MEC particularly beneficial for latency-sensitive and bandwidth-intensive applications such as autonomous driving, augmented reality, and smart manufacturing \cite{mao2017survey}. MEC offers several key benefits, including lower latency, reduced network congestion, and enhanced data privacy and security. By processing data locally, MEC minimizes the amount of information that needs to travel through the broader network, reducing delays and offloading traffic from core networks. This not only improves the performance of real-time applications but also optimizes resource usage across the network. Additionally, MEC's localized data processing enhances privacy by allowing sensitive data to remain closer to the end-user, thus reducing exposure risks that may arise from transmitting data to distant servers. Overall, MEC provides a robust framework to support the increasing demands of modern digital services and the growing proliferation of connected devices \cite{8060526,9039672}.
 
    MCC is a technology that combines mobile computing with cloud computing to enhance resource-constrained mobile devices, enabling them to handle high-performance tasks by leveraging cloud-based resources. By offloading computationally intensive tasks and large data processing requirements from mobile devices to powerful cloud servers, MCC addresses the limitations of mobile devices, such as limited battery life, processing power, and storage. This integration enables applications to perform complex functions, such as data analysis and real-time processing, without overburdening the device itself \cite{dinh2013survey}. The benefits of MCC are substantial, particularly in terms of scalability, flexibility, and resource efficiency. MCC allows applications to scale seamlessly since cloud resources can be allocated based on demand, enhancing user experience without requiring hardware upgrades on mobile devices. It also enables mobile applications to function with reduced energy consumption, as computational tasks are shifted to the cloud, extending device battery life. Additionally, MCC supports a wide range of applications, from mobile health monitoring to mobile gaming and social networking, providing users with high-quality, reliable services regardless of device constraints. By leveraging MCC, developers can deploy feature-rich applications that offer performance comparable to traditional desktop environments, supporting increasingly complex and data-driven mobile applications \cite{6553297,7057552}.
    
\comment{
 In MEC, a server is placed
	physically near the mobile device for computational tasks to be
	offloaded to the MEC-server for remote computation.
	MEC is similar to MCC; however, the server in the latter case is not necessarily physically close to the device.
	MEC is more suitable if the network latency or network congestion
	is a problem. In general, there are two scenarios when MEC is most appropriate:
	when the application has real-time constraints or when the user/wireless device has limited
	resources such as memory, storage, CPU, etc.
	Surveys on MEC and MCC can be found in \cite{mao2017survey} and \cite{dinh2013survey}, respectively.
 }
	}
	\fi

 \subsection{Related Works}

	\iftext {
	Exploiting the advantage of offloading systems, computational services requested by users can be either processed at local servers (e.g., MEC) or offloaded onto remote mobile servers (e.g., MCC) that have more computational capability \cite{ 8556474, 8360849, 8314121, 8057196}. This feature not only improves users' quality of experience by reducing the processing delay, latency, and power consumption but also allows different types of applications and services to be deployed on devices with low computational capability. Due to significant advances in practical applications, analyzing MEC and MCC systems has been attracting a lot of attention in the research literature. Different edge computing architectures, categorization of past research, and discussions on key models are covered in the survey of \cite{LIN2020102781}. Moreover, various approaches for task offloading such as optimization techniques, {Markov decision processes (MDPs)}, game theory, and machine learning are also explored in this survey along with future research directions and challenges. Additionally, key issues and methods in edge cloud offloading, proposed destination-based offloading model, load balancing, mobility, partitioning, and granularity are conveyed in \cite{3284387}. This survey also discusses important factors like environment constraints, cost models, and user configurations, providing a comprehensive overview of the current studies, with future research discussed based on emerging technologies.

    {\subsubsection{Missing Key Aspects of Edge Computing Systems in Existing Studies} In edge computing systems, optimizing task offloading requires accounting for several critical factors that influence performance and reliability. First, the randomness in the connection between both the remote server and the local device introduces uncertainty in task execution, as fluctuating network conditions can impact the feasibility and cost of offloading. Second, firm deadlines play a crucial role in real-time applications where task expiration may happen due to insufficient resource and unavailable task migration. The expiration of tasks incurs a significant penalty cost representing degraded quality of service, lost business opportunities, reduced user satisfaction, or operational inefficiencies. This raises an essential need to develop strategies that balance offloading costs and the risk of task expiration. Finally, the optimal stochastic control (dynamic programming) framework is well-suited for addressing the sequential decision-making nature of task offloading, where each decision affects future system states and overall performance. Jointly considering these three aspects can capture key practical challenges in edge computing, providing a more comprehensive and theoretically grounded approach to task offloading optimization. However, these issues have not been adequately addressed in existing works which typically assume that task offloading can always be controlled. As examples, some of the related research works are mentioned below.}

    Due to the importance of energy efficiency in implementing a real-life offloading system, it is considered as the optimization objective in various existing studies. For instance, \cite{zhang2014collaborative} addresses energy conservation for mobile applications through collaborative task execution. The problem is modeled as a constrained stochastic shortest path problem on an acyclic graph, considering both hard and probabilistic time deadlines. This study derives an optimal one-climb policy, an enumeration algorithm, and necessary conditions for optimal task scheduling. Based on application profiles and channel status, a practical rule of thumb is introduced together with an adapted Lagrangian relaxation-based aggregated cost algorithm for energy-efficient scheduling. The mentioned collaborative task execution enables mobile devices to consume significantly less energy, verified by simulation. From a different approach, \cite{huang2021deadline} investigates computational offloading in a real-time MEC environment where tasks have hard deadlines and formulate the problem as a partially observable MDP. In this work, task offloading and scheduling algorithms are proposed, committing to meeting real-time deadlines even with partial system observability. Similarly, an MDP-based approach is proposed in \cite{van2017optimization} to enable mobile users to make optimal offloading decisions in an ad-hoc mobile cloud environment. The effects of cloudlets' distance, mobility, and time-varying channel properties on the success of offloading actions are conveyed. This study offers an optimal offloading policy that determines how many tasks the mobile user should process locally and how many to offload to each nearby mobile cloudlet, with the goal of maximizing the user's overall utility while minimizing energy consumption and offloading costs. Furthermore, simulation across various scenarios is provided to show that the proposed method outperforms baseline schemes. Besides, the coexistence of MEC and MCC in a single system is worth investigating in this research area \cite{du2017computation}. In terms of this idea, the authors of \cite{du2017computation} consider a model constituted by a set of users, a fog node, and a remote cloud server. Tasks can be offloaded from users to either the fog node or the cloud server and from the fog node to the server. This work focuses on minimizing the weighted cost of task processing delay and energy consumption at the users' devices, for which a low-complexity suboptimal algorithm is proposed. Similarly, in the model of \cite{van2018deep}, the user searches for the cloudlets that are currently in its device-to-device communication range to offload its tasks. The authors formulate the offloading decision-making problem as an MDP and use a Deep Q-Network to learn the task offloading policy. This model takes the current system state, including the user's current location, the available resources of the cloudlets, and the user's computation task as input to make an offloading decision.
    
    {\subsubsection{Lack of Analytical and Theoretical Insights in Task Offloading in Existing Studies} Existing studies explore various task offloading models and aim to develop efficient offloading strategies, highlighting the importance of optimizing task offloading in edge computing systems. However, many of these works focus on heuristic or sub-optimal algorithms for complex models with multiple combined objectives. In addition, many of them focus on implementing deep learning-based decision-making models that lack theoretical comprehensiveness. While these approaches address practical challenges, they often overlook the theoretical foundations and fundamental properties of optimal offloading policies. Understanding these aspects is crucial for building a solid foundation that can guide the development of more advanced and efficient edge computing systems. Hereafter, we provide some examples of those existing research works.}

    Modeling a utility function that incorporates three factors - users' satisfaction, total computation amount, and energy consumption overhead - presents an interesting approach to optimizing both service quality and system efficiency \cite{fragkos2020artificial}. This type of function increases with respect to satisfaction, is characterized by tasks processed remotely, and decreases with respect to the other two factors. {\cite{9633644} addresses the task offloading problem in a multiuser, multiserver MEC system. It optimizes delay, energy consumption, and cost through joint task offloading, power allocation, and resource management. The objective prioritizes improving the offloading benefits of the worst-performing mobile users, formulating the problem as a multiobjective constrained optimization and a sub-optimal algorithm is proposed for the considered objective. From a different perspective, joint optimization of task offloading and resource allocation can be formulated as a mixed-integer non-linear programming problem \cite{9079564}. In this context, an offline solver called \textit{task learning-based feedforward neural network} is presented where the pre-trained model enables fast, low-cost online inference. \cite{9888096} proposes HyTOS, a hybrid task offloading scheme for urban Internet of vehicles (IoVs), which integrates vehicle-to-edge and vehicle-to-vehicle offloading to minimize task delay and energy consumption while leveraging distributed vehicle resources. A Deep Q-Network-based offloading method is introduced to meet computing requirements and ensure task delay constraints. Simulation results show that HyTOS outperforms single-scenario offloading methods and achieves better overall performance than game-theory-based hybrid approaches in terms of task delay and success rate. The study highlights the potential of HyTOS for dynamic, delay-constrained IoV scenarios, with future work focusing on incorporating vehicle mobility into the offloading model. The study in \cite{10048492} proposes a multi-agent meta-reinforcement learning strategy for efficient task offloading in MEC systems where mobile devices are capable of wireless power transfer. A system architecture integrating task offloading, energy harvesting, and meta-reinforcement learning is developed, where a meta-learner and base learner are deployed on the edge server and mobile devices, respectively. However, limitations include assumptions of stable networks and sufficient resources, and future work will explore broader MEC architectures such as cloud–edge–device and device-to-device frameworks. \cite{10234672} aims to minimize the distances between UAVs and users to ensure quality of service. This study takes into account the inter-dependency between sub-tasks in a given task and, by obtaining a distribution of ground devices via a DT network, constructs a heuristic greedy algorithm and a learning-based algorithm for task offloading. The former method serves as a low-complexity solution, while the latter requires an accurate training procedure and can be considered a more precise solution.}

	}
	\fi

	\iftext {

	}
	\fi

	Although the above studies address complex models with practical constraints, they often provide limited insight into the fundamental principles of edge computing systems and optimal task offloading strategies. The focus on heuristic or sub-optimal approaches in intricate settings can obscure the underlying system dynamics and key properties of an optimal solution. A deeper theoretical understanding is essential for developing more robust and efficient offloading policies that can generalize across different edge computing scenarios. Furthermore, despite several offloading algorithms proposing improvements for the  system performance under different contexts, to the best of our knowledge, there is a lack of studies for the characterization of the optimal policy for task offloading to enhance the foundation analysis and insight into system performance.
	
	\subsection{Motivation, Novelty, and Main Contributions}
	\iftext {
	{While existing research explores task offloading within complex, multifaceted models that reflect realistic system environments, there remains significant value in analyzing simpler, foundational setups. Moreover, a basic model that allows for theoretical insights is obscured in more complex configurations. Besides, a foundational analysis not only clarifies core dynamics within the task offloading problem but also serves as an essential basis for developing enhanced strategies in more intricate scenarios. Inspired by this motivation, our study focuses on a basic task offloading system that retains essential characteristics of real-world systems, such as task urgency, system load, computational capacity, and offloading capability. This approach allows us to examine core dynamics in a manageable setting, capturing critical aspects of task offloading performance while offering insights that can inform and support the development of more advanced models.} 
 
	In addition, our model also stands out from prior studies due to its unique \textit{combination of}: (a) the randomness in the connection between both the remote server and local device, (b) tasks that have firm deadlines, and (c) the formulation of the problem in the context of  an optimal stochastic control (dynamic programming) framework.
	As is well-known in DP formulations,  computing the solution of the DP Equation may become computationally prohibitive (also  known as the ``dimensionality curse'' of DP). In our earlier work \cite{DAKLV_conf}, we concisely presented the optimal task offloading scheme and its practical implementation. This extended version provides additional findings to the examined problem. Notably, we explore the convexity of the cost function, present a formal property governing the optimal offloading decision, and establish rigorous conditions that precisely determine when task offloading is required. The results are presented in Lemma \ref{Lem:Convexity}, Lemma \ref{Lem:Convex2L*}, and Proposition \ref{Pro:ONO_conds}, respectively. Furthermore, this work offers a comprehensive mathematical exposition, providing detailed proofs for theorems, lemmas, and propositions, enhancing the completeness and rigor of the presented material in \cite{DAKLV_conf}. In summary, the main contributions of this work can be summarized as follows.
	\begin{itemize}
		\item We provide the comprehensive optimal policy's structure by leveraging the relation between adjacent system state vectors where the concept of adjacency is formally presented. The finding is conveyed in a property provided in Theorem \ref{Theo:adjacent_L*}.
		\item We further study the structure of the optimal offloading scheme and present our finding in Theorem \ref{Theo:L*_thesmallest} which is built upon two crucial properties. Firstly, we prove the convexity of the achieved cost as a function of the offloading decision in Lemma \ref{Lem:Convexity}. Secondly, in Proposition \ref{Pro:ONO_conds}, we present the condition to determine \textit{non-offloading} and \textit{offloading} states \footnote{Non-offloading states refers to system states at which the task offloading will not be performed by the optimal policy. In contrast, a certain number of tasks will be offloaded at offloading states by the optimal policy. These concepts are formalized in Definition \ref{Def:Of_NOf}.}.
		\item Our third contribution facilitates the determination of the minimum expected (time average) cost by exclusively employing the computationally intensive DP Equation only on a finite set of states, known as the \textit{lean states}. This stands in contrast to the previous need to apply the DP Equation to an unbounded range of states. Subsequently, the optimal cost for any arbitrary state can be calculated via a considerably more straightforward algebraic computation. This finding is presented in Proposition \ref{Pro:SemiRedState}.

        \item Lastly, through simulation results, we verify the correctness of the derived theoretical results and demonstrate key advantages of the proposed optimal task offloading algorithm over baseline schemes.
	\end{itemize}}\fi
 
\subsection{Organization and Notation Simplification}

	\iftext {
	The subsequent sections of this paper are structured as follows: In the following section, we introduce the system model. In Section \ref{Sec:DP}, we present the formulation of the DP Equation to minimize the expected (time average) cost. In Section \ref{Sec:RS}, the concepts of a \textit{reduced} and \textit{lean state} space are introduced, leading to a significant reduction of the computational burden associated with calculating the optimal cost in our model. We study intrinsic properties inherent in the optimal policy and the minimum expected time-average cost in Section \ref{Sec:OptimalPolicyProp}. This in-depth examination leads to the characterization of the explicit optimal policy in Section \ref{Sec:OptimalPolicy}. Section \ref{Sec:NumericalResults} employs numerical results to illustrate and substantiate our findings. A conclusion of our work and potential future research directions are given in Section \ref{Sec:Conclusion}. Proofs for all the theorems, lemmas, and propositions are provided in the Appendix.
	}
	\fi

	\iftext {
	In this article, certain notations have been simplified to enhance presentation and comprehension. However, within the Appendix, more comprehensive notational forms are necessary to facilitate the presentation of the proofs therein. To prevent any confusion among readers, we will clearly indicate the notation changes at the beginning of the Appendix.
	}
	\fi

	\iftext {
	\section{System Model} \label{Sec:SystemModel}

	\subsection{System Components}

	\begin{figure}[t]
		\centering
		\includegraphics[scale=0.3]{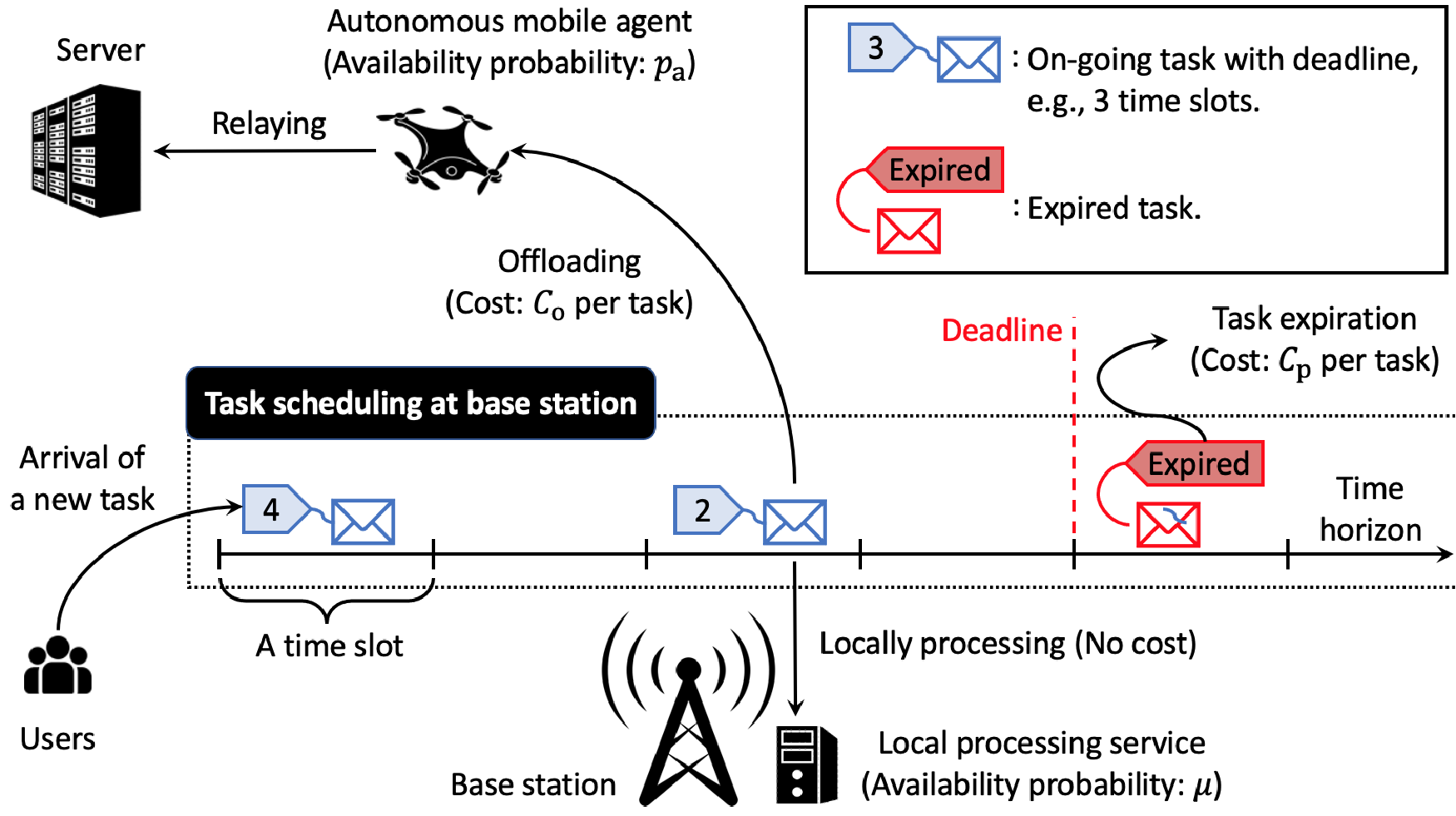}
		\caption{System model illustration where user tasks arrive at the BS with individual deadlines. The BS can process tasks locally or offload them to a remote server via an AMA, which intermittently arrives as a relay for task offloading.}
		\label{fig:SM}
	\end{figure}

    \ifbulletlist
	{\color{red}
	\begin{enumerate}
	    \item Description of how system operates and representation of system states with vectors $\mathbf{s}$
	\end{enumerate}
	}
	\fi

	The system we consider is depicted in Fig.~\ref{fig:SM}; it consists of a BS providing computational service to users in the area. Tasks sent by users will arrive at the BS's buffer and each of them is associated with a deadline indicating the number of time slots within which the task needs to be handled, otherwise, it will expire. There are two ways to handle a task: locally processing it at the BS or offloading it to a remote server. These two operations will be described in detail below. We assume that the maximum deadline for any task is $N$, a fixed positive integer. Our considered system operates in discrete time over $T$ time slots.

    The assumption that each task has an associated strict deadline reflects practical requirements in many real-world applications where tasks need to be processed within a specific time frame to ensure usability and relevance. 
	
	We define the system state vector at time slot $t$ by
	\begin{align}\label{state_vec}
		\mathbf{s}_t = \left(n_1^{(t)}, n_2^{(t)}, \ldots, n_N^{(t)}\right),  t \in \left\{0, 1, \ldots, T-1 \right\},
	\end{align}
	where $n_i^{(t)}\in \mathbb{N} = \left\{0, 1, 2, \ldots \right\}$, for $i = 1, 2, \ldots, N,$ represents the number of tasks having deadline $i$, buffered at the BS at time slot $t$. The state space for this system is $\mathbb{N}^N$, which is an $N$-dimensional infinite state space. There are a number of different events that take place and trigger the system to transit from one state to the next across time slots. Those events are listed in order as follows.
\begin{enumerate}
	\item \textbf{Connection to a remote server:} An external server, positioned beyond the communication range of the BS, provides a task offloading service. To bridge this spatial divide, an Autonomous Mobile Agent (AMA) functions as an intermediary. The AMA, characterized by its random arrival within the area, facilitates connectivity between the BS and the distant server. We denote by $p_{\sf a}$, the probability of the AMA's presence in each time slot. By leveraging the AMA, the BS in Fig.~\ref{fig:SM} can offload tasks to the server at cost $C_{\sf o} > 0$ per task. In a practical system, this offloading cost can be referred to as a combination of different types of costs such as cost for resource utilization at the server, network usage cost, and energy consumption for data transmission. Due to high computational resources, every offloaded task is guaranteed to be executed within its deadline by the server. The task offloading is a two-hop process, first from the BS to the AMA, and subsequently from the AMA to the server. We assume that the delay of the latter hop is negligible compared to the duration of a time slot as the AMA can freely reduce the distance between itself and the server.

    ~~In the proposed model, the involvement of the AMA in the task offloading operation is to reflect a practical reality that a remote server may not be continuously available due to resource demands from other services and applications. By assuming that the AMA arrives at the area with a probability $p_{\sf a}$, we capture the intermittent availability of offloading services, simulating the uncertainty of server resources in real-world scenarios. This approach not only models the stochastic nature of resource availability but also provides a realistic foundation for analyzing task offloading strategies under uncertain conditions.
    
	\item \textbf{The deadline shifting}: As the system transitions from one time slot to the next, the deadlines of all tasks are decremented by one unit. Any task that reaches a deadline of 0 expires and incurs a penalty cost $C_{\sf p} >C_{\sf o}$. In industries with service-level agreements, this penalty cost represents a compensation cost to the user due to service disruption or loss of functionality. It may also be referred to as opportunity cost resulting from lost business opportunities, lower system efficiency, or reduced overall performance. {We assume no cost for local processing to encourage the use of BS's resources, which are dedicated to user tasks, while the remote server may be shared with other services. The penalty cost $C_p$ reflects only the consequence of unmet deadlines and does not include local processing cost, as local processing is considered the preferred and cost-free option in our model.} 
	\item \textbf{The arrival of a new task}: There are incoming tasks from users throughout the system's operation at a rate of at most one new task per time slot. In each time slot, the probability of a new task arriving with a deadline of $i$ is represented by $p_i$; the probability of no task arrival is denoted as $p_0$. Hence, the task arrival probability vector is given by $\mathbf{p} = (p_0, p_1, \ldots, p_N)$ where $\sum_{i=0}^N p_i = 1$. We assume that task arrival events across distinct time slots remain independent of one another.
	\item \textbf{The local processing service}: The BS, in our model, is constrained to have limited computational capability. Therefore, it is assumed that at each time slot, at most one task can be processed with probability $\mu$.
\end{enumerate}
	}
	\fi

    The assumption of at most one task arriving per time slot is a simplification that facilitates theoretical analysis while still capturing task urgency and system load levels. Specifically, task urgency is reflected by each task's deadline. In addition, by adjusting the set of values of $p_i, i = 0, 1, \ldots, N$, we simulate different scenarios of system load without complicating the model, allowing for clearer insights into the task scheduling problem.

	\iftext{
	Fig. \ref{fig:order-events} summarizes the random events described above. The state $\mathbf{s}^{\sf in}$ in this figure is an \textit{intermediate state} of the system that will be discussed in Subsection \ref{subsec:representation_vectors}.

    {We would like to note that the cost for local resource consumption is simplified in our model for the sake of analysis. In subsequent section, we will show that only the DP equation needs to be modified to capture this type of cost, and the propose optimal offloading policy remains applicable.}

    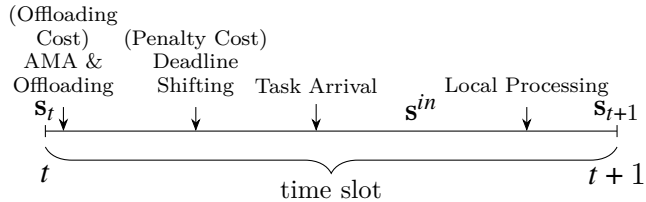
\begin{figure}[t]
		\centering
		\begin{tikzpicture}[>=Stealth,scale=0.8]
			\draw (0,1) -- (9.5,1);
			\draw (0,0.9) -- (0,1.1);
			\draw (9.5,0.9) -- (9.5,1.1);
			\node[above] at (0,1) {$\mathbf{s}_t$};
			\node[above] at (9.5,1) {$\mathbf{s}_{t+1}$};
			\draw[decorate,decoration={brace,amplitude=10pt,mirror,raise=6pt}]
			(0,1) -- node[black,below=0.5cm] {\small time slot} (9.5,1);
			\node[below=0.3cm] at (0,1) {$t$};
			\node[below=0.3cm] at (9.5,1) {$t+1$};
			\draw[<-] (0.3,1) -- (0.3,1.4) node[above,text width=2cm,align=center,font=\scriptsize\linespread{1}\selectfont] {(Offloading Cost)\\ AMA \& Offloading};
			\draw[<-] (2.5,1) -- (2.5,1.4) node[above,text width=4cm,align=center,font=\scriptsize\linespread{1}\selectfont] {(Penalty Cost)\\ Deadline\\ Shifting};
			\draw[<-] (4.5,1) -- (4.5,1.45) node[above] {\scriptsize Task Arrival};
			\node[above] at (6.25,1) {$\mathbf{s}^{in}$};
			\draw[<-] (8,1) -- (8,1.4) node[above] {\scriptsize Local Processing};
		\end{tikzpicture}
		\caption{Illustration of events occurring in a time slot. Among these events, the AMA's arrival, task arrival, and local processing are random events, while offloading and deadline shifting are deterministic events. $\mathbf{s}^{\sf in}$ is an intermediate state defined in Subsection \ref{subsec:representation_vectors}.}
        \label{fig:order-events}
	\end{figure}
	}
	\fi
	
	\subsection{Offloading Policy and Offloading Decision}

	\iftext{
	In our model, all tasks exhibit an identical offloading cost $C_{\sf o}$ and incur the same penalty $C_{\sf p}$ upon expiration. Trivially, it is optimal to process the task with the smallest deadline when the local processing service is available. Besides that, when the AMA is accessible/the local processing is available, it is optimal to offload/process  the most imminent tasks first. In other words, tasks will be offloaded or locally processed in an ascending order of their deadlines. Therefore, for a given system state $\mathbf{s}_t$ as in Eq. (\ref{state_vec}), we define an \textit{offloading decision} by an integer $L_t \ge 0$ representing the number of tasks that will be offloaded from $\mathbf{s}_t$ following the ascending order of deadlines. The set of feasible offloading decisions associated with state $\mathbf{s}_t$ is given by:
	\begin{align}\label{L_definition}
		L_t \in \mathbb{L}\left(\mathbf{s}_t\right) = \left\{0, 1, 2, \ldots, \sum_{i=1}^N n_i^{(t)} \right\}.
	\end{align}
	}
	\fi

	\iftext{
	From the above definition of  offloading decisions $L_t$, we define a task offloading policy as a rule that determines $L_t$ for every given system state $\mathbf{s}_t$ in time slot $t$. As we have previously mentioned that it is optimal to locally process and offload the most imminent tasks first. When the AMA arrives at time slot $t$ and an offloading decision $L_t$ is made, then, $L_t$ most imminent tasks will be offloaded.
	}
	\fi
	
	\subsection{Expected Instantaneous Cost}

	\iftext{
	Let us assume that, in a time slot $t$, a system state $\mathbf{s}_t$ is encountered and an offloading decision $L_t$ is made. We note that the offloading decision $L_t$ is made when the system state $\mathbf{s}_t$ is realized at the beginning of a time slot as in Fig. \ref{fig:order-events}. However, the offloading decision can only be applied in the presence of the AMA. As a result, when the AMA arrives, $L_t$ tasks can be offloaded at the cost of $C_{\sf o}L_t$. {Also in this case, if $L_t < n_1^{(t)}$, there are $n_1^{(t)} - L_t$ tasks having deadline 1 that will expire at the end of time slot $t$, incurring a cost of $C_{\sf p}(n_1^{(t)} - L_t)$.} Therefore, the \textit{instantaneous cost} incurred when the AMA arrives is defined by
    \begin{align}\label{eq:instant_cost2}
        \mathcal{C}^{\text{A}}\left(\mathbf{s}_t, L_t \right) = C_{\sf o}L_t + C_{\sf p}\max\left(n_1^{(t)} - L_t, 0 \right).
    \end{align}
    
    In the case when the AMA does not arrive, the offloading decision cannot be applied; hence, the instantaneous cost is the penalty cost due to the expiration of $n_1$ tasks having deadline 1, i.e.,
    \begin{align}\label{eq:instant_cost3}
        \mathcal{C}^{\overline{\text{A}}}(\mathbf{s}_t) = C_{\sf p}n_1^{(t)}.
    \end{align}
    
    As the AMA's arrival occurs with probability $p_{\sf a}$, the \textit{expected instantaneous cost} can be defined by
    \begin{align} \label{eq:instant_cost}
        \mathcal{C}\left(\mathbf{s}_t, L_t \right) = p_{\sf a}\mathcal{C}^{\text{A}}\left(\mathbf{s}_t, L_t \right) + \left(1 - p_{\sf a}\right)\mathcal{C}^{\overline{\text{A}}}\left(\mathbf{s}_t\right)
    \end{align}
    
    {We note that in our model, the cost of the local processing service is assumed to be significantly smaller than the cost for access resources in the remote server, i.e., the task offloading cost $C_{\sf o}$. Therefore, the local service cost is simplified; hence, the instantaneous cost is made up by only the task expiration cost $C_{\sf p}$ and offloading cost $C_{\sf o}$.}
    
	In the next section, we define the expected (time average) cost over a time horizon $T$ that we aim at minimizing. In addition, we formulate a DP Equation in order to derive an optimal offloading decision with respect to every given initial state.
	}
	\fi
    
	\section{Dynamic Programming Formulation } \label{Sec:DP}

	\subsection{Expected Time-Average Cost}

    \iftext{
    The expected time-average cost  over a horizon $T$ is given by
	\begin{equation}\label{eq:sum_cost}
         \frac{1}{T}
		\sum_{t=0}^{T-1} \mathbb{E}[\mathcal{C}\left(\mathbf{s}_t, L_t\right)].
	\end{equation}
    
  For the rest of this work, we will refer to the cost defined above as the \textit{average cost}. We aim to determine the optimal offloading policy that minimizes the above average cost for a given time horizon. For ease of presentation, we will drop the notation dependence on $t$. In the sequel, we will denote by $\mathbf{s} = \left(n_1, n_2, \ldots, n_N\right)$ the system state in a considered time slot and by $L$ the offloading decision on state $\mathbf{s}$.
    }
    \fi

	\subsection{Representation Vectors of System Components}\label{subsec:representation_vectors}

	\iftext{
	Consider a state $\mathbf{s}$ and an offloading decision $L$. Suppose a new task arrives with a deadline $k$. If  local processing is available, the system will transit to a new state denoted by  $\mathbf{s}'_{Lk}$; if not, the transitioned state will be another state denoted by $\mathbf{s}''_{Lk}$. 
	These state definitions facilitate the analysis of the DP Equation that will follow. They are formally defined in Eq. (\ref{next_state}) in this section. 	We need to introduce some additional notations to that effect. 
	}
	\fi

	\iftext{
Before doing this, we  provide the following example:
	\begin{example}
		Let $\mathbf{s} = (0,1,2,0,1)$ be the system state which is at the beginning of a time slot as illustrated in Fig. \ref{fig:order-events}. The first event shown in the figure is ``AMA \& Offloading''. We assume the AMA is available and $L=2$ tasks with the most imminent deadlines are to be offloaded. Therefore, the resulting intermediate state after offloading is $(0,0,1,0,1)$. The next event is deadline shifting to account for the new deadlines of the tasks in the next time slot, resulting in state $(0,1,0,1,0)$. Following Fig. \ref{fig:order-events}, we assume a new task with deadline 3 arrives, resulting in the intermediate state $(0,1,1,1,0)$. Finally, the next event is local processing, which processes the most imminent task, resulting in system state  $\mathbf{s}'_{23} = (0,0,1,1,0)$. If the local processing does not take place, the obtained system state would be $\mathbf{s}_{23}^{''}=(0, 1, 1, 1, 0)$. $\square$
	\end{example}

	}
	\fi

    \iftext{
    Vector $\mathbf{o}\left(\mathbf{s}, L\right) = \left(o_1, \ldots, o_N\right)$ represents the \textit{task offloading}. Let $d'$ be the index satisfying $d' = 1$ and $\sum_{i=1}^{d'} n_i \ge L$ or $d \ge 2$ and $\sum_{i=1}^{d'-1} n_i < L \le \sum_{i=1}^{d'} n_i$.
\comment{$d'$ is defined by:
\begin{itemize}
	\item If $L < n_1$, then, $d' = 1$.
	\item Otherwise, $d'$ satisfies the inequalities
		 \begin{align}\label{ineq:d'_cond2}
		        \sum_{i=1}^{d'-1} n_i < L \le \sum_{i=1}^{d'} n_i.
		    \end{align}
\end{itemize}}
   Then, $\mathbf{o}\left(\mathbf{s}, L\right)$ is defined as follows:
    \begin{itemize}
        \item Case 1: If $L < n_1$:
        \begin{align}
            o_i = 
            \begin{cases}
                L &, \text{ if } i = 1,\\
                0 &, \text{ if } 2 \le i \le N.
            \end{cases}
        \end{align}
    
        \item Case 2: Otherwise: \comment{If $d'$ satisfies Ineqs. (\ref{ineq:d'_cond2}):}
        \begin{align}
            o_i = 
            \begin{cases}
                n_i &, \text{ if } 1 \le i \le d'-1,\\
                L -  \sum_{j=1}^{d'-1} n_j &, \text{ if } i = d',\\
                0 &, \text{ if } d' + 1 \le i \le N.
            \end{cases}
        \end{align}
    \end{itemize}
    
    Let $\mathbf{a}_k = \left(a_1, \ldots, a_N \right)$ represent the task arrival vector where the arrived task has deadline $k$. $\mathbf{a}_k$ is defined as a one-hot vector with $a_k = 1$, $a_i=0$ for $i \ne k$. Then, the \textit{intermediate state}, $\mathbf{s}^{\sf in}$, of the system after $L$ most imminent tasks have been offloaded from $\mathbf{s}$, the deadline shifting has been performed, and the task arrival event has been realized, can be defined as follows:
    \begin{align*}
        \mathbf{s}^{\sf in} = \text{shift}\left(\mathbf{s} - \mathbf{o}\left(\mathbf{s}, L\right)\right) + \mathbf{a}_k, \text{ for } k \in \{0, 1, \ldots, N \}
    \end{align*}
    where the function $\text{shift}\left(\cdot\right)$ performs deadline shifting on the given vector. If $\mathbf{s} = \left(n_1, \ldots, n_N\right)$, $\text{shift}\left(\mathbf{s}\right)$ returns an $N$-dimensional vector $\left(n_2, \ldots, n_N, 0\right)$ in which the component $n_1$ is removed, representing task expiration. The position of $\mathbf{s}^{\sf in}$ in a time slot is shown in Fig. \ref{fig:order-events}. Let $\mathbf{l} = \left(l_1, \ldots, l_N\right)$ represent the local processing vector defined as follows:
	\begin{itemize}
		\item If $\mathbf{s}^{\sf in} \ne (0, \ldots, 0)$, and assuming $d$ is the deadline of the most imminent task in $\mathbf{s}^{\sf in}$:
			\begin{align}
			        l_i =
			        \begin{cases}
			            1 &, \text{ if } i = d,\\
			            0 &, \text{ otherwise}.
			        \end{cases}
			    \end{align}
		\item If $\mathbf{s}^{\sf in} = (0, \ldots, 0)$: $\mathbf{l}=(0, \ldots, 0)$.
	\end{itemize}
	}
	\fi

    \iftext{
    ~~The system state transition is defined as:
	\begin{align} \label{next_state}
        \mathbf{s}'_{Lk} &= \mathbf{s}^{\sf in} - \mathbf{l}, \text{ and } 
        \mathbf{s}''_{Lk} = \mathbf{s}^{\sf in}.
	\end{align}
	In Fig. \ref{fig:order-events}, assuming that $L$ most imminent tasks have been offloaded from the initial state and a new task with deadline $k$ has arrived ($k=0$ when there is no task arrival), $\mathbf{s}_{t+1} = \mathbf{s}'_{Lk}$ if the local processing happens, and $\mathbf{s}_{t+1} = \mathbf{s}''_{Lk}$ if the local processing does not happen.
	}
	\fi
	
	\subsection{Dynamic Programming Equation}	
	\iftext{
	To this end, let $J_T\left(\mathbf{s}\right)$ denote the minimum expected cost over $T$ time slots for a given initial state $\mathbf{s}$. We have the following DP Equation.
	\begin{align}
		\label{DP_Eq}
		\begin{split}
			&J_{T}\left( \mathbf{s} \right) = ~ \underset{L \in \mathbb{L}\left(\mathbf{s}\right)}{\min}~ \left\{\mathcal{C}\left(\mathbf{s}, L\right) + G_{T-1}\left(\mathbf{s}, L\right)\right\},
		\end{split}
	\end{align}
	with $G_0(\mathbf{s}, L) = 0$, and for $T \ge 1$,
	\begin{align} \label{G_func}
		G_T\left(\mathbf{s}, L\right) = p_{\sf a} G_T^{{\text{A}}}\left(\mathbf{s}, L\right) + \left(1-p_{\sf a}\right)G_T^{\overline{\text{A}}}\left(\mathbf{s}\right),
	\end{align}
	where 
	\begin{align}
		G_T^{{\text{A}}}\left(\mathbf{s}, L\right) &=   \mu \sum_{k=0}^N p_k J_T\left(\mathbf{s}'_{Lk} \right)   + \left(1-\mu\right)\sum_{k=0}^N p_k J_T\left(\mathbf{s}''_{Lk}\right), \label{G_AMA} \\
        G_T^{\overline{\text{A}}}\left(\mathbf{s}\right) &=   \mu \sum_{k=0}^N p_k J_T\left(\mathbf{s}'_{0k} \right)   + \left(1-\mu\right)\sum_{k=0}^N p_k J_T\left(\mathbf{s}'_{0k} \right). \label{G_NAMA}
	\end{align}
	$G_T^{{\text{A}}}\left(\mathbf{s}, L\right)$ denotes the minimum expected future cost, i.e., excluding the instantaneous cost in the current time slot, given that the AMA arrives at the current time slot and $L$ most imminent tasks are offloaded. $G_T^{\overline{\text{A}}}\left(\mathbf{s}\right)$  is defined similarly but without the arrival of the AMA; hence, it does not depend on the offloading decision $L$.
	}
	\fi

    {We highlight that in the equations \eqref{DP_Eq}-\eqref{G_NAMA}, the transitioned states in the next time slot, i.e., $\mathbf{s}_{Lk}^{'}$ and $\mathbf{s}_{Lk}^{''}$, depend on the state $\mathbf{s}$ in the current time slot and the offloading decision $L$. Therefore, the inter-dependency between time slots is conveyed in the relation between system states $\mathbf{s}$, $\mathbf{s}_{Lk}^{'}$, and $\mathbf{s}_{Lk}^{''}$.}
    
	\iftext{
 Eq. (\ref{DP_Eq}) is equivalent to
	\begin{align}\label{gen_U_noU}
		J_{T}\left(\mathbf{s}\right) = p_{\sf a}J_{T}^{\text{A}}\left(\mathbf{s}\right) + \left(1-p_{\sf a}\right)J_{T}^{\overline{\text{A}}}\left(\mathbf{s}\right),
	\end{align}
	in which
	\begin{align}\label{eq:JAMA}
		J_{T}^{\text{A}}\left(\mathbf{s}\right) = \underset{L \in \mathbb{L}\left(\mathbf{s}\right)}{\min}\left\{\mathcal{C}^{\text{A}}\left(\mathbf{s}, L\right) + G_{T-1}^{\text{A}}\left(\mathbf{s}, L\right) \right\},
	\end{align}
	and
	\begin{align}\label{eq:JNoAMA}
		J_{T}^{\overline{\text{A}}}\left(\mathbf{s}\right) = \mathcal{C}^{\overline{\text{A}}}\left(\mathbf{s}\right) + G_{T-1}^{\overline{\text{A}}}\left(\mathbf{s}\right).
	\end{align}
	}
	\fi

    {We adopt a DP approach because the task offloading decision in each time slot influences the system state in subsequent time slots, creating a temporal dependency. This sequential, state-dependent nature makes DP well-suited for optimizing the cumulative cost over a time horizon. In contrast, convex optimization methods are typically more appropriate for static problems without such dynamic state transitions.}

	\iftext{
		Subsequently, we take our initial step towards reducing the computational burden for solving the DP Eq. (\ref{DP_Eq}).
\comment{
	We observe that the DP Eq. (\ref{DP_Eq}) would result in high computational load due to the recursive nature. Therefore, in the next section, we introduce two types of system states called \textit{reduced states} and \textit{lean states} such that the consideration of the original infinite state space can be reduced to finite spaces of these states. We also present the concept of adjacency between states. Based on that, we derive important properties, helping to reduce the computational load.
}
	}
	\fi

	\section{Dynamic Programming Computational Load Reduction} \label{Sec:RS}

	\iftext{
	The recursive DP Eq. (\ref{DP_Eq}) in the previous section would result in a high computational load since it operates on countably infinite state space. In this section, we aim at reducing the computational load of the DP Equation by introducing two types of system states called \textit{reduced states} and \textit{lean states} so that the recursive computation in Eq. (\ref{DP_Eq}) occurs in a finite number of states. Throughout this work, we will use the term \textit{generic states} to refer to states that are neither reduced states nor lean states. For a generic state $\mathbf{s}$ and a lean state $\mathbf{s}^{(\ell)}$, the value of $J_T(\mathbf{s})$, in Eq. (\ref{DP_Eq}), can be computed with respect to $J_T(\mathbf{s}^{(\ell)})$ via a linear expression provided in Eq. (\ref{eq:gen2sem_eq}). As we will see the states $\mathbf{s}^{(\ell)}$ are finitely many.
	}
	\fi

	\subsection{Reduced State Space}\label{subsec:reduced_state}
	\iftext{
		We note that tasks having deadline 1 will expire at the deadline shifting event if they are not offloaded at the beginning of a time slot as illustrated in Fig.~\ref{fig:order-events}. In addition, there is at most one task can be locally processed per time slot. Therefore, for an initial state $\mathbf{s}$, there are at most $j-1$ tasks of $\mathbf{s}$ can be locally processed if $j$ time slots. As a result, there might be certain tasks that are {\em guaranteed to expire} if not offloaded; we will call such tasks \textit{excessive tasks}. For intuition, we provide the following example:
		\begin{example}
			Let us consider state $\mathbf{s} = (3, 3, 3, 3, 3)$ in time slot $t=0$ and assume that the AMA does not arrive in five consecutive time slots. According to the order of events presented in Fig. \ref{fig:order-events}, three tasks having a deadline of 1 in $\mathbf{s}$ will expire. At least, 2 tasks with deadlines 2, 3, 4, and 5 in the current time slot will expire in time slot $t=1,2,3$ and 4, respectively. These tasks are guaranteed to expire and are called excessive tasks. Trivially, all excessive tasks will be offloaded by the optimal policy whenever the AMA is available since the offloading cost is assumed to be smaller than the penalty cost, i.e., $C_{\sf o} < C_{\sf p}$. $\square$
		\end{example}
  
		We define the \textit{reduced states} as the ones having no excessive tasks. We assume that $\mathbf{s}^{(\sf r)} = \left(n_1^{(\sf r)}, \ldots, n_N^{(\sf r)}\right)$ is a reduced state which has no excessive task. According to the order of events in Fig.~\ref{fig:order-events}, all tasks having deadline 1 will expire if not offloaded. Hence, we must have $n_1^{(\sf r)}=0$. Next, at most one task can be processed in the next time slot. Therefore, we must have $n_1^{(\sf r)} + n_2^{(\sf r)} \le 1$. Subsequently, at most two tasks can be processed in the next two time slots, leading to $n_1^{(\sf r)} + n_2^{(\sf r)} + n_3^{(\sf r)} \le 2$. Otherwise, at least one task will expire  after two slots, and so on. Finally, at most $N-1$ tasks can be processed in $N$ slots, thus, we must have $\sum_{i=1}^{N} n_i^{(\sf r)} \le N-1$. Following this logic, the definition for a reduced state is as follows:
		\begin{definition}\label{Def:ReducedStates}
			\textit{
				A state $\mathbf{s}^{(\sf r)} = \left(n^{(\sf r)}_1, \ldots, n^{(\sf r)}_N\right)$ is a reduced state if and only if:
                \begin{align}\label{ineq:reduced_cond}
					\sum_{i=1}^j n_j^{(\sf r)} \le j - 1, \text{ for } j=1, 2, \ldots, N.
                \end{align}
			} 
		\end{definition}
	}
	\fi

	\iftext{
		As elements of a reduced state vector are bounded, the number of reduced vectors is finite. This number is equal to the Catalan number \cite{stanley2015catalan} as presented in the next lemma.
		\begin{lemma}\label{lem_Catalan}
			\textit{The number of reduced states having dimension $N$ is finite and equals to the Catalan number: $C_N = \binom{2N}{N}/(N+1)$.}
		\end{lemma}
		\textit{Proof}: See Section A of the Appendix. 
	}
	\fi

	\iftext{
		For the sake of simplicity,  by slight abuse of notation, we can associate a  corresponding reduced state $\mathbf{s}^{(\sf r)} = \left(n^{(\sf r)}_1, \ldots, n^{(\sf r)}_N\right)$ for any given state $\mathbf{s} = \left(n_1, \ldots, n_N\right)$, in the infinite state space, using Algorithm \ref{Gen2Red_Algo}. For $\mathbf{s}=\left(n_1, \ldots, n_N\right)$, it can be seen from Eqs. (\ref{eq:instant_cost3}), (\ref{eq:instant_cost}), and (\ref{DP_Eq}) that $n_i, i \ge 2$ do not contribute to the cost $J_1\left(\mathbf{s}\right)$. Similarly, $n_i, i\ge 3$ do not contribute to the cost $J_2\left(\mathbf{s}\right)$. Observe, therefore that in general, tasks having deadlines greater than the considered time horizon will not contribute to the cost in Eq. (\ref{DP_Eq}). Therefore, these tasks will not be considered, as reflected by line 4 of this algorithm. Algorithm \ref{Gen2Red_Algo} takes as input a general state $\mathbf{s}$, the number of state vector's dimensions $N$ and the considered time horizon $T$ to produce the corresponding reduced state $\mathbf{s}^{(\sf r)}$ and the number of most imminent tasks $L_{\sf g}$ to offload, in order to reach $\mathbf{s}^{(\sf r)}$ from $\mathbf{s}$.

        \renewcommand{\thealgorithm}{I}
	\begin{algorithm}[t]
		\caption{Derivation of Reduced States} \label{Gen2Red_Algo}
		\begin{algorithmic}[1] 
			\State \textbf{Input}: $\mathbf{s} = \left(n_1, n_2, \ldots, n_N\right)$, $N$, $T$.
			\State \textbf{Output}: $\mathbf{s}^{(\sf r)}$, $L_{\sf g}$. 
			\Comment{$L_{\sf g}$ is the number of most imminent tasks offloaded from state $\mathbf{s}$ to obtain the reduced state $\mathbf{s}^{(\sf r)}$.}
			\State \textbf{Initialize}: $L_{\sf g} \leftarrow 0$, $\tilde{\mathbf{s}} \leftarrow \mathbf{s}$.
			\For{$i=1,2,\ldots,\min\left(N, T\right)$}
			\State $L \leftarrow \sum_{j=1}^i n_j - i+1$.
                \State \textbf{If} $L > 0$:
                \State~~ Remove $L$ most imminent tasks from $\tilde{\mathbf{s}}$.
                \State~~ Increment $L_{\sf g}$ by $L$.
            \EndFor
			\State $\mathbf{s}^{(\sf r)} \leftarrow$ Offloading $L_{\sf g}$ most imminent tasks from $\mathbf{s}$.
			\State \textbf{return} $\mathbf{s}^{(\sf r)}$, $L_{\sf g}$.
		\end{algorithmic}
	\end{algorithm}
	}
	\fi

	\iftext{
		Determining the reduced states is the first step in reducing the computational burden of the DP Eq. (\ref{DP_Eq}). The next step is the determination of the set of \textit{lean states}, which is also a finite set and is directly used to compute the optimal cost $J_T\left(\mathbf{s}\right)$ according to the forthcoming Proposition~\ref{Pro:SemiRedState}.
	}
	\fi
	
	
	\subsection{Lean State Space} \label{sec:lean}

	\iftext{
	The intuition behind the concept of \textit{lean state} can be explained as follows. We consider two arbitrary states $\mathbf{s}$ and $\mathbf{s}^{(\ell)}$ as case 1 and 2, respectively, where the total number of tasks in $\mathbf{s}^{(\ell)}$ is less than that in $\mathbf{s}$. Let $\mathbf{s}$ and $\mathbf{s}^{(\ell)}$ transition through the same sequence of realizations of task arrival and local processing events without the arrival of the AMA. For instance, if a task arrives with a deadline $k$, we assume it arrives in both cases 1 and 2; if the local processing happens in case 1, it also happens in case 2; etc..  Assuming that the AMA arrives after a certain number of time slots, if the same reduced state is obtained via Algorithm \ref{Gen2Red_Algo} in both cases, then $\mathbf{s}^{(\ell)}$ is the lean state corresponding to $\mathbf{s}$. In other words, if $\mathbf{s}^{(\ell)}$ is the lean state of $\mathbf{s}$, after eliminating the excessive tasks, the optimal number of tasks to be offloaded more would be the same in both cases. This allows the minimum average costs of the two states to be related via a linear  Eq. (\ref{eq:gen2sem_eq}). We provide the following example to aid the understanding of the concept of lean states.
	\begin{example}
		We consider state $\mathbf{s}=(1, 2, 3, 4, 3)$ and its lean state $\mathbf{s}^{(\ell)}=(0, 1, 1, 1, 3)$ that can be derived via Definition \ref{Def:SemRedStates}. Assume that there is a task arriving with deadlines 3 and 2 in the current and the next time slots, respectively, and the local processing happens in both time slots. The two states will transition to states $(2, 6, 3, 0, 0)$ and (0, 3, 3, 0, 0), respectively. Algorithm \ref{Gen2Red_Algo} returns the same reduced state (0, 1, 2, 0, 0) in both cases. $\square$
	\end{example}
	}
	\fi

	\iftext{
            \begin{figure}[t]
		\centering
		\includegraphics[scale=0.3]{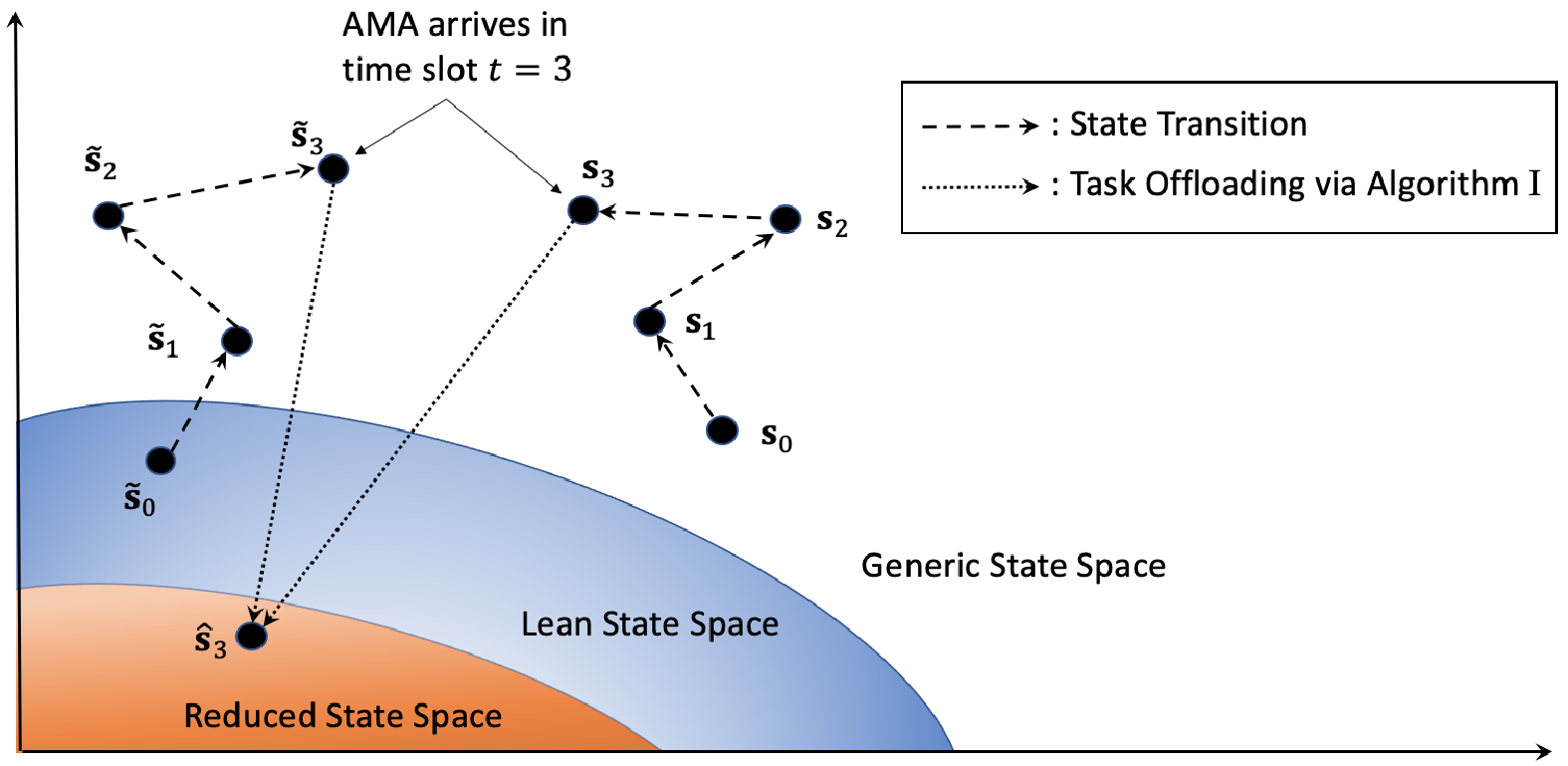}
		\caption{Illustrative example of the concept of lean states. In this figure, $\tilde{\mathbf{s}}_0$ is the corresponding lean state of $\mathbf{s}_0$ and $\hat{\mathbf{s}}_3$ denotes the reduced state that both $\tilde{\mathbf{s}}_3$ and $\mathbf{s}_3$ are mapped to by Algorithm I.}
    \label{fig:statespace_illus}
	\end{figure}
 
            In addition, we present an illustrative example in Fig.~\ref{fig:statespace_illus} for the relation between generic states, lean states, and reduced states. Next, we state the definition of lean states in Definition \ref{Def:SemRedStates}. 
		\begin{definition}\label{Def:SemRedStates}
			\textit{For a state $\mathbf{s} = \left(n_1, \ldots, n_N\right)$, let $\mathbf{s}^{(\sf r)} = \left(n_1^{(\sf r)}, \ldots, n_N^{(\sf r)}\right)$ a reduced state obtained from $\mathbf{s}$ by Algorithm \ref{Gen2Red_Algo}. We define the parameters $\gamma_j, j = 1, \ldots, N$ by:} 
			\begin{align}\label{eq:gamma_def}
				\gamma_j = 
				\begin{cases}
					0 ,  \text{ if } j = 1 \text{ or } j > \min\left(N, T\right), \\
					\min\left\{n_j, j - 1 - \sum_{i=1}^{j-1} \gamma_i \right\} ,  \text{ otherwise}.
				\end{cases}
			\end{align}
			\textit{Then, the lean state $\mathbf{s}^{(\ell)}= \left(n_1^{(\ell)}, \ldots, n_N^{(\ell)}\right)$ corresponding to $\mathbf{s}$ is given as follows:}
			\begin{align}\label{eq:semi_element_def}
			n^{(\ell)}_i = \max\{ \gamma_i, n^{(\sf r)}_i\}, ~ i=1, 2, \ldots, N.
			\end{align}
		\end{definition}
	}
	\fi

	\iftext{

		The relation in the minimum average cost of a given state $\mathbf{s}$ and that of its corresponding lean state is presented in Proposition \ref{Pro:SemiRedState}.
		
		\begin{proposition} \label{Pro:SemiRedState}
			\textit{Given state $\mathbf{s} = \left(n_1, \ldots, n_N\right)$ and its lean state $\mathbf{s}^{(\ell)} = \left(n_1^{(\ell)}, \ldots, n_N^{(\ell)}\right)$, the following equality holds:}
			\begin{equation} \label{eq:gen2sem_eq}
				J_T\left(\mathbf{s}\right) = J_T\left(\mathbf{s}^{(\ell)}\right) + C_{(\ell)}
			\end{equation}
			\textit{where }
			\begin{align}
				C_{\ell} &= C_{\sf o}\sum_{i=1}^{N} \left(n_i - n_i^{(\ell)} \right)\left(1 - \left(1 - p_{\sf a} \right)^i \right) \notag \\
                &+ C_{\sf p} p_{\sf a}\sum_{i=1}^{N-1} \left(1 - p_{\sf a} \right)^i \sum_{j=1}^{i}  \left(n_j - n_j^{(\ell)} \right) \notag \\
                &+ C_{\sf p} \left(1-p_{\sf a} \right)^{N} \sum_{i=1}^{N} \left(n_i- n_i^{(\ell)}\right)
			\end{align}
		\end{proposition}
		\textit{Proof}: See Section B of the Appendix. 
		
		As the lean state space is finite, if $J_T\left(\mathbf{s}^{(\ell)}\right)$ is computed for all lean states via the DP Eq. (\ref{DP_Eq}), $J_T\left(\mathbf{s}\right)$ can be computed for every state $\mathbf{s}$ via the simple algebraic manipulation of Eq. (\ref{eq:gen2sem_eq}). We would like to note that our presented definitions of reduced states, lean states, and the result in Proposition \ref{Pro:SemiRedState} above are not influenced by the number of tasks that can arrive in each time slot. Therefore, our result in Proposition \ref{Pro:SemiRedState} remains applicable in the context of the bulk arrival of tasks as emphasized in the following remark.

        \textbf{Remark 1. (Extension to the Multi-Task-Arrival Scenario)}: \textit{The computation of the formulated DP Eq. \eqref{DP_Eq} for a generic state $\mathbf{s}$ can always be converted to the computation with respect to its associated lean state $\mathbf{s}^{(\ell)}$ regardless of the number of tasks that can arrive per time slot. Therefore, Eq. \eqref{eq:gen2sem_eq} reduces the consideration of  infinite generic state space to finite lean state space for both cases of single arrival and bulk arrival of tasks in each time slot, alleviating the computational burden.}
    
	}
	\fi
	
	{To this end, we would like to emphasize that even when the system cost includes the cost of local resource consumption, i.e., a cost is incurred for local task processing, the concepts of reduced state and lean state remain valid and applicable. This is because reduced states are defined by eliminating excessive tasks that exceed the system's local processing capacity, and thus, their formation is independent of whether a local processing cost is considered. Similarly, the lean state $\mathbf{s}^{(\ell)}$ of a given state $\mathbf{s}$ is the state that maps to the same reduced state as $\mathbf{s}$ upon the arrival of an AMA, by removing excessive tasks. Moreover, the associated cost $C_{(\ell)}$ is expressed solely in terms of the offloading and expiration costs of excessive tasks in $\mathbf{s}$. Therefore, the definitions of reduced state, lean state, and the results of Proposition \ref{Pro:SemiRedState} remain applicable, even when a local processing cost is introduced.}

    \section{Properties of Optimal Offloading Policy and Cost Function} \label{Sec:OptimalPolicyProp}

	\iftext{
	In this section, we present key properties of the minimum average cost and the optimal offloading decisions. In addition to the reduction in the required state space discussed earlier, Theorem \ref{Theo:adjacent_L*}, being one of the main outcomes in this section, offers chances to further reduce the computational load of the DP Equation. This is achieved by characterizing the structure of the optimal offloading policy, thus, enabling optimal decisions for certain states to be inferred from those of their \textit{adjacent states}. The concept of adjacent states will be explicitly defined in Subsection \ref{subsec:adjacentstates}.
	}
	\fi

	\subsection{Convexity of the Minimum Average Cost with respect to the Offloading Decision} \label{SubSec:Convex}

	\iftext{
    We denote $\bar{\mathbf{s}}_{L}$ the state obtained by offloading, from $\mathbf{s}$, $L$ most imminent tasks. Below is an example clarifying the notation. 
	\begin{example}
		For a given state $\mathbf{s} = \left(0, 5, 6, 7, 8\right)$ and an offloading decision $L=7$, the resulting state after offloading will be $\bar{\mathbf{s}}_{L} = \left(0, 0, 4, 7, 8\right)$. $\square$
	\end{example}
	}
	\fi

	\iftext{
	For a given time horizon $T$, and an initial state $\mathbf{s} = \left(n_1, \ldots, n_N\right)$, we define the function
	\begin{align} \label{F_def}
		F\left(\mathbf{s},  L\right) = J_T^{\overline{\text{A}}}\left(\bar{\mathbf{s}}_{L}\right) + LC_{\sf o}.
	\end{align}
	Function $F\left(\mathbf{s},  L\right)$ takes $\mathbf{s}$ as a parameter and $L$ as its variable. {The interpretation of this function is provided as follows. We assume that the system state $\mathbf{s}$ is encountered in the current time slot with the arrival of the AMA, the function $F(\mathbf{s}, L)$ denotes the minimum expected cost over the entire considered time horizon that can incur, given the condition that exactly $L$ task are offloaded from $\mathbf{s}$ in the current time slot. Accordingly, the first term in Eq. \eqref{F_def} is the minimum expected cost over the considered time horizon that can be incurred, after exactly $L$ tasks have been offloaded from $\mathbf{s}$, resulting in $\bar{\mathbf{s}}_L$. The second term is the cost of offloading $L$ tasks.}

    {We would like to highlight that the first term in Eq. \eqref{F_def} is $J_T^{\overline{\text{A}}}\left(\bar{\mathbf{s}}_{L}\right)$ and not $J_T^{\text{A}}\left(\bar{\mathbf{s}}_{L}\right)$ for the following reasons. $J_T^{\overline{\text{A}}}\left(\bar{\mathbf{s}}_{L}\right)$ is the minimum expected cost incurred  without the AMA's arrival. This implies that $J_T^{\overline{\text{A}}}\left(\bar{\mathbf{s}}_{L}\right)$ does not involve the cost for task offloading on state $\bar{\mathbf{s}}_{L}$ in the current time slot. In contrast, $J_T^{\text{A}}\left(\bar{\mathbf{s}}_{L}\right)$ denotes the minimum expected cost incurred given the AMA's arrival. Therefore, $J_T^{\text{A}}\left(\bar{\mathbf{s}}_{L}\right)$ may involve the cost for additional task offloading on state $\bar{\mathbf{s}}_{L}$ if $L$ is not the optimal offloading decision of the original state $\mathbf{s}$. The cost $J_T\left(\bar{\mathbf{s}}_{L}\right)$, as defined in Eq. \eqref{gen_U_noU}, also includes the aforementioned additional offloading cost.} 

 The domains of function $F\left(\mathbf{s}, L\right)$ is defined by
	\begin{align}\label{F_domain2_simple}
	\mathbb{L}\left(\mathbf{s} \right) \backslash \left\{0, 1, \ldots, \max(n_1-1, 0) \right\}.
	\end{align}
	We note that the set defined above is a convex set. Also, since all tasks having deadline 1 are excessive tasks and must always be offloaded, the optimal offloading decision will be conveyed in the set $\left\{n_1, \ldots, \sum_{i=1}^N n_i \right\}$. Hence, in the definition of the above domain, we remove $L <  n_1$ to obtain a simpler form of this function while still maintaining the properties presented later on. An example is provided below.
\begin{example}
	For a $N=3$ and $\mathbf{s} = \left(2, 3, 4\right)$, the domain of function $F(\mathbf{s}, L)$ is given by $\left\{2, 3, \ldots, 9 \right\}$. $\square$
\end{example}
	}
	\fi

	\iftext{Hereafter, we formally provide the definition of a discrete convex property of function $F(\mathbf{s}, L)$ as follows:
		
		\begin{definition}\label{def:discreteconvex}
			\textit{For a given state $\mathbf{s} = (n_1, \ldots, n_N)$, if $\sum_{i=1}^N n_i \ge 2$, a function $F(\mathbf{s}, L)$ is a discrete convex function with respect to $L$ if the following inequality holds for every offloading decision $L$ that belongs to the convex set defined in (\ref{F_domain2_simple}):}
			\begin{align}
				F(\mathbf{s}, L) + F(\mathbf{s}, L+2) \ge 2F(\mathbf{s}, L+1).
			\end{align}
		
		\textit{If $\sum_{i=1}^N n_i = 1$, $F(\mathbf{s}, L)$ is a discrete linear, and thus, a discrete convex function with respect to $L \in \{0, 1\}$.}
		\end{definition}
		}
		\fi

	\iftext{
	We state the convexity property of function $F\left( \mathbf{s},  L\right)$ in the next lemma.
	\begin{lemma}\label{Lem:Convexity}
		\textit{For every given time horizon $T$ and an initial state $\mathbf{s}$, the function $F\left( \mathbf{s},  L\right)$ as defined in Eq. (\ref{F_def}) is a discrete convex function with respect to $L \in \mathbb{L}\left(\mathbf{s} \right) \backslash \left\{0, 1, \ldots, \max(n_1-1, 0) \right\}$.}
	\end{lemma}
	\textit{Proof}: See Section C of the Appendix. 
	}
	\fi

	\iftext{
	In the next lemma, we present the relation between the minimum of function $F\left(\mathbf{s},  L\right)$ and the optimal offloading decision associated with state $\mathbf{s}$ and a time horizon $T$.
	\begin{lemma}\label{Lem:Convex2L*}
		\textit{ Assuming that the function $F\left(\mathbf{s},  L\right)$ attains its minimum at $L^*$, then, $L^*$ is the optimal offloading decision associated with state $\mathbf{s}$.}
	\end{lemma}
	\textit{Proof}: See Section D of the Appendix. 
	
	Other important properties of the optimal offloading decisions will be presented in the next subsection based on the concept of \textit{adjacent states}.
	}
	\fi

	\subsection{Concept of Adjacent States}\label{subsec:adjacentstates}

	\iftext{
	The goal of this subsection is to introduce the concept of adjacency among states, and related properties. These properties facilitate the design of the optimal policy presented later on. The definition of adjacent states is given below.
	\begin{definition}\label{Def:adjacent_states}
		\textit{Consider a state $\mathbf{s} = \left(n_1, \ldots, n_N \right) \ne \left(0, \ldots, 0 \right)$, with $d$ as the smallest deadline satisfying $n_d > 0$. Then, state $\mathbf{s}^{(\sf a)} = \left(n^{(\sf a)}_1, \ldots, n^{(\sf a)}_N\right)$ is an adjacent state to $\mathbf{s}$ if there exists a deadline $j \in \left\{1, \ldots, d \right\}$ such that:}
        \begin{align}
            n_i^{(\sf a)} =
            \begin{cases}
                n_i + 1 &, \text{ if } i = j,\\
                n_i &, \text{ otherwise}.
            \end{cases}
        \end{align}
		
		\textit{If a state $\mathbf{s}^{(\sf a)}$ has only one task with an arbitrary deadline, it is adjacent to state $\left(0, \ldots, 0 \right)$.}
	\end{definition}
	}
	\fi

	\iftext{
	The following two examples facilitate the understanding of the adjacent state concept. 
	\begin{example}
	In the first example, we assume that a state $\mathbf{s} = \left(0, 0, 1, 4, 4 \right)$ is given in which the deadline of the most imminent task in $\mathbf{s}$ is 3. Therefore, an adjacent state $\mathbf{s}^{(\sf a)}$ of $\mathbf{s}$ can be obtained by adding a task with deadline less than or equal to 3, e.g., $\mathbf{s}^{(\sf a)} = \left(0, 1, 1, 4, 4\right)$.
	
	In the second example, we assume that $\mathbf{s}^{(\sf a)} = \left(0, 2, 1, 3, 3\right)$. A state $\mathbf{s}$ for which $\mathbf{s}^{(\sf a)}$ is adjacent to, can be obtained by offloading the most imminent task in $\mathbf{s}^{(\sf a)}$, i.e., $\mathbf{s} = \left(0, 1, 1, 3, 3\right)$. $\square$
	\end{example} 

	We denote by $\mathbb{S}_{\sf adj}\left(\mathbf{s}\right)$ the set of all adjacent states of $\mathbf{s}$. For a given time horizon, the optimal offloading decision of $\mathbf{s}$ can be inferred from that of $\mathbf{s}^{(\sf a)} \in \mathbb{S}_{\sf adj}\left(\mathbf{s}\right)$ and vice versa, as described in Theorem \ref{Theo:adjacent_L*}.
	\begin{theorem}\label{Theo:adjacent_L*}
        \textit{Given two states $\mathbf{s}$, $\mathbf{s}^{(\sf a)} \in \mathbb{S}_{\sf adj}\left(\mathbf{s}\right)$, and a time horizon $T$. We call $L^*$ and $L^*_{\sf a}$ the optimal offloading decision of $\mathbf{s}$ and $\mathbf{s}^{(\sf a)}$, respectively. We have the following properties:}
		\begin{enumerate}
			\item \textit{If $L_{\sf a}^*$ is known and $L_{\sf a}^* \ge 1$, then, $L^*$ can be computed by $L^* = L_{\sf a}^*-1$.}
			\item \textit{If $L_{\sf a}^*$ is known and $L_{\sf a}^* = 0$, then, $L^*$ can be computed by $L^*=0$.}
			\item \textit{If $L^*$ is known and $L^* \ge 1$, then, $L^*_{\sf a}$ can be computed by  $L^*_{\sf a} = L^* + 1$.}
		\end{enumerate}
	\end{theorem}
	\textit{Proof}: See Section E of the Appendix. 
	}
	\fi

	\subsection{Offloading and Non-Offloading Conditions}\label{Subsec:Conditions}

	\iftext{
	Firstly, we specify the concepts of \textit{offloading states} and \textit{non-offloading states} below.
	\begin{definition}\label{Def:Of_NOf}
		\textit{For a given state $\mathbf{s}$ and a time horizon $T$, $\mathbf{s}$ is called an offloading state if the associated optimal offloading decision is a positive integer.}
  
        \textit{State $\mathbf{s}$ is called a non-offloading state if the associated optimal offloading decision is 0.}
	\end{definition}
	}
	\fi

	\iftext{
	Subsequently, we identify the offloading and non-offloading conditions for a given system state and time horizon. This is stated in Proposition \ref{Pro:ONO_conds}.
	\begin{proposition}\label{Pro:ONO_conds}
		\textit{Assume that two states $\mathbf{s}$, $\mathbf{s}^{(\sf a)} \in \mathbb{S}_{\sf adj}\left(\mathbf{s}\right)$, and a time horizon $T$ are given. Then, state $\mathbf{s}^{(\sf a)}$ is a non-offloading state if and only if the following inequality holds:}
        \begin{align}
            J_T\left(\mathbf{s}^{(\sf a)}\right) - J_T\left(\mathbf{s}\right) < C_{\sf o}.
        \end{align}
        \textit{Otherwise, $\mathbf{s}^{(\sf a)}$ is an offloading state.}
	\end{proposition}
	\textit{Proof}: See Section F of the Appendix. 
	}
	\fi

	\iftext{
	The next property of the optimal offloading decision is stated in Theorem \ref{Theo:L*_thesmallest}.	
    \begin{theorem}\label{Theo:L*_thesmallest}
		\textit{For a given state $\mathbf{s}$, let $\bar{\mathbf{s}}_L$ denote the state obtained by removing the $L$ most imminent tasks from $\mathbf{s}$. The offloading decision $L$ is optimal for $\mathbf{s}$ if $L$ is the smallest decision such that $\bar{\mathbf{s}}_{L}$ is a non-offloading state}.
	\end{theorem}
	\textit{Proof}: See Section G of the Appendix. 

    The optimal offloading policy will be described in the next section based on our presented properties.
	}
	\fi

	\section{Optimal Offloading Policy} \label{Sec:OptimalPolicy}

	\iftext{
	Based on the results of Sections \ref{Sec:RS} and \ref{Sec:OptimalPolicyProp}, this section presents the optimal task offloading strategy. The details of every steps of this policy is provided in Algorithm \ref{Optim_Algo}.
	}
	\fi

	\iftext{
	
	Given an initial state $\mathbf{s}$ and a time horizon $T$, whenever the local processing service is available, the most imminent task will be processed. When the AMA is present, the optimal policy consists of two steps.
	
	\textbf{Step 1.} Tasks are offloaded from $\mathbf{s}$ following Algorithm \ref{Gen2Red_Algo} to reach a reduced state $\mathbf{s}^{(\sf r)}$. We remind that Algorithm \ref{Gen2Red_Algo} defines the number of most imminent tasks $L_{\sf g}$ to be offloaded from the given state. The offloaded tasks are excessive tasks that are guaranteed to expire if not offloaded.
	
	\textbf{Step 2.} In this step, the DP Eq. \eqref{DP_Eq} with respect to state $\mathbf{s}^{(\sf r)}$ is solved recursively and
    \begin{align}\label{eq:compute_Lr}
		L_{\sf r} = {\argmin_{L \in \mathbb{L}\left(\mathbf{s}^{(\sf r)}\right)}}\left\{\mathcal{C}\left(\mathbf{s}^{(\sf r)}, L\right) + G_{T-1}\left(\mathbf{s}^{(\sf r)}, L\right) \right\}
	\end{align}
	tasks are offloaded from $\mathbf{s}^{(\sf r)}$. Note that as described in Eq. (\ref{eq:gen2sem_eq}), the recursion involves evaluation of only a finite number of states in each one of the terms $J_{T-1}\left(\cdot\right), J_{T-2}\left(\cdot\right)$ etc. for any state $\mathbf{s}$. The optimal number of tasks that are then offloaded from $\mathbf{s}$ is
    \begin{align}\label{eq:Lr_compute}
        L^* = L_{\sf g} + L_{\sf r}.
    \end{align}
	 
	The computational load required to solve the recursive DP Eq. (\ref{DP_Eq}) can be further reduced as follows. We emphasize that $L_{\sf r}$ is used to denote the optimal offloading decision of the reduced state $\mathbf{s}^{(\sf r)}$, and we denote by $L_{\ell}$ the optimal decision for the lean state $\mathbf{s}^{(\ell)}$ where both $\mathbf{s}^{(\sf r)}$ and $\mathbf{s}^{(\ell)}$ are associated with the given initial state $\mathbf{s}$. $L_{\sf r}$ in Eq. \eqref{eq:Lr_compute} in Step 2 can be computed by recursively solving the DP Eq. \eqref{DP_Eq}. This recursive process may requires to compute $J_{T-t}(\mathbf{s}), t \in \{0, \ldots, T-1\}$, for some states $\mathbf{s}$. By using Eq. \eqref{eq:gen2sem_eq}, the computation of $J_T(\mathbf{s})$ can be altered by that of $J_T(\mathbf{s}^{(\ell)})$. We note that solving the DP Eq. \eqref{DP_Eq} to compute $J_T(\mathbf{s}^{(\ell)})$ allows us to obtain $L_{\ell}$. Then, every time $J_{T-t}(\mathbf{s}^{(\ell)})$ and $L_{\ell}$ are computed for a lean state $\mathbf{s}^{(\ell)}$, the quadruplet $\left(\mathbf{s}^{(\ell)}, T-t, J_{T-t}(\mathbf{s}^{(\ell)}), L_{\ell}\right)$ is saved. At the end of the recursive process for solving Eq. \eqref{DP_Eq}, $J_T(\mathbf{s}^{(\sf r)})$ and $L_{\sf r}$ are obtained and the quadruplet $\left(\mathbf{s}^{(\sf r)}, T-t, J_{T-t}(\mathbf{s}^{(\sf r)}), L_{\sf r}\right)$ is saved, then, and loaded later when needed to reduce the reliance on the DP Eq. \eqref{DP_Eq}.
 
    {In summary, only the quadruplets associated with the reduced and lean states are saved. This is firstly because any generic state $\mathbf{s}$ can be mapped to a corresponding reduced state via Algorithm \ref{Gen2Red_Algo}, then, the optimal decision $L^*$ is computed via Eq. \eqref{eq:Lr_compute} where $L_{\sf g}$ is provided by Algorithm \ref{Gen2Red_Algo} and $L_{\sf r}$ is retrieved from the memory if it has been saved. Secondly, $J_T(\mathbf{s})$ can be computed from $J_T(\mathbf{s}^{(\ell)})$ of the corresponding lean state $\mathbf{s}^{(\ell)}$ via Eq. \eqref{eq:gen2sem_eq}; hence, $J_T(\mathbf{s}^{(\ell)})$ can be retrieved from the memory if it has been saved. The aforementioned saving process can be progressively performed for all the lean states and reduced states, as the number of these two types of states is finite. As a result, the computational burden is reduced. In addition, we would like to highlight that although the optimal offloading decision on $\mathbf{s}$ can be directly inferred from that of $\mathbf{s}^{(\sf r)}$, if the entry associated with $\mathbf{s}^{(\sf r)}$ is not saved and $J_{T-t}(\mathbf{s})$ needs to be computed, the entry associated with $\mathbf{s}^{(\ell)}$ would simplify the use of the DP Eq. Therefore, both types of state are useful in the reduction of computational burdens.}
 

	}
	\fi

	\iftext{
	
	To further alleviate the computational burden of Eq. (\ref{eq:compute_Lr}) (or equivalently, the DP Eq. \eqref{DP_Eq}), we exploit the  properties presented in Theorem \ref{Theo:adjacent_L*}. Let us consider a sequence of adjacent states: $\mathbf{s}_1, \ldots, \mathbf{s}_i, \ldots$ 
  in which $\mathbf{s}_{i+1} \in \mathbb{S}_{\sf adj}\left(\mathbf{s}_i\right)$. Assume the optimal decision of state $\mathbf{s}_i, i \ge 1$ is known, and denoted by $L^*_i$. From Theorem \ref{Theo:adjacent_L*}, the optimal decisions $L^*_{i-u}$ of states $\mathbf{s}_{i-u}, u = 1, \ldots, i-1$, can be inferred as follows
	\begin{align}
		L^*_{i-u} &= \max\left(L^*_i - u, 0 \right), \text{ for } u = 1, \ldots, i-1. \label{eq:tr_down}
	\end{align}
	In the case when $L^*_i \ge 1$ for state $\mathbf{s}_i$, the optimal decision for states $\mathbf{s}_{i+v}, v = 1, 2, \ldots$ are computed by
	\begin{align}
	L^*_{i+v} &= L^*_i + v, \text{ for } v = 1, 2 \ldots    \label{eq:tr_up}
	\end{align}
	In general, the optimal offloading decisions of all the states $\mathbf{s}_{i+v}$ and $\mathbf{s}_{i-u}$ mentioned above can be obtained without explicitly solving the DP Eq. (\ref{DP_Eq}).

	}
	\fi

	\iftext{
	A complete presentation of the optimal offloading policy is presented in Algorithm \ref{Optim_Algo} in which the set $\mathbb{M}$ represents the memory storing the quadruplets $\left(\mathbf{s}^{(\ell)}, T-t, J_{T-t}(\mathbf{s}^{(\ell)}), L_{\ell}\right)$ and $\left(\mathbf{s}^{(\sf r)}, T-t, J_{T-t}(\mathbf{s}^{(\sf r)}), L_{\sf r}\right)$ described above.

        \renewcommand{\thealgorithm}{II}
	\begin{algorithm} [t]
		\caption{Optimal Task Offloading Policy} \label{Optim_Algo}
		\begin{algorithmic}[1] 
			\State \textbf{Input}: $\mathbf{s}$, $N$, $T$, $\mathbb{M}$.
			\State \textbf{Output}: $\mathbf{s}^*$, $L^*$. \Comment{$L^*$ is the optimal offloading decision for $\mathbf{s}$; $\mathbf{s}^*$ is the resulting state by offloading $L^*$ most imminent tasks from $\mathbf{s}$.}
			\State $\mathbf{s}^{(\sf r)}, L_{\sf g} \leftarrow$ Algorithm \ref{Gen2Red_Algo}.
                \State \textbf{If} $(\mathbf{s}^{(\sf r)}, T, J_{T}(\mathbf{s}^{(\sf r)}), L_{\sf r}) \in \mathbb{M}$:
                \State~~ $L^* \leftarrow L_{\sf r} + L_{\sf g}$.
                \State~~ $\mathbf{s}^* \leftarrow$ Offloading $L_{\sf r}$ most imminent tasks from $\mathbf{s}^{(\sf r)}$.
                \State~~ \textbf{return} $\mathbf{s}^*$, $L^*$ and \textbf{terminate}.
                \State Assign $L \leftarrow 0$ and $\tilde{\mathbf{s}} \leftarrow \mathbf{s}$.
                \State \textbf{Loop}:
                \State~~ Remove the most imminent task from $\tilde{\mathbf{s}}$.
                \State~~ \textbf{If} $\tilde{\mathbf{s}}=(0,\ldots,0)$: break the loop.
                \State~~ Increment $L$ by 1.
                \State~~ \textbf{If} $(\tilde{\mathbf{s}},T, J_T(\tilde{\mathbf{s}}), \tilde{L}) \in \mathbb{M}$:
                \State~~~~ $L^* \leftarrow \tilde{L}^* + L$.
                \State~~~~ $\mathbf{s}^* \leftarrow$ Offloading $L_{\sf r}$ most imminent tasks from $\mathbf{s}^{(\sf r)}$.
                \State~~~~ \textbf{return} $\mathbf{s}^*$, $L^*$ and \textbf{terminate}.
                \State Solve Eq. \eqref{DP_Eq} for $J_T(\mathbf{s}^{(\sf r)})$ and  $L_{\sf r}$. In this recursive process, $J_{T-t}(\mathbf{s})$ for a generic state $\mathbf{s}$ is computed with respect to $J_{T-t}(\mathbf{s}^{(\ell)})$ via Eq. \eqref{eq:gen2sem_eq}, $t \in \{0, \ldots, T-1 \}$.
                \State In the recursive process of solving Eq. \eqref{DP_Eq}:
                \State~~\parbox[t]{\dimexpr\linewidth-\algorithmicindent}{\textbf{If} $(\mathbf{s}^{(\ell)}, T-t, J_{T-t}(\mathbf{s}^{(\ell)}), L_{\ell}) \in \mathbb{M}$: load $J_{T-t}(\mathbf{s}^{(\ell)})$.}
                \State~~\parbox[t]{\dimexpr\linewidth-\algorithmicindent}{\textbf{Else}: compute $J_{T-t}(\mathbf{s}^{(\ell)})$ via Eq. \eqref{DP_Eq} and update $\mathbb{M} \leftarrow \mathbb{M} \cup \{(\mathbf{s}^{(\ell)}, T, J_{T-t}(\mathbf{s}^{(\ell)}), L_{\ell}) \}$.}
                \State Update $\mathbb{M} \leftarrow \mathbb{M} \cup \{(\mathbf{s}^{(\sf r)}, T, J_{T-t}(\mathbf{s}^{(\sf r)}), L_{\sf r}) \}$.
                \State $L^* \leftarrow L_{\sf r} + L_{\sf g}$.
                \State $\mathbf{s}^* \leftarrow$ Offloading $L_{\sf r}$ most imminent tasks from $\mathbf{s}^{(\sf r)}$.
                \State \textbf{return} $\mathbf{s}^*$, $L^*$ and \textbf{terminate}.
		\end{algorithmic}
	\end{algorithm}
	}
	\fi

    {As discussed earlier, the concepts of reduced state, lean state, and Proposition \ref{Pro:SemiRedState} remain valid even when a local processing service cost is introduced. Consequently, although the DP Eq. \eqref{DP_Eq} must be modified to account for the local service cost, affecting lines 17, 18, and 20 of Algorithm \ref{Optim_Algo}, the rest of the algorithm remains applicable without change.}
    
	\section{Numerical Results} \label{Sec:NumericalResults}

	\iftext{
	In this section, we present numerical examples that help visualize the presented properties and equations related to the optimal offloading policy. We also show the advantage of using Eq. (\ref{eq:gen2sem_eq}) in terms of memory saving. Different parameter configurations are used in illustrating the system performance. The details of parameter configurations are presented in Table \ref{tbl:sys-param-lin-eq}, \ref{tbl:sys-param-lin-eq2}, and III for the simulations associated with Fig. \ref{fig:n3}, \ref{fig:n3N5}, and \ref{fig:convex}, respectively. Besides, the simulation of Fig. \ref{fig:memory_savings} relies on the set of  parameters in Table \ref{tbl:sys-param-lin-eq} with $N$ alternatively takes values 3, 4, and 5.
	}
	\fi
	
\subsection{Optimal Offloading Decision Visualization} \label{seC:opt-off-d}

	\iftext{
    \begin{table}
		\centering
        \caption{{System parameters for Fig. \ref{fig:n3}.}}
		\label{tbl:sys-param-lin-eq}
        {
		\begin{tabular}{|c|c|c|c|c|c|c|c|c|}
			 \hline
             \textbf{Parameters} & $T$   & $p_{\sf a}$ & $C_{\sf p}$ & $C_{\sf o}$ & $\mu$ & $p_0$ & $N$ & $p_i$\\ \hline
			\textbf{Values} & $1000$ & $0.7$ & $ 3 $ & $ 1 $ & $0.7$ & $0.5$ & $3$ & $1/6$ \\
            \hline
		\end{tabular}
        }
	\end{table}

 \begin{table}
		\centering
        \caption{{System parameters for Fig. \ref{fig:n3N5}.}}
		\label{tbl:sys-param-lin-eq2}
        {
	\begin{tabular}{|c|c|c|c|c|c|c|c|c}
    \hline
			  \textbf{Parameters} & $T$   & $p_{\sf a}$ & $C_{\sf p}$ & $C_{\sf o}$ & $\mu$& $N$ & $p_i$\\ \hline
			\textbf{Values} & $1000$ & $0.8$ & $ 3 $ & $ 1 $ & $0.6$ & $5$ & $1/6$ \\
            \hline
		\end{tabular}
        }
	\end{table}

  
    \begin{figure*}
  
	    \begin{subfigure}[b]{0.295\textwidth}
		    \centering
		    \includegraphics[width=1.1\textwidth]{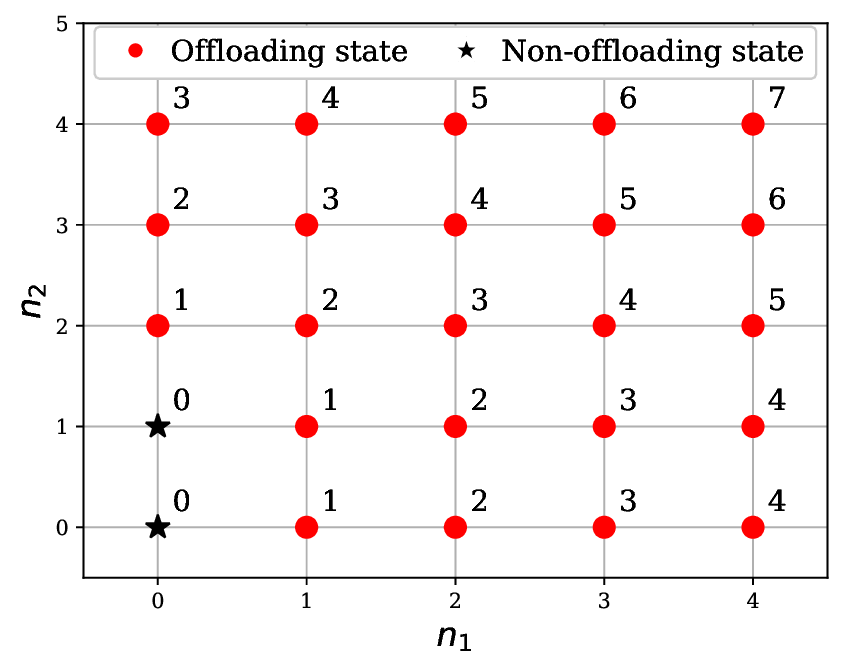}
		    \caption{$n_3 = 0$.}
		    \label{fig:n30}
	    \end{subfigure}
	    \hspace{0.8em}
    	\begin{subfigure}[b]{0.295\textwidth}
    		\centering
	    	\includegraphics[width=1.1\textwidth]{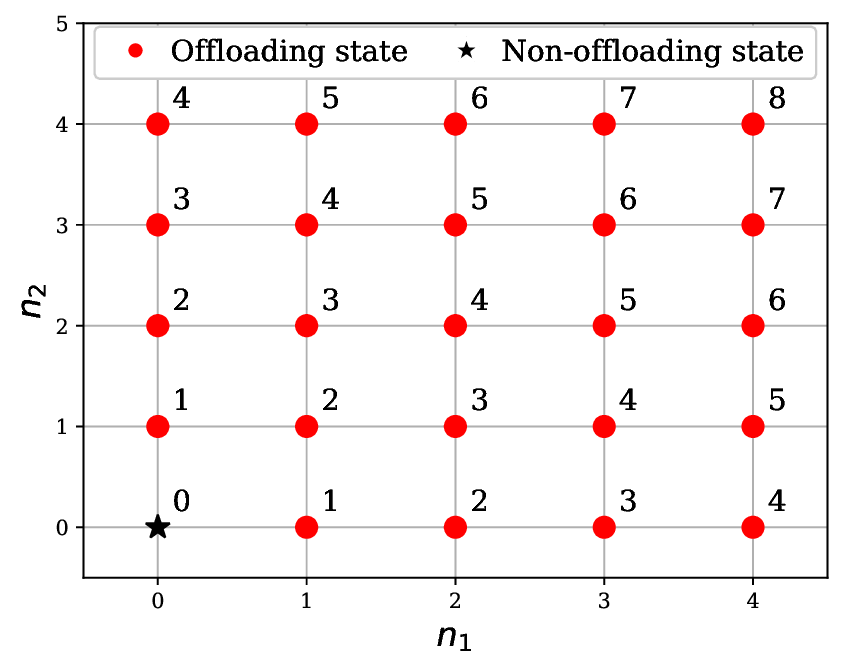}
		    \caption{$n_3 = 1$.}
		    \label{fig:n31}
	    \end{subfigure}
	    \hspace{0.8em}
    	\begin{subfigure}[b]{0.295\textwidth}
    		\centering
    		\includegraphics[width=1.1\textwidth]{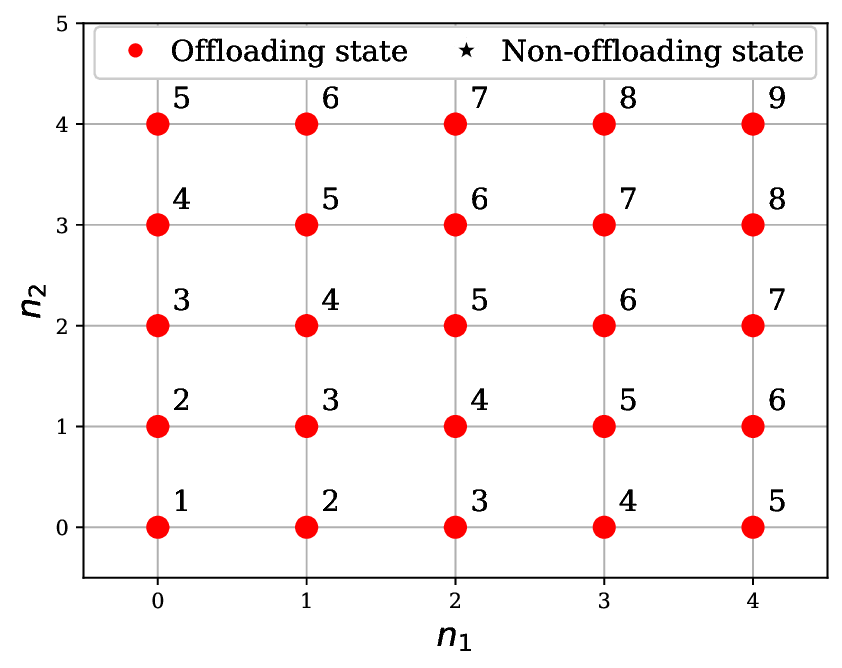}
    		\caption{$n_3 = 2$.}
    		\label{fig:n33}
    	\end{subfigure}
    	\caption{Visualisation of Theorem \ref{Theo:L*_thesmallest} for the state vector $\mathbf{s} = \left(n_1, n_2, n_3\right)$. Offloading and non-offloading states are represented by red and black stars, respectively. The number beside each state is the optimal number of tasks, $L^*$, to be offloaded to achieve the minimum expected cost; $L^*=0$ implies that, optimally, no task is offloaded.}
     \label{fig:n3}
	\end{figure*}
 
    \begin{figure*}
	    \begin{subfigure}[b]{0.295\textwidth}
		    \centering
		    \includegraphics[width=1.1\textwidth]{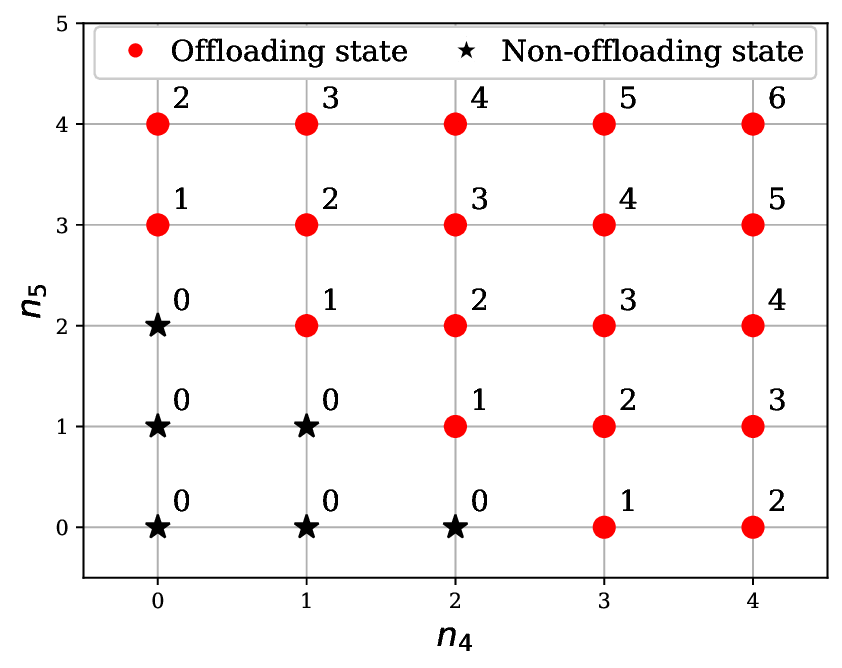}
		    \caption{$n_3 = 0$.}
		    \label{fig:n30N5}
	    \end{subfigure}
	    \hspace{0.8em}
    	\begin{subfigure}[b]{0.295\textwidth}
    		\centering
	    	\includegraphics[width=1.1\textwidth]{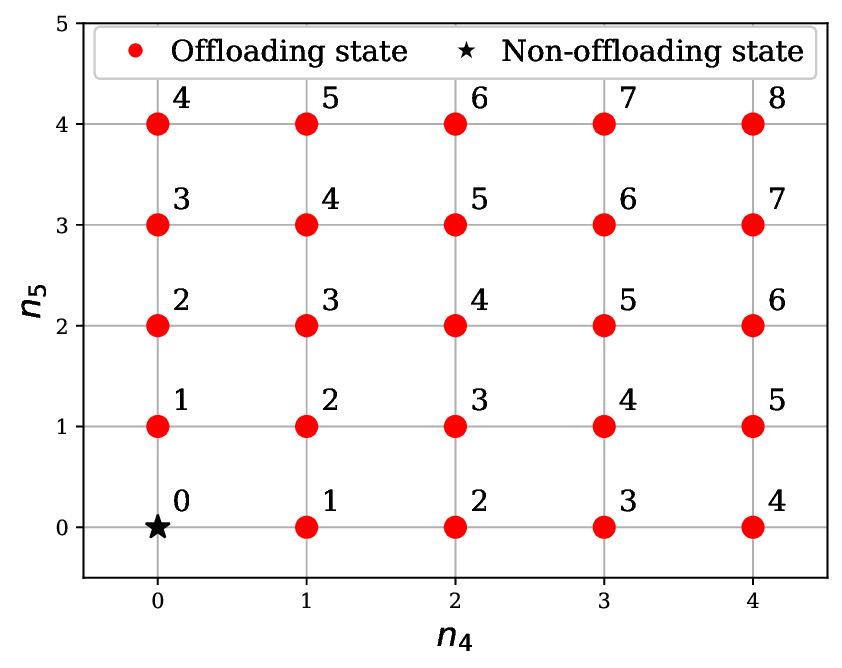}
		    \caption{$n_3 = 1$.}
		    \label{fig:n31N5}
	    \end{subfigure}
	    \hspace{0.8em}
    	\begin{subfigure}[b]{0.295\textwidth}
    		\centering
    		\includegraphics[width=1.1\textwidth]{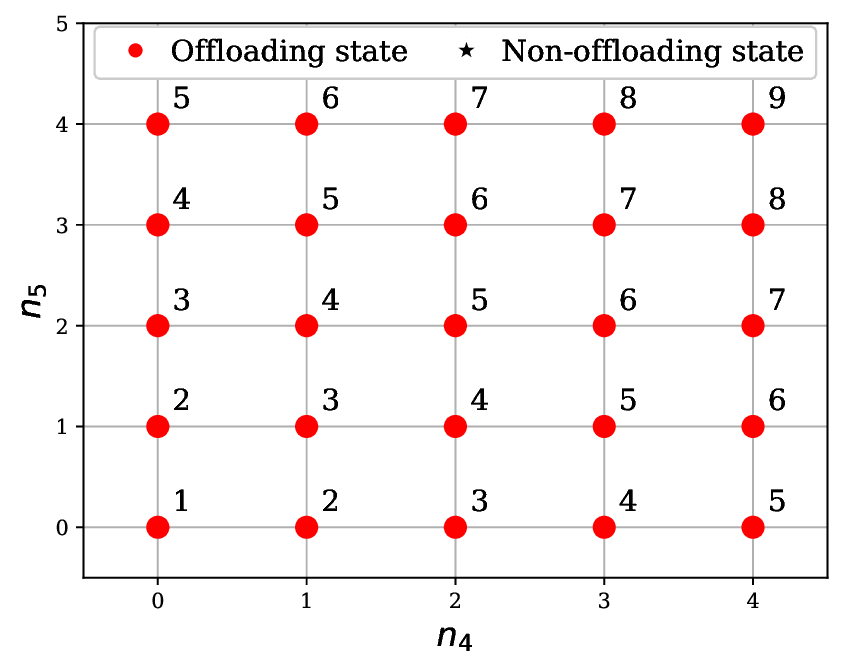}
    		\caption{$n_3 = 2$.}
    		\label{fig:n33N5}
    	\end{subfigure}
    	\caption{Visualisation of Theorem \ref{Theo:L*_thesmallest} for the state vector $\mathbf{s} = \left(0, 0, n_3, n_4, n_5\right)$. Offloading and non-offloading states are represented by red and black stars, respectively. The number beside each state is the optimal number of tasks, $L^*$, to be offloaded to achieve the minimum expected cost; $L^*=0$ implies that, optimally, no task is offloaded.}
     \label{fig:n3N5}
	\end{figure*}

	In this example, we illustrate the result of Theorem \ref{Theo:L*_thesmallest} visually for a system with a 3-dimensional state vector $s = (n_1, n_2, n_3)$. Then, with $n_3 = 0, 1, 2$, for visualization, we consider the 2-D slices of the state space and depict them as Figs.~\ref{fig:n3}(a)-\ref{fig:n3}(c) correspondingly where we based our simulation on the system parameters described in Table \ref{tbl:sys-param-lin-eq} with $p_i = 1/6$ for $i=1,\ldots,N$. In these figures, red dots represent offloading states (states associated with positive optimal offloading decisions), and black stars represent non-offloading states (states associated with optimal offloading decision 0). The number next to each state $\mathbf{s}$, as shown in the figures, is the optimal number of tasks, $L^*$, to be offloaded to achieve the minimum expected cost $J_T(\mathbf{s})$ defined in Eq. \eqref{DP_Eq}. For non-offloading states, $L^*=0$, suggesting that no task is offloaded as the optimal decision. The figures with $n_3 \ge 2$ would contain all red dots as in Fig.~\ref{fig:n3}(c). Moreover, if the optimal offloading decision of a state $(n_1, n_2, 2)$ is $L^*$ in Fig.~\ref{fig:n3}(c), then, the optimal offloading decision of state $(n_1, n_2, n_3)$ for $n_3 \ge 2$ would be $L^* + n_3 - 2$. 
    
    In addition to Theorem \ref{Theo:L*_thesmallest}, we recall that the optimal policy offloads tasks from the most imminent to the least imminent deadlines; hence, the optimal decision can be obtained following the rule: the optimal decision is 0 if starting at a non-offloading state. Otherwise,
\begin{enumerate}
	\item Moving to the left (decreasing $n_1$). Stop when reaching a non-offloading state (black star).
	\item If $n_1=0$ before reaching a non-offloading state, moving down (decreasing $n_2$). Stop when reaching a non-offloading state.
	\item If $n_2=0$ before reaching a non-offloading state, decreasing $n_3$ by 1 and repeating step 1.
\end{enumerate}
   As we reach a non-offloading state following this rule, the number of steps is equal to the optimal offloading decision.

	From states with component $n_2 \geq 1$ in Fig. \ref{fig:n3}(a), such as $\left(0, 2, 0\right)$, $\left(1, 2, 0\right)$, $\left(1, 1, 0\right)$, $\left(2, 1, 0\right)$, etc., we can reach the non-offloading state $\left(0, 1, 0\right)$ with a smaller number of offloaded tasks  than state $\left(0, 0, 0\right)$.
    A similar argument applies for states with component $n_2 = 0$, like $\left(1, 0, 0\right)$, $\left(2, 0, 0\right)$, etc.,  whose ``nearest'' non-offloading state is  $\left(0, 0, 0\right)$.
    
    In Fig. \ref{fig:n3}(b), only the state $\left(0, 0, 1\right)$ is non-offloading. The optimal offloading decisions of all the offloading states shown are the smallest number of most imminent tasks to be offloaded to reach state $\left(0, 0, 1\right)$.
    Fig. \ref{fig:n3}(c) does not have any non-offloading state. For example, the optimal offloading decision for the state $\left(0, 0, 2\right)$ is 1 to reach the non-offloading state $\left(0, 0, 1\right)$ which is shown in Fig. \ref{fig:n3}(b).
	}
	\fi

	\iftext{
	We conduct a similar experiment for the parameter setting of Table \ref{tbl:sys-param-lin-eq2} and present the result in Figs.~\ref{fig:n3N5}(a)-\ref{fig:n3N5}(c). In this setup, the system has a 5-dimensional state, $\mathbf{s} = (n_1, n_2, n_3, n_4, n_5)$. Since state vectors that have either $n_1 > 0$ or $n_2 > 0$ are all offloading states, we only demonstrate states associated with $n_1 = n_2 = 0$. By applying the same rule described above with $n_1$ is now $n_4$ and $n_2$ is now $n_5$, it can be seen that the smallest number of steps taken to reach a non-offloading state is also the optimal offloading decision attached to each dot.
	}
\fi

\subsection{Optimal Offloading Decisions for Adjacent States}

	\iftext{
 \begin{table}
		\centering
        \caption{{System parameters for Fig. \ref{fig:convex} in which $p_i = \frac{1-p_0}{N}, i=1,\ldots, N$.}}
		\subcaption{{Parameters for Fig. \ref{fig:convex}(a).}}
        {
		\begin{tabular}{|c|c|c|c|c|c|}
        \hline
			\textbf{Parameters} & $p_{\sf a}$ & $C_{\sf p}$ & $C_{\sf o}$ & $\mu$ & $p_0$ \\ \hline
			\textbf{Values} & $0.5$ & $ 3 $ & $ 1 $ & $0.5$ & $0.5$ \\
            \hline
		\end{tabular}
        }
            \bigskip
		
		\subcaption{{Parameters for Fig. \ref{fig:convex}(b).}}
        {
        \begin{tabular}{|c|c|c|c|c|c|}
        \hline
			\textbf{Parameters} & $p_{\sf a}$ & $C_{\sf p}$ & $C_{\sf o}$ & $\mu$ & $p_0$\\ \hline
			\textbf{Values} & $0.4$ & $ 4 $ & $ 1 $ & $0.3$ & $0.3$\\
            \hline
		\end{tabular}
        }
            \bigskip
  
		\subcaption{{Parameters for Fig. \ref{fig:convex}(c).}}
        {
		\begin{tabular}{|c|c|c|c|c|c|}
        \hline
			\textbf{Parameters} & $p_{\sf a}$ & $C_{\sf p}$ & $C_{\sf o}$ & $\mu$ & $p_0$\\ \hline
			\textbf{Values} & $0.7$ & $ 2 $ & $ 1 $ & $0.6$ & $0.6$\\
            \hline
		\end{tabular}
        }
        
	\end{table}

    
\begin{figure*}
	\begin{subfigure}[b]{0.295\textwidth}
        \captionsetup{justification=centering}
		\includegraphics[width=1.12\textwidth]{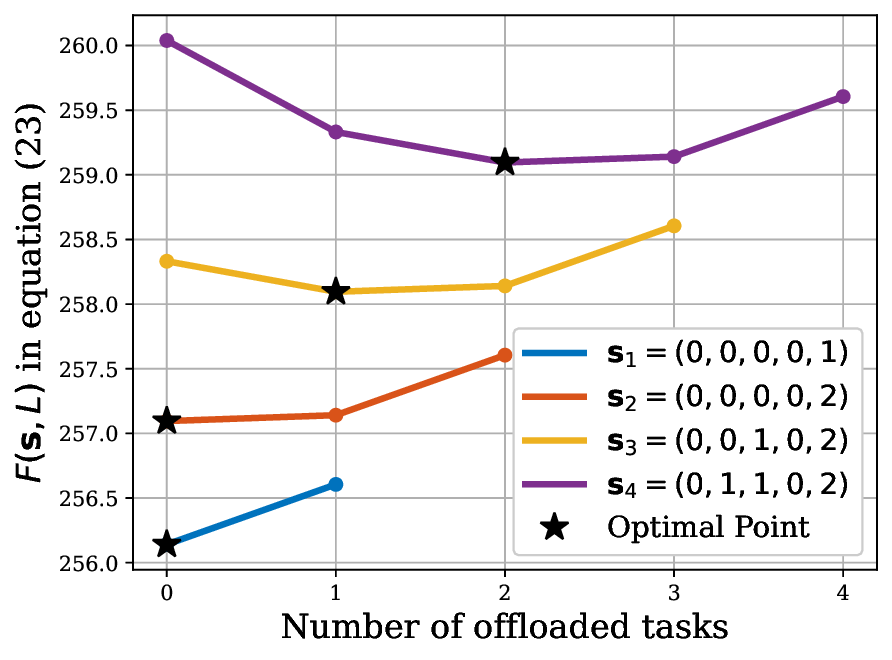}
		\caption{Parameters in Table III(a).}
		\label{fig:convex-dp-n}
	\end{subfigure}
	\hspace{1.5em}
	\begin{subfigure}[b]{0.295\textwidth}
        \captionsetup{justification=centering}
		\includegraphics[width=1.12\textwidth]{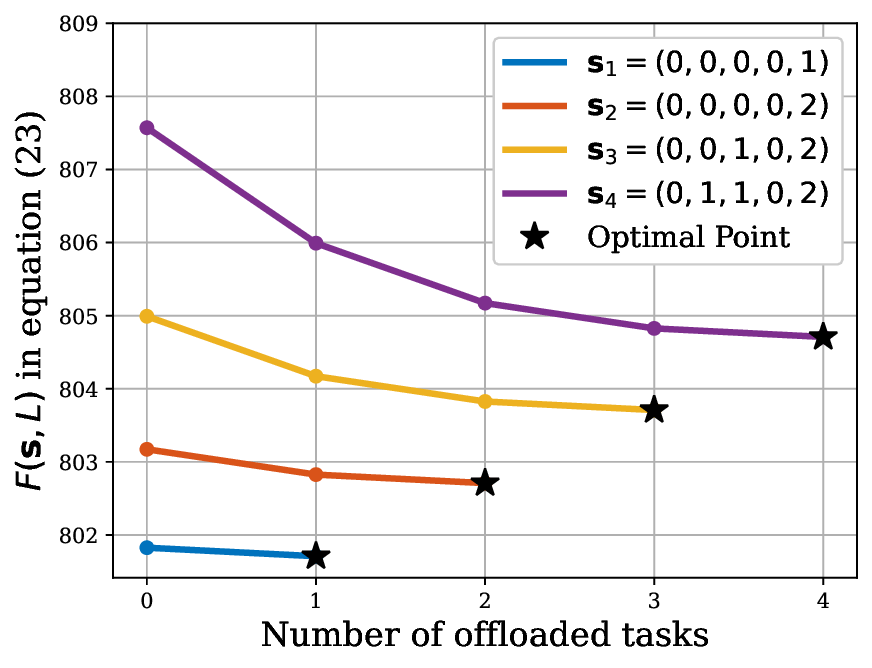}
        \caption{Parameters in Table III(b).}
		\label{fig:convex-dp-o}
	\end{subfigure}
	\hspace{1.5em}
	\begin{subfigure}[b]{0.295\textwidth}
        \captionsetup{justification=centering}
		\includegraphics[width=1.12\textwidth]{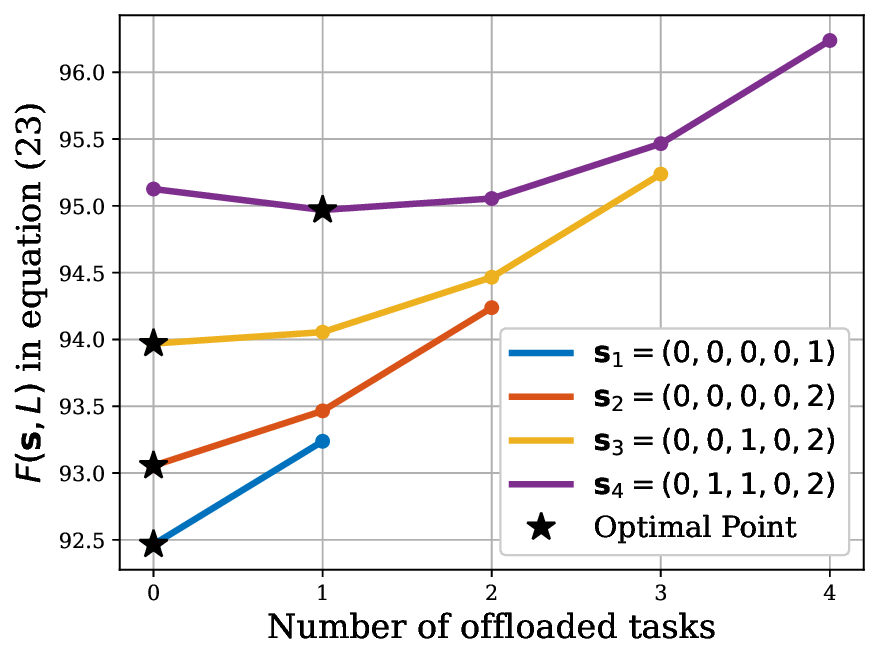}
        \caption{Parameters in Table III(c).}
		\label{fig:convex-dp-l}
	\end{subfigure}
	\caption{Visual interpretation of Eqs. (\ref{eq:tr_down})-(\ref{eq:tr_up}) and Theorem \ref{Theo:adjacent_L*}. The star markers ($\mathlarger{\mathlarger{\boldsymbol{\star}}}$) mark the optimal points. The figures demonstrate the convexity of the system cost with respect to the offloading decision, and also, verify the correctness of Eqs. (\ref{eq:tr_down})-(\ref{eq:tr_up}).}
    \label{fig:convex}
	\end{figure*}
 
    In Fig.~\ref{fig:convex}(a)-\ref{fig:convex}(c), we visualize the results of Eqs. (\ref{eq:tr_down}), (\ref{eq:tr_up}) and Theorem \ref{Theo:adjacent_L*}. In addition, we also aim to verify the convexity of the function $F(\mathbf{s}, L)$ defined in Eq. \eqref{F_def} as stated in Lemma \ref{Lem:Convexity}. For these examples, all figures are associated with $N=5$ as the dimension of state vectors over a time horizon $T = 1,000$. Figs.~\ref{fig:convex}(a), \ref{fig:convex}(b), and \ref{fig:convex}(c) use the parameters listed in Tables III(a), III(b), and III(c), respectively. In these figures, the optimal points attained at the optimal offloading decisions are marked by star symbols. On the vertical axis, we graph the minimum expected cost $F(\mathbf{s}_i, L)$ defined in Eq. \eqref{F_def} attained by offloading $L$ most imminent tasks from state $\mathbf{s}_i$ given that the AMA is available in the instant time slot. \comment{This is introduced in Eq. (\ref{eq:JsL_def}).}
    
    In these three figures, for $i = 1, 2, 3, 4$, we consider the states   $\mathbf{s}_1 = \left(0, 0, 0, 0, 1\right)$, $\mathbf{s}_2 = \left(0, 0, 0, 0, 2\right)$, $\mathbf{s}_3 = \left(0, 0, 1, 0, 2\right)$, and $\mathbf{s}_4 = \left(0, 1, 1, 0, 2\right)$. These states are chosen such that $\mathbf{s}_{i}$ is adjacent to $\mathbf{s}_{i+1}, i = 1, 2, 3$. We note that, for example, in Fig.~\ref{fig:convex}(a), the optimal points of $\mathbf{s}_1$, $\mathbf{s}_2$, $\mathbf{s}_3$, and $\mathbf{s}_4$ represented by the star symbols are attained by offloading 0, 0, 1, and 2 most imminent tasks, respectively. This implies that these are the optimal offloading decisions for the considered states, respectively. A similar note applies to Figs.~\ref{fig:convex}(b) and \ref{fig:convex}(c). The presented results indicate that the optimal offloading decision of a state differs from that of its adjacent state by 1, or both are capped at 0, as  Eqs. (\ref{eq:tr_down})-(\ref{eq:tr_up}) and Theorem \ref{Theo:adjacent_L*} suggest. For example, in Fig. \ref{fig:convex}(a), the optimal decision of $\mathbf{s}_4$ is 2, and that of $\mathbf{s}_3$ is 1; hence, the difference is 1. The same observation applies for the pair $\mathbf{s}_3$ and $\mathbf{s}_2$. The optimal decision of $\mathbf{s}_2$ is 0. Therefore, that of $\mathbf{s}_1$ is also 0. Figs~\ref{fig:convex}(b) and \ref{fig:convex}(c) present the same properties. 
	}
	\fi
	
\subsection{Memory Savings Using Equation (\ref{eq:gen2sem_eq})}

	\iftext{
   \comment{In order to numerically compute the DP Eq. (\ref{DP_Eq}),
	we store the computed value of $J_T(\mathbf{s})$ in memory for a given arbitrary state $\mathbf{s}$ and time horizon $T$.
	By using Eq. (\ref{eq:gen2sem_eq}), the size of memory required is reduced.  This is because saving $J_T(\cdot)$ values is only required for lean states $\mathbf{s}^{(\ell)}$ where the number of lean states is smaller than that of ``generic'' states (regular states). Note that these savings are achieved at the expense of calculating the term $C_{\ell}$ in Eq.  (\ref{eq:gen2sem_eq}).} 
	
	We would like to highlight that, in the computational load reduction process described in Section \ref{Sec:OptimalPolicy}, the quadruplet $(\mathbf{s}^{(\ell)}, T, J_T(\mathbf{s}^{(\ell)}), L_{\ell})$ are stored for different lean states $\mathbf{s}^{(\ell)}$. Let us call each such quadruplet that needs to be saved an \textit{entry}. {In Figs.~\ref{fig:memory_savings}(a)-\ref{fig:memory_savings}(c), we demonstrate the memory-saving efficiency of Eq. \eqref{eq:gen2sem_eq} by reducing the number of entries that need to be saved. In particular, instead of an infinite generic state space, Eq. \eqref{eq:gen2sem_eq} allows us to store only entries associated with lean states in the finite lean state space, while still being able to address every generic state. We emphasize that saving the aforementioned quadruplets as entries in the memory allows to retrieve the system cost and optimal offloading decisions instead of completely relying on the recursive DP Eq. \eqref{DP_Eq} as detailed in Section \ref{Sec:OptimalPolicy}, thus, reducing the computational burden.} We consider three cases with $N$ set to 3, 4, and 5 to provide a broader illustration and highlight the increase in memory savings as the dimension $N$ of the state vectors grows. This improvement in memory efficiency is attributed to the fact that larger state vector dimension $N$ requires more iterations to solve the DP Eq. \eqref{DP_Eq}, thereby demanding more memory for computation. Consequently, the difference in memory usage between scenarios where the derived lean state transformation Eq. \eqref{eq:gen2sem_eq} is applied and where it is not becomes more pronounced. This observation suggests that the shown results imply a lower bound in the memory saving performance of Eq. \eqref{eq:gen2sem_eq}; hence, more significant improvement is expected for larger values of $N$.
    
    We note that different parameters other than $N$ and $T$ will not affect the memory savings. In the first case, Eq. (\ref{DP_Eq}) is used, while in the second case, Eq. (\ref{DP_Eq}) is used with the aid of Eq. (\ref{eq:gen2sem_eq}).
	The line in blue represents the values saved using only Eq. (\ref{DP_Eq}), while the line in red
	utilizes both Eqs. (\ref{DP_Eq}) and (\ref{eq:gen2sem_eq}). The results suggest that the proposed method offers 95\%, 97\%, and 98\% reduction in memory usage for $T=15$ time slots when $N=3$, 4, and 5, respectively.
	
  
	\begin{figure*}
	\begin{subfigure}[b]{0.295\textwidth}
		\centering
		\includegraphics[width=1.12\textwidth]{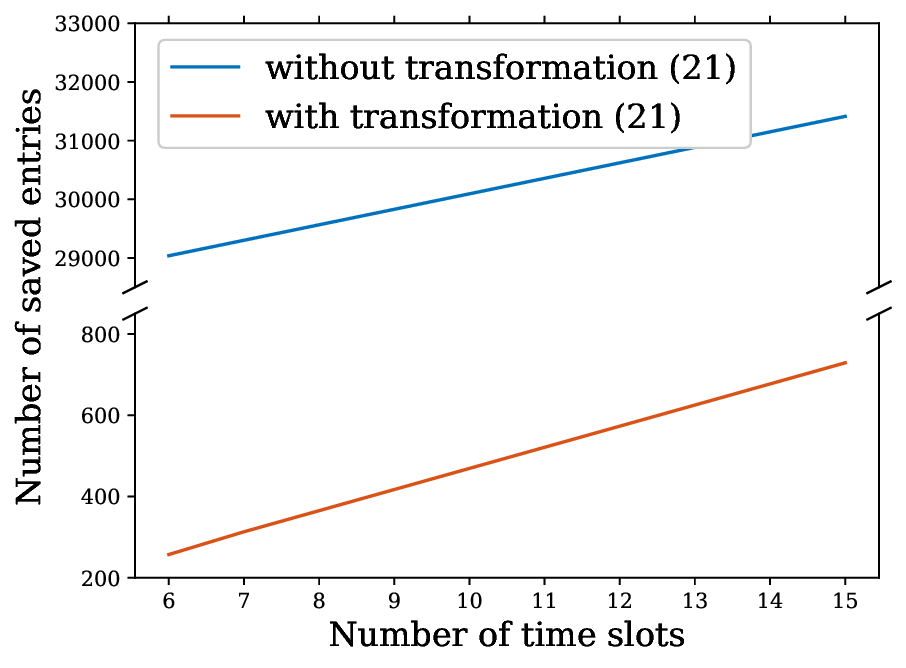}
		\caption{$N=5$.}
		\label{fig:memory-saving-n5}
	\end{subfigure}
    \hspace{1.5em}
	\begin{subfigure}[b]{0.295\textwidth}
		\centering
		\includegraphics[width=1.12\textwidth]{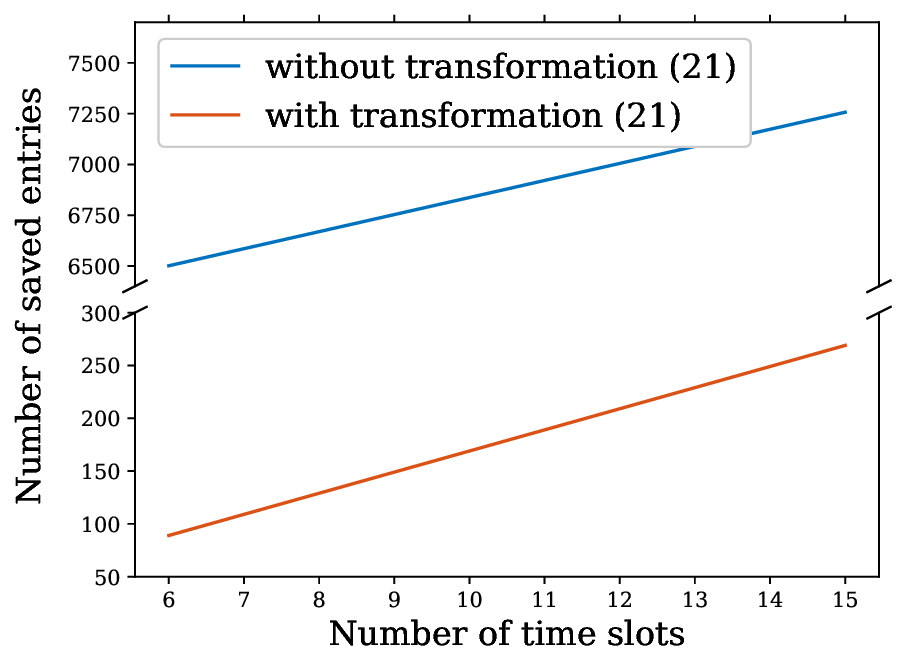}
		\caption{$N=4$.}
		\label{fig:memory-saving-offload-n4}
	\end{subfigure}
    \hspace{1.5em}
	\begin{subfigure}[b]{0.295\textwidth}
		\centering
		\includegraphics[width=1.12\textwidth]{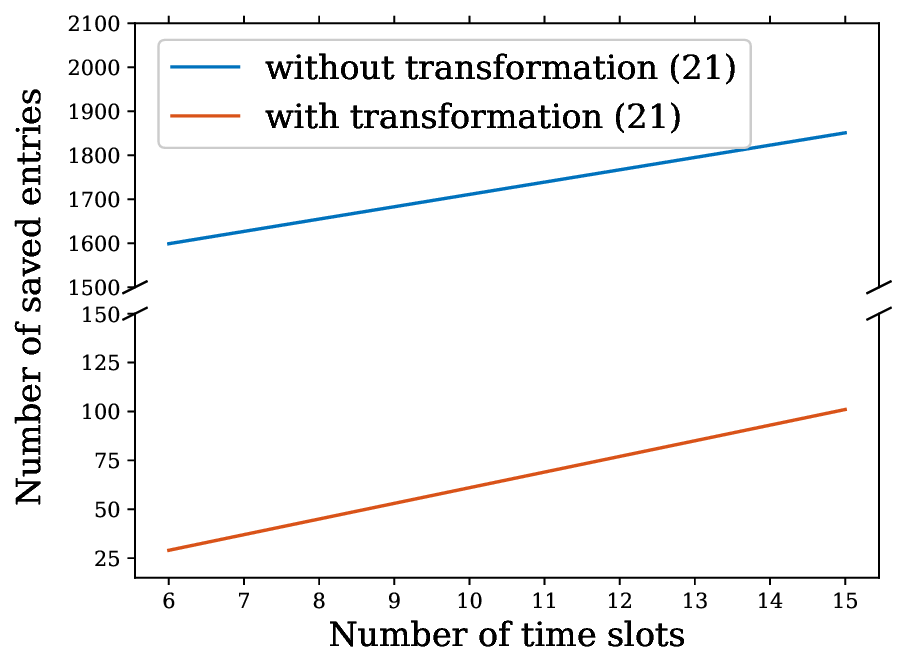}
		\caption{$N=3$.}
		\label{fig:memory-saving-process-n3}
	\end{subfigure}
	\caption{Memory savings achieved by using the transformation Eq. \eqref{eq:gen2sem_eq}. The memory usage is reflected via the number of saved entries; each entry is a quadruplet $(\tilde{\mathbf{s}}, T-t, J_{T-t}(\tilde{\mathbf{s}}), \tilde{L}^*), t\in \{0, \ldots, T-1\}$, where $\tilde{\mathbf{s}}$ is either a reduced or lean state, and $\tilde{L}^*$ is the corresponding optimal offloading decision.}
	\label{fig:memory_savings}
	\end{figure*}
    }
	\fi

\subsection{Performance Comparison Against Baseline Methods}

    This subsection presents simulation results aimed at verifying the optimality of the proposed algorithm and illustrating its performance relative to baseline methods.

    \begin{table}
    \label{tab:baseline_compare}
		\centering
        \caption{{System parameters for Fig. \ref{fig:baseline_compare}.}}
		\label{tbl:sys-param-baseline}
            {
		\begin{tabular}{|c|c|c|c|c|c|c|c|c}
        \hline
			  \textbf{Parameters} & $T$   & $p_{\sf a}$ & $C_{\sf p}$ & $C_{\sf o}$ & $N$ & $p_i,  i=0,\ldots,N$\\ \hline
			\textbf{Values} & $1000$ & $0.1$ & $ 3 $ & $ 1 $ & $10$ & $1/11$ \\
            \hline
		\end{tabular}
        }
	\end{table}

    \captionsetup[figure]{
  textfont=normal, 
  labelfont={color=blue}, 
  skip=10pt}
    \begin{figure}
		\centering
		\includegraphics[scale=0.5]{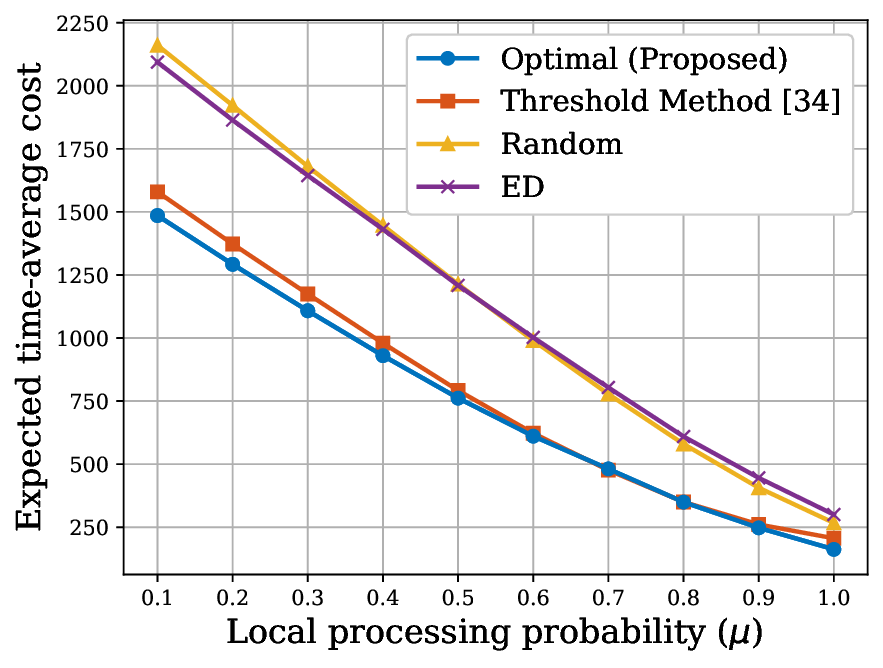}
		\caption{{Overall cost (defined in Eq. \eqref{eq:sum_cost}) versus the local processing probability, $\mu$, for the proposed optimal algorithm, threshold method \cite{9488821}, ED, and random methods. For each value of $\mu$, the optimal threshold is used for the threshold method.}}
		\label{fig:baseline_compare}
	\end{figure}
 
    The system parameter configuration for the simulation of Fig.~\ref{fig:baseline_compare} is provided in Table IV. We compare the proposed optimal task offloading algorithm presented in Algorithm \ref{Optim_Algo} against the following baseline schemes:
    \begin{itemize}
        \item \textit{Threshold Method}: This method is described in \cite{9488821} where multiple users manage their own queues of computational tasks and perform task offloading independently. Each user stabilizes its queue by defining a separated threshold $B$ and offloading tasks to maintain the queue length, i.e., the number of tasks in the queue, to be less than or equal to $B$. The goal of \cite{9488821} is to derive an optimal threshold value $B$ for each user to minimize the average processing delay and the cost of using cloud services.
        
        ~~In our context, the task processing delay is fixed (1 time slot for the local processing service and no delay for the remote server). Therefore, to adapt the threshold method in our considered scenario, the optimal threshold $B$ is derived only to minimize the system cost constituted by the task offloading cost ($C_{\sf o}$) and the task expiration cost ($C_{\sf p}$).
        \item \textit{Expiry-Driven (ED) Task Offloading Method:} This method only focuses on offloading tasks that are going to expire in the next time slot.
        \item \textit{Random Method:} This method uniformly randomly selects tasks to be offloaded where the number of offloaded tasks is also randomly defined between 0 and the total number of tasks in the queue.
    \end{itemize}

    \comment{
    \begin{itemize}
        \item this method is described in \cite{6353239} as a task offloading scheme for mobile devices through WiFi hotspots. This method uses spontaneous connectivity to WiFi and transfers data on the spot. By adopting this method in our context, whenever the AMA is available to support the task offloading, the BS transfers all un-executed tasks in the queue. If the AMA does not arrive and the local processing service is available, the most imminent task is processed.
        
        \item \textit{Threshold method}: in this method, a parameter $\theta \in \{1, \ldots, N\}$ is predefined as a deadline threshold, and tasks having deadlines less than or equal to $\theta$ are offloaded with the present of the AMA.
        \item \textit{Random method}: this method randomly uniformly selects the number of tasks to be offloaded. Then, tasks are offloaded from the most imminent ones.
    \end{itemize}
    }

    {We note that the ED and Random methods do not incorporate the ability to offload excessive tasks—i.e., tasks that clearly exceed the local processing capacity of the system, as described in Subsection \ref{subsec:reduced_state}. To ensure a fair comparison, we constrain the system state generation to a reduced state space where no excessive tasks occur throughout the operation. The results are presented in Fig.~\ref{fig:baseline_compare}, where the y-axis represents the expected cost, which corresponds to the minimization objective defined in Eq. \eqref{eq:sum_cost}.} In this figure, the performance of the threshold method is associated with the optimal threshold for each parameter $\mu$ along the x-axis. It is important to emphasize that although the threshold method achieves its optimality, it is constrained to the family of threshold-type policies, leading to a higher average cost per task compared to our proposed optimal algorithm. In addition, the other two baselines, ED and random algorithms result in significantly higher average cost per task compared to the two formerly mentioned methods. As $\mu$ becomes large, i.e., the local service is offered with high probability, the system relies less on the offloading operation, enclosing the gaps among all considered schemes.

    \begin{figure}
		\centering
    \includegraphics[scale=0.5]{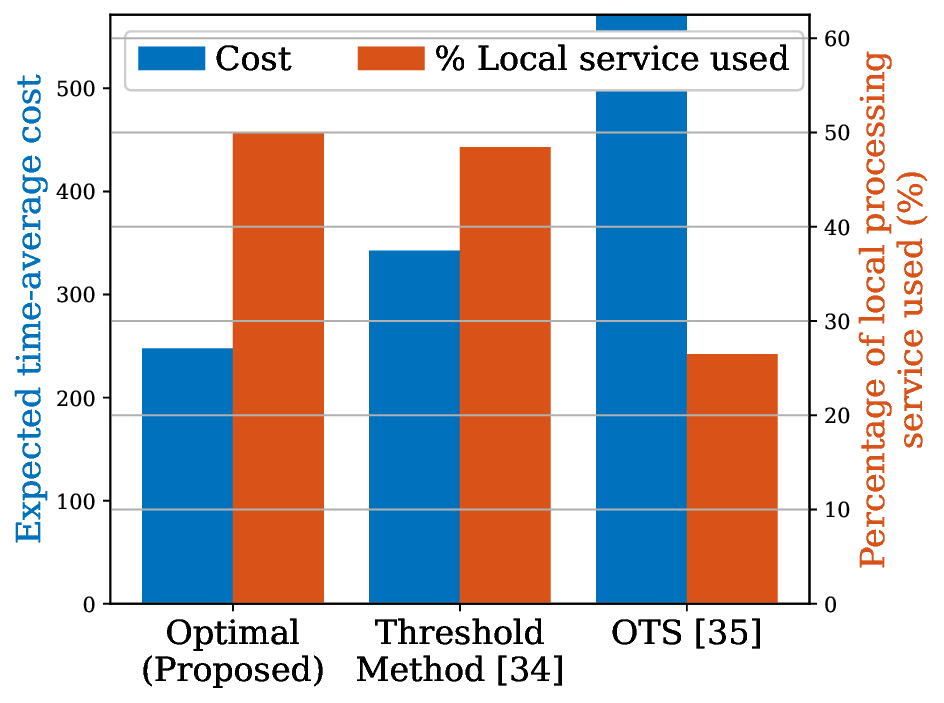}
		\caption{{Evaluation of the average cost per task incurred and the percentage of available local processing services that can be used by the proposed method, the threshold method \cite{9488821} at its optimal threshold value, and the OTS scheme \cite{6353239}.}}
		\label{fig:baseline_compare_bar}
	\end{figure}
    
    {The next simulation, based on the setup in Table IV with $\mu = 0.9$, is presented in Fig.~\ref{fig:baseline_compare_bar}. Our goal is to evaluate the effectiveness of the proposed algorithm in minimizing the objective cost defined in Eq. \eqref{eq:sum_cost}. For this purpose, we report two metrics: the average cost per task (left y-axis) and the utilization rate of local processing services (right y-axis). The latter quantity is calculated as the ratio of the number of tasks processed locally to the total number of local processing opportunities that occurred during system operation.} It is worth noting that a local processing session is not used when it is available in a time slot while there is no task presenting in the queue. An empty queue could be due to over-offloading decisions of certain algorithms or due to low task arrival probability where the latter reason is not influenced by an employed algorithm. Therefore, to accurately evaluate the considered methods, we set the task arrival probability, i.e., $1-p_0$, to be approximately 0.9. As a result, the percentage of local processing services used as shown on the right y-axis suggests how efficiently each of the involved methods can make use of the free-of-cost local service.

    In addition to the threshold method that has been discussed previously, we consider another policy called {\textit{On-the-spot} (OTS)} offloading policy. OTS is described in \cite{6353239} as a method that uses spontaneous connectivity to WiFi and transfers all the computational tasks on the spot. By adopting this method in our context, whenever the AMA arrives, the BS transfers all un-executed tasks, leaving an empty queue.

    As in the previous simulation, the performance of the threshold method has been optimized in Fig.~\ref{fig:baseline_compare_bar}. At its optimal performance, the threshold method still incurs a higher average cost per task and utilizes the local processing service slightly less effectively than our proposed approach. In terms of the OTS scheme, although this method safely offloads all tasks when the AMA is available to minimize the number of expired tasks, it is almost five times more expensive to handle a task than the proposed algorithm. This discrepancy arises from the OTS method's ineffective use of the cost-free local service, demonstrated by a 25\% lower utilization of local resources compared to our approach.
    
    \section{Conclusion and Future Research} \label{Sec:Conclusion}

	\iftext{
	In this work, we studied a mobile edge computing system with dynamic user demand. In the context of an optimal stochastic control framework for serving user tasks, we  considered the following features: tasks with firm deadlines, their random offloading to a remote server (AMA) or their processing by a local server (BS) with intermittent service. We considered an expected time-average cost over a finite time horizon and formulated a DP problem toward the minimization of this cost. In order to tackle the ``Curse of Dimensionality'', we studied important characteristics of the optimal policy and reduced the computational load for its calculation. In particular, we proved that the DP Equation can be evaluated for every given state (in the infinite state space of our model) by considering a specific finite space called \textit{lean state} space. Further reduction in the computational load was achieved by using the concept of \textit{adjacent states}. This allowed us to evaluate the optimal cost for all such states from knowledge of the cost in only one state. Finally, 
	based on these properties, we described an optimal task offloading  policy. Our future research aims to extend the theoretical results to a broader context that accommodates multiple task arrivals per time slot over the considered time horizon. Additionally, developing a heuristic approach to balance cost minimization and execution time could be a promising and practical solution to the considered problem. {Moreover, extending our model to include the cost of local resource consumption would enhance its ability to reflect real-world scenarios. This addition allows the system to more accurately capture the trade-offs among task expiration, offloading, and local processing, thereby providing a more comprehensive understanding of optimal system operation.}
	}
	\fi
		
\section*{Appendix}
For the presentation of the proofs of theorems, lemmas and propositions,  hereafter, we will modify some of the notations introduced earlier in the paper. Specifically, in Lemma \ref{Lem:Convexity}, we present the convexity of function $F(\mathbf{s}, L)$ that is defined in Eq. (\ref{F_def}). In order to prove this lemma, we will need to prove a more general result where tasks can be offloaded starting from an arbitrary deadline $d = 1, 2, \ldots, N$, and following the ascending order of tasks' deadlines. Therefore, we will modify the notations to a more general form to facilitate the proofs presented in this appendix.
	
	We denote by $\bar{\mathbf{s}}_{dL}$ the state obtained by offloading, from $\mathbf{s}$, $L$ most imminent tasks having deadline greater than or equal to $d$. To support the interpretation of our notations, we provide the following example: 
	\begin{example}
		Given state $\mathbf{s} = \left(0, 5, 6, 7, 8\right)$, with $d=3$ and $L=7$, the state $\bar{\mathbf{s}}_{37}$ is obtained by offloading 7 most imminent tasks starting from deadline 3, thus, $\bar{\mathbf{s}}_{37} = \left(0, 5, 0, 6, 8 \right)$. Similarly, with $d=5$, we have $\bar{\mathbf{s}}_{57} = \left(0, 5, 6, 7, 1 \right)$. $\square$
	\end{example}
	In addition, we define the function $F(T, \mathbf{s}, d, L)$ by
	\begin{align}\label{F_def_gen}
		F\left(T, \mathbf{s}, d, L\right) = J_T^{\overline{\text{A}}}\left(\bar{\mathbf{s}}_{dL}\right) + LC_{\sf o}.
	\end{align}
	The right-hand side of Eq. (\ref{F_def}) can be interpreted as the minimum expected cost over $T$ time slots that can be achieved, given that $L$ most imminent tasks starting from deadline $d$ are offloaded from state $\mathbf{s}$ in the initial time slot.

	We define the domain of function $F\left(T, \mathbf{s}, d, L\right)$ as follows:
	\begin{align}\label{F_domain2}
    \begin{split}
	&\mathbb{L}_1\left(\mathbf{s}\right) = \left\{n_1, n_1+1, \ldots, \sum_{i=1}^N n_i \right\}, \\
    &\mathbb{L}_d\left(\mathbf{s}\right) = \left\{0, 1, \ldots, \sum_{i=d}^N n_d \right\}, \text{ for } d = 2, \ldots, N. 
    \end{split}
	\end{align}
	
	From the newly introduced notation, $\mathbb{L}_1\left(\mathbf{s}\right)$ is equivalent to the domain of function $F\left(\mathbf{s}, L\right)$ defined in (\ref{F_domain2_simple}), and
	\begin{align}
		F\left(\mathbf{s}, L\right) \equiv F\left(T, \mathbf{s}, d = 1, L\right),
	\end{align}
	where function $F\left(\mathbf{s}, L\right)$ is given in Eq. (\ref{F_def}).

	We define 
	\begin{align}\label{eq:JsL_def}
		J_T^{\text{A}}\left(\mathbf{s}, L\right) = \mathcal{C}^{\text{A}}\left(\mathbf{s}, L\right) + G_T^{\text{A}}\left(\mathbf{s}, L\right).
	\end{align}
as the minimum average cost over the horizon $T$ given that $L$ most imminent tasks are offloaded from $\mathbf{s}$ in the initial time slot with the AMA's presence. Then, we have the expression:
	\begin{align}\label{rel_orig}
		J_T^{\text{A}}\left(\mathbf{s}, L\right) = J_T^{\text{A}}\left(\bar{\mathbf{s}}_{1L}, 0\right) + LC_{\sf o},~ L \in \mathbb{L}\left(\mathbf{s}\right)
	\end{align}
	which can be explained as follows. We note that by offloading $L$ most imminent tasks from $\mathbf{s}$, we pay a cost $LC_{\sf o}$, and reach state $\bar{\mathbf{s}}_{1L}$. Therefore, if we wish to describe the offloading of $L$ tasks on the left-hand side of Eq. (\ref{rel_orig}), this is equivalent to removing the $L$ most imminent tasks from state $\mathbf{s}$ to reach state $\bar{\mathbf{s}}_{1L}$, and offloading 0 task from $\bar{\mathbf{s}}_{1L}$. Finally, we further add the offloading cost $LC_{\sf o}$ to the overall cost. Eq. (\ref{rel_orig}) describes this idea.
	
	We note that the minimum average cost when offloading 0 task is not affected by the presence of the AMA, i.e.,
	\begin{align}\label{equi_noAMA}
		J_T^{\text{A}}\left(\bar{\mathbf{s}}_{1L}, 0\right) = J_T^{\overline{\text{A}}}\left(\bar{\mathbf{s}}_{1L}\right), \text{ for } L \in \mathbb{L}\left(\mathbf{s}\right).
	\end{align}
	Thus, we have
	\begin{align}\label{JsL_intro}
		J_T^{\text{A}}\left(\mathbf{s}, L\right) = J_T^{\overline{\text{A}}}\left(\bar{\mathbf{s}}_{1L}\right) + LC_{\sf o}, \text{ for } L \in \mathbb{L}\left(\mathbf{s}\right).
	\end{align}
	
	Furthermore, considering two offloading decisions $L_1$ and $L_2$ where $L_1+L_2, L_1, L_2 \in \mathbb{L}\left(\mathbf{s}\right)$. We describe the minimum average cost attained by offloading $L_1+L_2$ tasks from $\mathbf{s}$ as follows:  
	\begin{enumerate}
		\item The cost $L_1C_{\sf o}$ to offload $L_1$ tasks from $\mathbf{s}$ and reach state $\bar{\mathbf{s}}_{1L_1}$.
		\item The minimum average cost attained by offloading $L_2$ tasks from $\bar{\mathbf{s}}_{1L_1}$, i.e., $J_T^{\text{A}}\left(\bar{\mathbf{s}}_{1L_1}, L_2\right)$.
	\end{enumerate}
	Therefore, we have the following expression:
	\begin{align}\label{JsL_intro1}
		J_T^{\text{A}}\left(\mathbf{s}, L_1+L_2\right) = J_T^{\text{A}}\left(\bar{\mathbf{s}}_{1L_1}, L_2\right) + L_1C_{\sf o}.
	\end{align}
	
	The proofs of results introduced through out the paper will be presented in the subsequent sections starting with the proof of Lemma \ref{lem_Catalan}.
	

\section*{Appendix A}
\section*{Proof of Lemma \ref{lem_Catalan}}\label{proof_lem_Catalan}
	A sequence $(a_1,a_2, \ldots, a_N)$ of non-negative integers is called {\em a Catalan sequence} \cite{stanley2015catalan} if:
	\begin{align}
		\label{eq:catalan}
		1 \leq a_1 \leq a_2 \leq \cdots \leq a_N \mbox{ and } a_i \leq i, \mbox{ for all $1 \leq i \leq N$.}
	\end{align}
	
	In the proof we show that there is a one-to-one correspondence between reduced sequences and Catalan sequences of the same length $N$. The correspondence is defined as follows.
	
	Given a reduced sequence $(n_1,n_2, \ldots, n_N)$, which satisfies inequalities \eqref{ineq:reduced_cond}, define a sequence $(a_1,a_2, \ldots, a_N)$ as follows: $a_i =  (n_1+1) +n_2+\cdots+n_i$. It is clear that the resulting sequence satisfies the Catalan sequence property~\eqref{eq:catalan}. Conversely, given a Catalan sequence $(a_1,a_2, \ldots, a_N)$, which satisfies property~\eqref{eq:catalan}, define the sequence $(n_1,n_2, \ldots, n_N)$ as follows: $n_1 = 0$, and $n_i = a_i - a_{i-1}$ for $2 \le i \le N$. The resulting sequence contains elements of a reduced state vector because 
	\begin{align*}
    \begin{split}
		&n_1 + n_2 + n_3+ \cdots + n_i 
		\\
        &=  0 + a_2 -a_1 + a_3-a_2 + \cdots + a_i - a_{i-1} \\
        &= a_i -a_1 \leq i-1 ,
    \end{split}
	\end{align*}
	since $a_1 = 1$. Also observe that the resulting correspondence between reduced and Catalan sequences of the same length $N$ is one-to-one.
	
	The proof of Lemma \ref{lem_Catalan} is now complete since in exercise 78 from \cite{stanley2015catalan}, the number of Catalan sequences of length $N$ is equal to the Catalan number $C_N$.

\section*{Appendix B}
	\section*{Proof of Proposition \ref{Pro:SemiRedState}} \label{proof_Pro:SemiRedState}
	Considering a state $\mathbf{s} = \left(n_1, \ldots, n_N\right)$ and a time horizon $T$. Let $\mathbf{s}^{(\sf r)} = \left(n^{(\sf r)}_1, \ldots, n^{(\sf r)}_N\right)$ be the corresponding reduced state obtained via Algorithm \ref{Gen2Red_Algo}, and let $\mathbf{s}^{(\ell)} = \left(n^{(\ell)}_1, \ldots, n^{(\ell)}_N\right)$ denote the corresponding lean state obtained according to Definition \ref{Def:SemRedStates}. As $n_i^{(\ell)}$, $i = 1, \ldots, N$, are defined by Eq. (\ref{eq:semi_element_def}) with the parameters $\gamma_i$, $i = 1, \ldots, N$, are given in Eq. (\ref{eq:gamma_def}), we have: $n_i^{(\sf r)} \le n_i^{(\ell)} \le n_i, ~ i = 1, \ldots, N$.

Moreover, we recall that $n_i - n^{(\sf r)}_i$ tasks having deadline $i$ in $\mathbf{s}$, $ i = 1, \ldots, N$ are excessive tasks, i.e., they are guaranteed to expire if not offloaded within the first $i$ time slots. Therefore, $n_i - n^{(\ell)}_i$ tasks having deadline $i$ in $\mathbf{s}$, $ i = 1, \ldots, N$ are also excessive tasks. In other words, for each deadline, tasks that $\mathbf{s}$ has more than $\mathbf{s}^{(\ell)}$ are excessive tasks. Let's illustrate this with the following example:
	\begin{example}
    For $\mathbf{s} = \left(0, 3, 4, 0, 5 \right)$, the corresponding lean state would be $\mathbf{s}^{(\sf r)}= \left(0, 1, 1, 0, 4 \right)$. Then, for deadline 1, there are $3-1=2$ excessive tasks. For deadline 2, there are $4-1=3$ excessive tasks. For deadline 4 and 5, there is $0$ excessive task. $\square$
\end{example}

	Now let us consider the following cases in which we use the notations $J_{T,t}\left(\mathbf{s}\right)$ and $J_{T,t}\left(\mathbf{s}^{(\ell)}\right)$ to
	denote the minimum average cost associated with the initial states $\mathbf{s}$ and $\mathbf{s}^{(\ell)}$, respectively, given that the AMA arrives
	for the first time at time slot $t$. Correspondingly, we denote by $J_{T,t\ge N}\left(\mathbf{s}\right)$ and $J_{T,t\ge N}\left(\mathbf{s}^{(\ell)}\right)$ the minimum average cost associated with the initial states $\mathbf{s}$ and $\mathbf{s}^{(\ell)}$, respectively, given that the first AMA's arrival is at time slot $N$ or later. Then, there are following cases.
	\begin{itemize}
		\item Case 0: The AMA is available for the first time at the current time slot, $t=0$. As we mentioned, for every deadline, tasks that state $\mathbf{s}$ has more than state $\mathbf{s}^{(\ell)}$ are excessive tasks which should be offloaded by the optimal policy whenever the AMA is available. Therefore, in this case, if $L^*_{\ell}$ denotes the optimal number of tasks to offload of $\mathbf{s}^{(\ell)}$, that of $\mathbf{s}$ will be $L^*_{\ell} + y$, where $y = \sum_{i=1}^N (n_i-n_i^{(\ell)})$ is the partial number of excessive tasks in $\mathbf{s}$. Hence,
		\begin{align}
        J_{T,0}\left(\mathbf{s}\right) = J_{T,0}\left(\mathbf{s}^{(\ell)}\right) + C_{\sf o}\sum_{i=1}^N\left(n_i-n_i^{(\ell)}\right).
		\end{align}
		
        \item Case 1: If the AMA is available for the first time at time slot 1, the number of tasks having deadline 1 expiring from $\mathbf{s}$ is more than that from $\mathbf{s}^{(\ell)}$ by $n_{1} - n_{1}^{(\ell)}$. All the other remaining excessive
		tasks can be offloaded from both states $\mathbf{s}$ and $\mathbf{s}^{(\ell)}$. Hence,
		\begin{align}
			J_{T,1}\left(\mathbf{s}\right) &= J_{T,1}\left(\mathbf{s}^{(\ell)}\right) + C_{\sf o}\sum_{i=2}^N\left(n_i-n_i^{(\ell)}\right) \notag \\
            &+ \left(n_{1} - n_{1}^{(\ell)}\right)C_{\sf p}.
		\end{align}
		
		The same logic can be applied to other cases when the AMA
		first arrives at time slot $2, 3, \ldots, N$. Therefore, we present next the last case.
        
		\item Case $N$:  If the AMA is available for the
		first time at time slot $N$, the number of tasks having deadline $i$ expiring from $\mathbf{s}$ is more than that from $\mathbf{s}^{(\ell)}$ by $n_i - n_i^{(\ell)}$. Therefore,
		\begin{align}
			J_{T,N}\left(\mathbf{s}\right) = J_{T,N}\left(\mathbf{s}^{(\ell)}\right) + C_{\sf p}\sum_{i=1}^N\left(n_i-n_i^{(\ell)}\right).
		\end{align}
	\end{itemize}
	
	The probability that the AMA's first arrival is at time slot $t$ is computed as
	\begin{align}\label{eq:Pt_moved}
		P_t = p_{\sf a}\left(1-p_{\sf a} \right)^{t}.
	\end{align}
	Also, the probability that the AMA does not arrive within the first $N$ time slots is
	\begin{align}\label{eq:PN_moved}
		P_{t \ge N} = \left(1-p_{\sf a} \right)^{N}.
	\end{align}

	From the above logic, $J_T\left(\mathbf{s} \right)$ can be expressed in terms of $J_T\left(\mathbf{s}^{(\ell)}\right)$ as follows:
	\begin{align} \label{proof_expand}
		\begin{split}
			&J_T\left(\mathbf{s}\right) =  P_{0}\left(J_{T,0}\left(\mathbf{s}^{(\ell)}\right) + C_{\sf o}\sum_{i=1}^{N} \left(n_i - n_i^{(\ell)} \right) \right)\\
			& + P_{1}\left( J_{T,1}\left(\mathbf{s}^{(\ell)}\right)  + C_{\sf o}\sum_{i=2}^{N} \left(n_i - n_i^{(\ell)} \right) + C_{\sf p}\left(n_1 - n_1^{(\ell)} \right)\right)\\
			& + \ldots \\
			& + P_{t \ge N} \left(J_{T,t\ge N}\left(\mathbf{s}^{(\ell)}\right)  + C_{\sf p}\sum_{i=1}^{N} \left(n_i - n_i^{(\ell)} \right) \right).
		\end{split}
	\end{align}
	$J_T\left(\mathbf{s}^{(\ell)}\right)$ is the minimum cost averaged over
	all cases, hence, can be expressed by
	\begin{align} \label{proof_expla}
		J_T\left(\mathbf{s}^{(\ell)}\right) = \sum_{i=0}^{N-1} P_{i} J_{T,i}\left(\mathbf{s}^{(\ell)}\right) + P_{t \ge N} J_{T,t\ge N}\left(\mathbf{s}^{(\ell)}\right).
	\end{align}
	From the aid of Eq. (\ref{proof_expla}), the equation (\ref{proof_expand}) can be simplified to 
	\begin{align}
				C_{\ell} &= C_{\sf o}\sum_{i=1}^{N} \left(n_i - n_i^{(\ell)} \right)  \sum_{j=0}^{i-1} P_{j} \notag \\
                &+ C_{\sf p}\sum_{i=1}^{N-1} P_{i} \sum_{j=1}^{i}  \left(n_j - n_j^{(\ell)} \right) \notag \\
                &+ C_{\sf p}P_{t \ge N}\sum_{j=1}^{N} \left(n_j- n_j^{(\ell)}\right)
	\end{align}
	By plugging the expressions of $P_t$ and $P_{t \ge N}$ in Eqs. (\ref{eq:Pt_moved}) and (\ref{eq:PN_moved}), respectively,  into the above expression, we obtain Eq. (\ref{eq:gen2sem_eq}).
	
\section*{Appendix C}
	\section*{Proof of Lemma \ref{Lem:Convexity}}\label{proof_Lem:Convexity}
	\subsection{Overview}
	We recall that the definition of a discrete convex function is provided in Definition \ref{def:discreteconvex}. From the definition of function $F\left(T, \mathbf{s}, d, L \right)$ in Eq. (\ref{F_def_gen}), we notice that $LC_{\sf o}$ is discrete linear, and hence, discrete convex with respect to (wrt.) $L$. Therefore, in order to prove that $F\left(T, \mathbf{s}, d, L \right)$ is discrete convex wrt. $L$, we need to prove that for $J_T^{\overline{\text{A}}}\left(\bar{\mathbf{s}}_{dL} \right)$.
	
	For the original state $\mathbf{s}=(n_1, \ldots, n_N)$, Eqs. \eqref{eq:instant_cost3}, \eqref{G_NAMA}, and \eqref{eq:JNoAMA} allow us to present $J_T^{\overline{\text{A}}}\left(\bar{\mathbf{s}}_{dL} \right)$ by
	\begin{align}\label{eq:J_barA_repeat}
	J_T^{\overline{\text{A}}}\left(\bar{\mathbf{s}}_{dL} \right) &= C_{\sf p}n_1 + \mu\sum_{k=1}^N p_k J_{T-1}(\mathbf{s}_{dLk}^{'}) \notag \\
	&+ (1-\mu)\sum_{k=1}^N p_k J_{T-1}(\mathbf{s}_{dLk}^{''})
	\end{align}
	where $\mathbf{s}_{dLk}^{'}$ is obtained by removing the most $L$ imminent tasks having deadline greater than or equal to $d$ from the original state $\mathbf{s}$, then, performing deadline shifting, adding a task with deadline $k$, and removing the most imminent task. $\mathbf{s}_{dLk}^{''}$ is obtained in a similar way but without removing the most imminent task. Also, $\mu$ the local processing probability and $p_k$ is the probability that there is a new task arrives with deadline $k$ where $p_0$ is the probability of no task arrival.
	
	To prove that $J_T^{\overline{\text{A}}}\left(\bar{\mathbf{s}}_{dL} \right)$ is a discrete convex function wrt. $L$, we do the following. Firstly, since $\bar{\mathbf{s}}_{dL}$ is obtained from a given original state $\mathbf{s}$, we represent $J_T^{\overline{\text{A}}}\left(\bar{\mathbf{s}}_{dL} \right)$ by a function $f(T, \mathbf{s}, d, L)$, i.e., 
	\begin{align}\label{eq:f_recursive}
	f(T, \mathbf{s}, d, L) &= C_{\sf p}n_1 + \mu\sum_{k=1}^N p_k J_{T-1}(\mathbf{s}_{dLk}^{'}) \notag \\
	&+ (1-\mu)\sum_{k=1}^N p_k J_{T-1}(\mathbf{s}_{dLk}^{''}).
	\end{align}
	Secondly, we show that $J_{T-1}(\mathbf{s}_{dLk}^{'})$ and $J_{T-1}(\mathbf{s}_{dLk}^{''})$ on the right-hand side of Eq. \eqref{eq:J_barA_repeat} can be expressed in terms of functions $f(T-1, \tilde{\mathbf{s}}_{k}^{'}, d^{'}, L)$ and $f(T-1, \tilde{\mathbf{s}}_{k}^{''}, d^{''}, L)$ where $\tilde{\mathbf{s}}_{k}^{'}$ and $\tilde{\mathbf{s}}_{k}^{''}$ are fixed states obtained from $\mathbf{s}$ via deterministic operators and $d^{'}, d^{''} \in \{1, \ldots, N \}$. Subsequently, by induction, we assume that $f(T-1, \cdot, \cdot, L)$ is a discrete convex function wrt. $L$ and prove that the convexity also holds for $f(T, \cdot, \cdot, L)$ with a base case provided. This allows us to prove the convexity of $J_T^{\overline{\text{A}}}\left(\bar{\mathbf{s}}_{dL} \right)$ and $F\left(T, \mathbf{s}, d, L \right)$.
	
	In the next subsection, we will presented detailed proof that is generally described above.
	
	\subsection{Proof}
	To begin, we will show that $J_{T-1}(\mathbf{s}_{Lk}^{'})$ and $J_{T-1}(\mathbf{s}_{Lk}^{''})$ on the right-hand side of Eq. \eqref{eq:J_barA_repeat} can be represented by functions $f(T-1, \tilde{\mathbf{s}}_{k}^{'}, d^{'}, L)$ with $L \in \mathbb{L}_d(\tilde{\mathbf{s}}_{k}^{'})$ and $f(T-1, \tilde{\mathbf{s}}_{k}^{''}, d^{''}, L)$ with $L \in \mathbb{L}_d(\tilde{\mathbf{s}}_{k}^{''})$, respectively, for $d^{'}, d^{''} \in \{1, \ldots, N \}$. In general, there are two different cases as follows.
	\begin{itemize}
	\item \textbf{Case 1.} $k+1 < d$ which implies $d \ge 2$.
	
	In this case, the state $\mathbf{s}_{dLk}^{'}$ can be obtained by removing $L \in \mathbb{L}_{d-1}(\mathbf{s}_{k}^{'})$ most imminent tasks having deadline greater than or equal to $d-1$ from a state $\mathbf{s}_{k}^{'}$. State $\mathbf{s}_{k}^{'}$ is obtained by performing deadline shifting on the original state $\mathbf{s}$, adding a task with deadline $k$ if $k\ge 1$, and removing the most imminent task. 
	
	Similarly, the sequence of states $\mathbf{s}_{dLk}^{''}$, for $L \in \mathbb{L}_d(\mathbf{s})$  can be obtained by removing $L \in \mathbb{L}_{d-1}(\mathbf{s}_{k}^{''})$ most imminent tasks having deadline greater than or equal to $d-1$ from a state $\mathbf{s}_{k}^{''}$. State $\mathbf{s}_{k}^{''}$ is obtained by performing deadline shifting on the original state $\mathbf{s}$ and adding a task with deadline $k$ if $k\ge 1$.
	
	An intuitive example for this point is as follows.
	\begin{example}
	Given the original state $\mathbf{s} = (2, 3, 4, 5, 6)$, we let $k=2$ and $d=4$. Then, $\bar{\mathbf{s}}_{dLk}^{'}$, $L \in \mathbb{L}_4(\mathbf{s}) = \{0, \ldots, 11 \}$, are states: $(2, 5, 5, 6, 0)$, $(2, 5, 4, 6, 0)$, $(2, 5, 3, 6, 0)$, etc..
	
	The above sequence of states, can be generated by removing $L \in \mathbb{L}_3(\tilde{\mathbf{s}}_{k}^{'})$ most imminent tasks having deadline greater than or equal to $d-1$ from state $\tilde{\mathbf{s}}_{k}^{'} = (2, 5, 5, 6, 0)$. Therefore, $J_{T-1}(\mathbf{s}_{dLk}^{'})$ can be represented by the function $f(T-1, \tilde{\mathbf{s}}_{k}^{'}, 3, L)$. Similarly, $J_{T-1}(\mathbf{s}_{dLk}^{''})$ can be represented by the function $f(T-1, \tilde{\mathbf{s}}_{k}^{''}, 3, L)$ where $\tilde{\mathbf{s}}_{k}^{''} = (3, 5, 5, 6, 0)$. $\square$
	\end{example}
	Therefore, $J_{T-1}(\mathbf{s}_{Lk}^{'})$ and $J_{T-1}(\mathbf{s}_{Lk}^{''})$ on the right-hand side of Eq. \eqref{eq:J_barA_repeat} can be represented by functions $f(T-1, \tilde{\mathbf{s}}_{k}^{'}, d-1, L)$ and $f(T-1, \tilde{\mathbf{s}}_{k}^{''}, d-1, L)$, respectively, where the mappings from the original state $\mathbf{s}$ to states $\tilde{\mathbf{s}}_{k}^{'}$ and $\tilde{\mathbf{s}}_{k}^{''}$ are described above.
	
	\item \textbf{Case 2.} $k + 1 \ge d$ and $d=1$.
	
	We let $\beta_{dk} = \sum_{i=d}^{k+1} n_i$. This case can be separated in the following two subcases.
	\begin{itemize}
	\item \textbf{Subcase 2.1.} $k + 1 \ge d$, $d=1$, and $L \le \beta_{dk}$.
	
	The state $\mathbf{s}_{dLk}^{'}$  can be obtained by removing $L \in \mathbb{L}_1(\mathbf{s}_{k}^{'}) \uplus \{L\le \beta_{dk}\}$ most imminent tasks having deadline greater than or equal to $1$ from a state $\mathbf{s}_{k}^{'}$. State $\mathbf{s}_{k}^{'}$ is obtained by performing deadline shifting on the original state $\mathbf{s}$, adding a task with deadline $k$ if $k\ge 1$, and removing the most imminent task.
	
	Similarly, the sequence of states $\mathbf{s}_{dLk}^{''}$, for $L \in \mathbb{L}_d(\mathbf{s})$ and $L\le \beta_{dk}$,  can be obtained by removing $L \in \mathbb{L}_1(\mathbf{s}_{k}^{''})$ most imminent tasks having deadline greater than or equal to $1$ from a state $\mathbf{s}_{k}^{''}$. State $\mathbf{s}_{k}^{''}$ is obtained by performing deadline shifting on the original state $\mathbf{s}$ and adding a task with deadline $k$ if $k\ge 1$.
	
	We provide the following example for this case.
	\begin{example}
	Given the original state $\mathbf{s} = (2, 3, 4, 5, 6)$, we let $k=2$ and $d=1$; hence, $\beta_{dk} = 9$. Then, $\mathbf{s}_{dLk}^{'}$, for $L \in \mathbb{L}_1(\mathbf{s}) = \{2, 3, \ldots, 9\}$, are states: $(2, 5, 5, 6, 0)$, $(1, 5, 5, 6, 0)$, $(0, 5, 5, 6, 0)$, $(0, 4, 5, 6, 0)$, etc.
	
	The above sequence of states, can be generated by removing $L \in \mathbb{L}_1(\mathbf{s}_{k}^{'} )$ most imminent tasks having deadline greater than or equal to $1$ from state $\mathbf{s}_{k}^{'} = (2, 5, 5, 6, 0)$. Therefore, $J_{T-1}(\mathbf{s}_{dLk}^{'})$ can be represented by the function $f(T-1, \tilde{\mathbf{s}}_{k}^{'}, 1, L)$. Similarly, $J_{T-1}(\mathbf{s}_{dLk}^{''})$ can be represented by the function $f(T-1, \tilde{\mathbf{s}}_{k}^{''}, 1, L)$ where $\tilde{\mathbf{s}}_{k}^{''}$ is given by $\tilde{\mathbf{s}}_{k}^{''} = (3, 5, 5, 6, 0)$. $\square$
	\end{example}
	
	\item \textbf{Subcase 2.2.} $k + 1 \ge d$, $d=1$, and $L > \beta_{dk}$.
	
	In this subcase, the sequence of states $\mathbf{s}_{dLk}^{'}$, for $L \in \mathbb{L}_1(\mathbf{s})$ and $L> \beta_{dk}$,  can be obtained by removing $L-\beta_{dk}$ most imminent tasks having deadline greater than or equal to $k+1$ from a state $\tilde{\mathbf{s}}_{k}^{'}$. State $\tilde{\mathbf{s}}_{k}^{'}$ is obtained by removing the most imminent $\beta_{dk}$ tasks having deadline greater than or equal to $k+1$ from the original state $\mathbf{s}$, then, performing deadline shifting, adding a task with deadline $k$ if $k\ge 1$, and removing the most imminent task.
	
	Similarly, the sequence of states $\mathbf{s}_{dLk}^{''}$, for $L \in \mathbb{L}_d(\mathbf{s})$ and $L> \beta_{dk}$,  can be obtained by removing $L-\beta_{dk}$ most imminent tasks having deadline greater than or equal to $k+1$ from a state $\tilde{\mathbf{s}}_{k}^{''}$. State $\tilde{\mathbf{s}}_{k}^{''}$ is obtained by removing the most $\beta_{dk}$ tasks having deadline greater than or equal to $k+1$ from the original state $\mathbf{s}$, then, performing deadline shifting, and adding a task with deadline $k$ if $k\ge 1$.
	
	Let us consider the following intuitive example.
	\begin{example}
	Given the original state $\mathbf{s} = (2, 3, 4, 5, 6)$, we let $k=2$ and $d=1$; hence, $\beta_{dk}=9$. $\mathbf{s}_{dLk}^{'}$, for $L \in \{10, \ldots, 15\}$, is the following sequence of states: $(0, 0, 4, 6, 0)$, $(0, 0, 3, 6, 0)$, $(0, 0, 2, 6, 0)$, $\ldots$.
	
	The above sequence of states, can be generated by removing $L-9$ ($\beta_{dk}=9$ in this example) most imminent tasks having deadline greater than or equal to $1$ from state $\tilde{\mathbf{s}}_{k}^{'} = (0, 0, 5, 6, 0)$. Therefore, $J_{T-1}(\mathbf{s}_{dLk}^{'})$ can be represented by the function $f(T-1, \tilde{\mathbf{s}}_{k}^{'}, 1, L-9)$. Similarly, $J_{T-1}(\mathbf{s}_{dLk}^{''})$ can be represented by the function $f(T-1, \tilde{\mathbf{s}}_{k}^{''}, 1, L-9)$ where $\tilde{\mathbf{s}}_{k}^{''} = (0, 1, 5, 6, 0)$. $\square$
	\end{example}
	\end{itemize}
	
	\item \textbf{Case 3.} $k + 1 \ge d$ and $d \ge 2$.
	
	We define $\beta_{dk} = \sum_{i=d}^{k+1} n_i$. Similar to case 2, this case can also be separated into the following two cases.
	\begin{itemize}
	\item \textbf{Subcase 3.1.} $k + 1 \ge d$, $d \ge 2$, and $L \le \beta_{dk}$.
	
	The sequence of states $\mathbf{s}_{dLk}^{'}$, for $L \in \mathbb{L}_d(\mathbf{s})$ and $L\le \beta_{dk}$,  can be obtained by removing $L$ most imminent tasks having deadline greater than or equal to $d-1$ from a state $\mathbf{s}_{k}^{'}$. State $\mathbf{s}_{k}^{'}$ is obtained by performing deadline shifting on the original state $\mathbf{s}$, adding a task with deadline $k$ if $k\ge 1$, and removing the most imminent task.
	
	Similarly, the sequence of states $\mathbf{s}_{dLk}^{''}$, for $L \in \mathbb{L}_d(\mathbf{s})$ and $L\le \beta_{dk}$,  can be obtained by removing $L$ most imminent tasks having deadline greater than or equal to $d-1$ from a state $\mathbf{s}_{k}^{''}$. State $\mathbf{s}_{k}^{''}$ is obtained by performing deadline shifting on the original state $\mathbf{s}$ and adding a task with deadline $k$ if $k\ge 1$.
	
	\item \textbf{Subcase 3.2.} $k + 1 \ge d$, $d\ge 2$, and $L > \beta_{dk}$.
	
	In this subcase, the sequence of states $\bar{\mathbf{s}}_{dLk}^{'}$, for $L \in \mathbb{L}_d(\mathbf{s})$ and $L> \beta_{dk}$,  can be obtained by removing $L-\beta_{dk}$ most imminent tasks having deadline greater than or equal to $k+1$ from a state $\mathbf{s}_{k}^{'}$. State $\mathbf{s}_{k}^{'}$ is obtained by removing the most $\beta_{dk}$ tasks having deadline greater than or equal to $k+1$ from the original state $\mathbf{s}$, then, performing deadline shifting, adding a task with deadline $k$ if $k\ge 1$, and removing the most imminent task.
	
	Similarly, the sequence of states $\bar{\mathbf{s}}_{dLk}^{''}$, for $L \in \mathbb{L}_d(\mathbf{s})$ and $L> \beta_{dk}$,  can be obtained by removing $L-\beta_{dk}$ most imminent tasks having deadline greater than or equal to $k+1$ from a state $\mathbf{s}_{k}^{''}$. State $\mathbf{s}_{k}^{''}$ is obtained by removing the most $\beta_{dk}$ tasks having deadline greater than or equal to $k+1$ from the original state $\mathbf{s}$, then, performing deadline shifting, and adding a task with deadline $k$ if $k\ge 1$.
	
	Since case 3 is the same as case 2 except that, when $L \le \beta_{dk}$, in the process of mapping state $\mathbf{s}$ to states $\mathbf{s}_{k}^{'}$ and $\mathbf{s}_{k}^{''}$ tasks are removed starting from deadline $d-1$ instead of 1 as in case 2. Therefore, we omitted the example of this case.
	\end{itemize}
	
	\end{itemize}
	
	In summary, the terms $J_{T-1}\left(\mathbf{s}'_{d L k}\right)$ and $J_{T-1}\left(\mathbf{s}''_{d L k}\right)$ on the right-hand side of Eq. \eqref{eq:f_recursive} can be expressed in terms of functions $f$ as follows:
    \begin{itemize}
\item If $k + 1 < d$:
\begin{align}
J_{T-1}\left(\mathbf{s}'_{d Lk}\right) &= f\left(T-1, \mathbf{s}'_k, d-1, L\right), \label{eq:Jf1}\\
J_{T-1}\left(\mathbf{s}''_{d L k}\right) &= f\left(T-1, \mathbf{s}''_k, d-1, L\right). \label{eq:Jf2}
\end{align}
The state $\mathbf{s}'_k$ is obtained from $\mathbf{s}$ through the following steps: deadline shifting, adding a new task with deadline $k$ if $k \ge 1$, and removing the most imminent task. State $\mathbf{s}''_k$ is defined similarly as state $\mathbf{s}'_k$ but without removing the most imminent task.

\item If $k + 1 \ge d$, by denoting $\beta_{dk} = \sum_{i=d}^{k+1} n_i$, we obtain the following expressions:
\begin{align}\label{eq:Jf3}
\begin{split}
&J_{T-1}\left(\mathbf{s}'_{d L k}\right) = g'_{dk}\left(L\right) \\
&=
\begin{cases}
f\left(T-1, \mathbf{s}'_{k1}, \max(d-1, 1), L \right) &,  \text{ if } L \le \beta_{dk},\\
f\left(T-1, \mathbf{s}'_{k2}, k+1, L - \beta_{dk}\right)&,  \text{ if } L > \beta_{dk},
\end{cases}
\end{split}
\end{align}

\begin{align}\label{eq:Jf4}
\begin{split}
&J_{T-1}\left(\mathbf{s}''_{d Lk}\right) = g''_{dk}\left(L\right) \\
&=
\begin{cases}
f\left(T-1, \mathbf{s}''_{k1}, \max(d-1, 1), L \right) &,  \text{ if } L \le \beta_{dk},\\
f\left(T-1, \mathbf{s}''_{k2}, k+1, L - \beta_{dk}\right)&,  \text{ if } L > \beta_{dk},
\end{cases}
\end{split}
\end{align}

The states $\mathbf{s}_{k1}^{'}$ and $\mathbf{s}_{k1}^{''}$ are obtained in the same way as $\mathbf{s}_{k}^{'}$ and $\mathbf{s}_{k}^{''}$ described above. $\mathbf{s}_{k2}^{'}$ is obtained as follows: removing $\beta_{dk}$ most imminent tasks having deadline greater than or equal to $d$ from the initial state $\mathbf{s}$, performing deadline shifting, adding a task with deadline $k$ if $k\ge 1$, and removing the most imminent task. State  $\mathbf{s}_{k2}^{''}$ is defined in a similar way but without removing the most imminent task.

\end{itemize}
In the cases above, functions $g'_{dk}\left(L\right)$ and $g''_{dk}\left(L\right)$ are characterized by parameters $d$ and $k$, and take $L$ as their variables.

Next, we will prove the convexity of $f$ using induction on $T$, starting with an initial induction step as follows.
 
	\textbf{Initial step:} We consider state $\mathbf{s} = \left(n_1, \ldots, n_N\right)$, and a time horizon $T=1$. In this case, if $N = 1$, we have $\mathbf{s} = \left(n_1\right)$, then, there is only one valid offloading decision $\mathbb{L}_1\left(\mathbf{s}\right) = \left\{n_1\right\}$. For $N \ge 2$, it is trivially that all tasks having deadline 1 are excessive tasks, and should be offloaded whenever the AMA is available, which results in a cost $n_1C_{\sf o}$. If the AMA is not available at the initial time slot, tasks with deadline 1 will expire and result in a cost $n_1C_{\sf p}$. Since in this case, we are considering a time horizon with only one time slot, the minimum average cost can be computed straightforwardly. Thus, we have the following expression:
	\begin{align}
		f\left(1, \mathbf{s}, 1, L\right) &= J_1\left(\bar{\mathbf{s}}_{1L}\right) \notag \\
        &= \left(p_{\sf a}C_{\sf o} + \left(1-p_{\sf a}\right)C_{\sf p}\right)n_1 \notag\\
        &+ \left(L-n_1\right)C_{\sf o}, \text{ for } L \in \mathbb{L}_1\left(\mathbf{s}\right), \label{eq:induct_init1}
	\end{align}
    and
    \begin{flalign}
		f\left(1, \mathbf{s}, d, L\right) &= J_1\left(\bar{\mathbf{s}}_{dL}\right) \notag \\
        &= n_1C_{\sf p} + LC_{\sf o}, \text{ for } d \ge 2 \text{ and } L \in \mathbb{L}_d\left(\mathbf{s}\right). \label{eq:induct_init2}
    \end{flalign}
	We recall that the smallest offloading decision in the set $\mathbb{L}_1\left(\mathbf{s}\right)$ is $n_1$. From Eqs. (\ref{eq:induct_init1}) and (\ref{eq:induct_init2}), $f\left(1, \mathbf{s}, d, L\right), d = 1, \ldots, N$, are discrete linear function wrt. $L$, hence, they are discrete convex function wrt. $L$. Next is an inductive step where we prove the convexity of $f\left(T, \mathbf{s}, d, L\right)$ given that of $f\left(T-1, \mathbf{s}, d, L\right)$ for every parameters $\mathbf{s}$ and $d$.

	\textbf{Inductive step:} We assume that the functions $f\left(T-1, \mathbf{s}, d, L\right)$ is discrete convex wrt. $L$ for every given state $\mathbf{s}$ and deadline $d$.
	
	It can be inferred from the above assumption, $f\left(T-1, \mathbf{s}'_k, d-1,  L + \hat{L} \right)$ in Eq. (\ref{eq:Jf1}), and $f\big(T-1, \mathbf{s}''_k, d-1, L + \hat{L} \big)$ in Eq. (\ref{eq:Jf2}) are discrete convex wrt. $L$. Subsequently, we will prove that $g'_{dk}\left(L \right)$ in Eq. (\ref{eq:Jf3}) is discrete convex with respect wrt. $L$. Then, the convexity of functions $g''_{dk}\left(L \right)$ in Eq. (\ref{eq:Jf4}) can also be proved in a very similar way.
	
	Let us consider function $g'_{dk}\left(L\right)$ in Eq. (\ref{eq:Jf3}). From our assumption above, $f\big(T-1, \mathbf{s}'_k, d-1, L\big)$ for $L \le \beta_{dk}$, and $f\left(T-1, \mathbf{s}'_{d \beta_{dk} k}, k+1, L - \beta_{dk}\right)$ for $L  > \beta_{dk}$ are discrete convex wrt. $L$. Therefore, in order to prove the convexity of $g'_{dk}\left(L \right)$ we will prove that the discrete Jensen's inequality holds at the connecting point, i.e., $L=\beta_{dk}$, of the two mentioned convex functions, specifically, we prove that
	\begin{align}\label{ineq:target_convex}
	f\left(T-1, \mathbf{s}'_{k1}, d-1, \beta_{dk}\right) &+ f\left(T-1, \mathbf{s}'_{k2}, k+1, 2\right) \notag \\
    &\ge 2f\left(T-1, \mathbf{s}'_{k2}, k+1, 1\right).
	\end{align}
	
 From the definition of function $f$, we have the following equalities:
 \begin{align}
  &f\left(T-1, \mathbf{s}'_{k1}, d-1, \beta_{dk}\right) = J_{T-1}\left( \mathbf{s}'_{k2}\right),\\
  &f\left(T-1, \mathbf{s}'_{k2}, k+1, 0\right) = J_{T-1}\left( \mathbf{s}'_{k2}\right).
 \end{align}
	Hence, the following holds:
	\begin{align}\label{eq:f_link}
	f\left(T-1, \mathbf{s}'_{k1}, d-1, \beta_{dk}\right) = f\left(T-1, \mathbf{s}'_{k2}, k+1, 0\right).
	\end{align}
	Moreover, since $f\left(T-1, \mathbf{s}'_{k2}, k+1, L + \hat{L} \right) $ is discrete convex wrt. $L$ due to our assumption, the following inequality holds:
	\begin{align}\label{ineq:link_hold}
        f\left(T-1, \mathbf{s}'_{k2}, k+1, 0\right)  &+ f\left(T-1, \mathbf{s}'_{k2}, k+1, 2\right) \notag \\
        &\ge 2f\left(T-1, \mathbf{s}'_{k2}, k+1, 1\right).
	\end{align}
	By combining Eq. (\ref{eq:f_link}) and Ineq. (\ref{ineq:link_hold}), we show that Ineq. (\ref{ineq:target_convex}) is true. Therefore, $g'_{dk}\left(L\right)$ in Eq. (\ref{eq:Jf3}) is discrete convex wrt. $L$. Similarly, $g''_{dk}\left(L\right)$ in Eq. (\ref{eq:Jf4}) is also discrete convex wrt. $L$, which can be proved similarly. 
	
	To this end, we can conclude that the terms $J_{T-1}\left(\mathbf{s}^{'}_{d Lk}\right)$ and $J_{T-1}\left(\mathbf{s}^{''}_{d Lk}\right) $ are discrete convex functions wrt. $L$ for every given original state $\mathbf{s}$ and parameters $d$ and $k$. Then, $f(T, \mathbf{s}, d, L)$ is the linear combination of convex functions as presented in  Eq. \eqref{eq:J_barA_repeat}; hence, $f(T, \mathbf{s}, d, L)$ is also a discrete convex function wrt. $L$, completing this inductive step.
	
	Finally, the convexity of $f(T, \mathbf{s}, d, L)$ suggests that $J_T(\bar{\mathbf{s}}_{dL})$ is convex, and from the definition of function $F\left(T, \mathbf{s}, d, L \right)$ in Eq. (\ref{F_def_gen}), we can conclude that $F\left(T, \mathbf{s}, d, L \right)$ is a discrete convex function wrt. $L$ for every given original state $\mathbf{s}$ and parameter $d$. This completes our proof.	
	
	\section*{Appendix D}
	\section*{Proof of Lemma \ref{Lem:Convex2L*}}\label{proof_Lem:Convex2L*}
	If we assume that function $F\left(T, \mathbf{s}, d, L\right)$ attains its minimum at $L^*$ for $d=1$, the following set of inequalities hold
	\begin{align}\label{optim_fromF}
		J_T^{\overline{\text{A}}}\left(\bar{\mathbf{s}}_{1L^*}\right) + L^*C_{\sf o} \le J_T^{\overline{\text{A}}}\left(\bar{\mathbf{s}}_{1L}\right) + LC_{\sf o}, \text{ for all } L \in \mathbb{L}_1\left(\mathbf{s}\right).
	\end{align}
	
	From Eq. (\ref{JsL_intro}) and Ineqs. (\ref{optim_fromF}), we have:
     \begin{align}
     J_T^{\text{A}}\left(\mathbf{s}, L^*\right) \le J_T^{\text{A}}\left(\mathbf{s}, L\right), \text{ for all } L \in \mathbb{L}_1\left(\mathbf{s}\right),
     \end{align}
 which is equivalent to
	\begin{align} \label{eq:JAL*_asmin}
		J_T^{\text{A}}\left(\mathbf{s}, L^*\right) = \underset{L \in \mathbb{L}_1\left(\mathbf{s}\right)}{\min} \left\{J_T^{\text{A}}\left(\mathbf{s}, L\right)\right\}.
	\end{align}
	Observe in Eq. (\ref{eq:JAL*_asmin}) that $\mathbb{L}_1(\mathbf{s})$, as defined in Eq. (\ref{F_domain2}), does not include offloading decisions that are less than $n_1$ (the first component of $\mathbf{s}$). However, we recall a fact that the optimal offloading decision must not be less than $n_1$ which are all excessive tasks. Thus, $L^*$ is the optimal offloading decision of $\mathbf{s}$ for the time horizon $T$.


\section*{Appendix E}
\section*{Proof of Theorem \ref{Theo:adjacent_L*}}\label{proof_Theo:adjacent_L*}
	From Lemma \ref{Lem:Convexity}, we have that function $F\left(T, \mathbf{s}, d, L\right)$ is discrete convex wrt. $L$ for every given $T$, $d$, and $\mathbf{s} = \left(n_1, \ldots, n_N\right)$. Let us consider the case when $d=1$, and we call $L^* \in \mathbb{L}_1\left(\mathbf{s}\right)$ the value at which $F\left(T, \mathbf{s}, 1, L\right)$ attains its minimum, where $\mathbb{L}_1(\mathbf{s})$ is defined in the first expression of (\ref{F_domain2}). From Lemma \ref{Lem:Convex2L*}, $L^*$ is also the optimal offloading decision for $\mathbf{s}$. Hence,
	\begin{align} \label{optim_at_s}
		J_T^{\text{A}}\left(\mathbf{s}, L\right) \ge J_T^{\text{A}}\left(\mathbf{s}, L^*\right), \text{ for all } L \in \mathbb{L}_1\left(\mathbf{s}\right).
	\end{align}
	
	Let us consider the following cases:\\
	$\bullet$ If $L^* \ge 1$ which means that $\mathbf{s} \not\equiv \left(0, \ldots, 0\right)$,  the set of inequalities (\ref{optim_at_s}) becomes
		\begin{align}\label{ineq:convex2L*_1}
			J_T^{\text{A}}\left(\mathbf{s}, 1 + L-1\right) \ge J_T^{\text{A}}\left(\mathbf{s}, 1 + L^*-1\right),  L \in \mathbb{L}_1\left(\mathbf{s}\right) \backslash \left\{0\right\}.
		\end{align}
		Applying Eq. (\ref{JsL_intro1}) with $L_1=1$, $L_2 = L-1$ to the left-hand side of Ineqs. (\ref{ineq:convex2L*_1}), and with $L_2=L^*-1$ to the right-hand side of Ineqs. (\ref{ineq:convex2L*_1}), we have
		\begin{align}\label{optim_s11}
			J_T^{\text{A}}\left(\bar{\mathbf{s}}_{11}, L-1\right) \ge J_T^{\text{A}}\left(\bar{\mathbf{s}}_{11}, L^*-1\right),  L \in \mathbb{L}_1\left(\mathbf{s}\right) \backslash \left\{0\right\}
		\end{align}
		 in which we recall that $\bar{\mathbf{s}}_{11}$ is obtained by offloading the most imminent task from $\mathbf{s}$, thus, $\mathbf{s} \in \mathbb{S}_{\sf adj}\left(\bar{\mathbf{s}}_{11}\right)$. Since $L \in \mathbb{L}_1\left(\mathbf{s}\right) \backslash \left\{0\right\}$, it is guaranteed that  $L-1 \in \mathbb{L}_1\left(\bar{\mathbf{s}}_{11}\right)$. Therefore, the set of inequalities (\ref{optim_s11}) suggests that $L^*-1$ is the optimal decision for $\bar{\mathbf{s}}_{11}$. This proves the first point of Theorem \ref{Theo:adjacent_L*}.\\
	$\bullet$ If $L^*=0$, i.e., $\mathbf{s}$ is a non-offloading state. Then, if $\mathbf{s}$ has only one task, we have $\bar{\mathbf{s}}_{11} \equiv \left(0, \ldots, 0\right)$. Therefore, $\bar{\mathbf{s}}_{11}$ is a non-offloading state trivially.
		
		If $\mathbf{s}$ has at least 2 tasks. From the convexity of function $F\left(T, \mathbf{s}, d, L\right)$ in Lemma \ref{Lem:Convexity}, and the condition that $L^*=0$, we have the following inequalities for $L \in \mathbb{L}_1\left(\mathbf{s}\right)\backslash \left\{0, 1\right\}$:
		\begin{align} \label{ineq:convex&L*=0}
			J_T^{\overline{\text{A}}}\left(\mathbf{s}\right) \le J_T^{\overline{\text{A}}}\left(\bar{\mathbf{s}}_{11}\right) + C_{\sf o} \le J_T^{\overline{\text{A}}}\left(\bar{\mathbf{s}}_{1L}\right) + LC_{\sf o},
		\end{align}
		where the two inequalities in Ineq. (\ref{ineq:convex&L*=0}) are from the optimality of $L^*$, and the convexity of cost functions proven in Subsec. \ref{proof_Lem:Convexity}, respectively. Applying Eq. (\ref{equi_noAMA}) to the second inequality of Ineqs. (\ref{ineq:convex&L*=0}), we have
		\begin{align} \label{optim_L*=0}
			J_T^{\text{A}}\left(\bar{\mathbf{s}}_{11}, 0\right) + C_{\sf o} \le J_T^{\text{A}}\left(\bar{\mathbf{s}}_{1L}, 0\right) + LC_{\sf o}, ~ L \in \mathbb{L}_1\left(\mathbf{s}\right)\backslash \left\{0, 1\right\}.
		\end{align}
		Using Eq. (\ref{JsL_intro1}) with $L_1=1$ and $L_2 = L-1$ gives us $J_T^{\text{A}}\left(\mathbf{s}, L\right) = J_T^{\text{A}}\left(\bar{\mathbf{s}}_{11}, L-1\right) + C_{\sf o}$. Also, using Eq. (\ref{JsL_intro1}) with $L_1=L$ and $L_2 = 0$ yields $J_T^{\text{A}}\left(\mathbf{s}, L\right) = J_T^{\text{A}}\left(\bar{\mathbf{s}}_{1L}, 0\right) + LC_{\sf o}$.
		From these two equalities, we have
		\begin{align}\label{mani2bis}
			J_T^{\text{A}}\left(\bar{\mathbf{s}}_{11}, L-1\right) + C_{\sf o} = J_T^{\text{A}}\left(\bar{\mathbf{s}}_{1L}, 0\right) + LC_{\sf o}.
		\end{align}
		Combining Eq. (\ref{mani2bis}) with Ineqs. (\ref{optim_L*=0}) gives the following set of inequalities:
		\begin{align} 
			J_T^{\text{A}}\left(\bar{\mathbf{s}}_{11}, 0\right) \le J_T^{\text{A}}\left(\bar{\mathbf{s}}_{11}, L-1\right), ~ L \in \mathbb{L}_1\left(\mathbf{s}\right)\backslash \left\{0, 1\right\}.
		\end{align}
		By replacing $L-1$ with $\tilde{L}$ in the above inequalities, we have
		\begin{align} 
			J_T^{\text{A}}\left(\bar{\mathbf{s}}_{11}, 0\right) \le J_T^{\text{A}}\left(\bar{\mathbf{s}}_{11}, \tilde{L}\right), ~ \tilde{L} \in \mathbb{L}_1\left(\bar{\mathbf{s}}_{11}\right)\backslash \left\{0\right\}.
		\end{align}
		Therefore, $\bar{\mathbf{s}}_{11}$ is a non-offloading state. This proves the second point of Theorem \ref{Theo:adjacent_L*}.
	
	Finally, from the first two points of Theorem \ref{Theo:adjacent_L*}, we can conclude that, for given time horizon $T$, if $L^* \ge 1$ is optimal state $\mathbf{s}$, $L^*_{\sf a} = L^* + 1$ is optimal for every state $\mathbf{s}^{(\sf a)} \in \mathbb{S}_{\sf adj}\left(\mathbf{s}\right)$. To prove this, we first assume $L^*_{\sf a} \ne L^* + 1$ and $L^*_{\sf a} \ge 1$. Then, from the first point of Theorem \ref{Theo:adjacent_L*}, the optimal decision for $\mathbf{s}$ is $L^* = L^*_{\sf a} - 1$ which is a contradiction to our assumption. Next, we assume $L^*_{\sf a} = 0$. From the second point of Theorem \ref{Theo:adjacent_L*}, we must have $L^*=0$, leading to another contradiction. Therefore, the third statement of Theorem \ref{Theo:adjacent_L*} is true.
	
\section*{Appendix F}
	\section*{Proof of Proposition \ref{Pro:ONO_conds}}\label{proof_Pro:ONO_conds}
	Assuming that a time horizon $T$, the system state $\mathbf{s}$ and $\mathbf{s}^{(\sf a)}$ are given in which $\mathbf{s}^{(\sf a)}$ is an adjacent state of $\mathbf{s}$. Due to the notation complexity, we would like to recall that, in this appendix section, we use $\bar{\mathbf{s}}_{dL}$ to denote the system state obtained by offloading, from state $\mathbf{s}$, $L$ most imminent tasks having deadlines greater than or equal to $d$. Following this rule, the notation $\bar{\mathbf{s}}_{\sf{a}\textit{dL}}$ denotes the state obtained by offloading, from state $\mathbf{s}^{(\sf a)}$ - an adjacent state of $\mathbf{s}$, $L$ most imminent tasks having deadlines greater than or equal to $d$.

    If $\mathbf{s}^{(\sf a)}$ is a non-offloading state, the corresponding optimal offloading decision is 0, we have an inequality:
    \begin{align}
        J_T^{\text{A}}\left(\mathbf{s}^{(\sf a)}, 0\right) < J_T^{\text{A}}\left(\mathbf{s}^{(\sf a)}, 1\right).
    \end{align}
    By using Eq. (\ref{JsL_intro}), this inequality can be re-written as
	\begin{align}\label{ineq_NO}
		J_T^{\overline{\text{A}}}\left(\mathbf{s}^{(\sf a)}\right) < J_T^{\overline{\text{A}}}\left(\mathbf{s}\right) + C_{\sf o}.
	\end{align}
	We assume that $\mathbf{s}$ is a state in which $\mathbf{s}^{(\sf a)} \in \mathbb{S}_{\sf adj}\left(\mathbf{s}\right)$. According to Theorem \ref{Theo:adjacent_L*}, if $\mathbf{s}^{(\sf a)}$ is a non-offloading state, $\mathbf{s}$ is also a non-offloading state. Hence, trivially, we have
    \begin{align}
        J_T^{\text{A}}\left(\mathbf{s}^{(\sf a)}\right) &= J_T^{\overline{\text{A}}}\left(\mathbf{s}^{(\sf a)}\right) \\
         J_T^{\text{A}}\left(\mathbf{s}\right) &= J_T^{\overline{\text{A}}}\left(\mathbf{s}\right).
    \end{align}
    Combining theses two equalities with Eq. (\ref{gen_U_noU}) and Ineq. (\ref{ineq_NO}), we have
	\begin{align}\label{optimcond_NO}
		J_T\left(\mathbf{s}^{(\sf a)}\right) - J_T\left(\mathbf{s}\right) < C_{\sf o}.
	\end{align}
	At this point, we make a conclusion that: For two states $\mathbf{s}$ and $\mathbf{s}^{(\sf a)} \in \mathbb{S}_{\sf adj}\left(\mathbf{s}\right)$, if $\mathbf{s}^{(\sf a)}$ is a non-offloading state, Ineq. (\ref{optimcond_NO}) holds.
	
	Now, we will prove the reverse, which is proving the following: Given two states $\mathbf{s}$ and $\mathbf{s}^{(\sf a)} \in \mathbb{S}_{\sf adj}\left(\mathbf{s}\right)$, if Ineq. (\ref{optimcond_NO}) holds, $\mathbf{s}^{(\sf a)}$ is a non-offloading state. We assume, in contradict, that $\mathbf{s}^{(\sf a)}$ is an offloading state. By applying Eq. (\ref{gen_U_noU}) to Ineq. (\ref{optimcond_NO}), we have
	\begin{align}\label{ineq:NOconds_rev}
	p_{\sf a}\left( J_T^{\text{A}}\left(\mathbf{s}^{(\sf a)}\right) - J_T^{\text{A}}\left(\mathbf{s}\right) \right) &+ \left(1-p_{\sf a}\right)\left( J_T^{\overline{\text{A}}}\left(\mathbf{s}^{(\sf a)}\right) - J_T^{\overline{\text{A}}}\left(\mathbf{s}\right) \right) \notag \\
	&< C_{\sf o}.
	\end{align}
	We denote $L_{\sf a}^*  \ge 1$ the optimal offloading decision of $\mathbf{s}^{(\sf a)}$. From Theorem \ref{Theo:adjacent_L*}, the optimal offloading decision of $\mathbf{s}$ would be $L_{\sf a}^*-1$. Combining this with Eq. (\ref{rel_orig}), we have
	\begin{align}
	J_T^{\text{A}}\left(\mathbf{s}^{(\sf a)}\right) &= J_T^{\text{A}}\left(\mathbf{s}^{(\sf a)}, L_{\sf a}^*\right) \notag \\
	&= J_T^{\text{A}}\left(\bar{\mathbf{s}}_{1\textit{L}_{\sf a}^*}^{(\sf a)}, 0\right) + L_{\sf a}^*C_{\sf o},
	\end{align}
	and
	\begin{align}
	J_T^{\text{A}}\left(\mathbf{s}\right) &= J_T^{\text{A}}\left(\mathbf{s}, L_{\sf a}^*-1\right) \notag \\
	&= J_T^{\text{A}}\left(\bar{\mathbf{s}}_{1\left(L_{\sf a}^*-1\right)}, 0\right) + \left(L_{\sf a}^*-1\right)C_{\sf o},
	\end{align}
	where state $\bar{\mathbf{s}}_{1\textit{L}_{\sf a}^*}^{(\sf a)}$ is obtained by offloading $L_{\sf a}^*$ most imminent tasks from $\mathbf{s}^{(\sf a)}$, and $\bar{\mathbf{s}}_{1\left(L_{\sf a}^*-1\right)}$ is obtained by offloading $L_{\sf a}^*-1$ most imminent tasks from $\mathbf{s}$. 
	
	We recall that
    \begin{align}
        J_T^{\text{A}}\left(\mathbf{s}, L\right) = \mathcal{C}^{\text{A}}\left(\mathbf{s}, L\right) + G_T^{\text{A}}\left(\mathbf{s}, L\right)
    \end{align}
    where $J_T^{\text{A}}\left(\mathbf{s}, L\right)$
	  denotes the minimum average cost attained over $T$ time slots by offloading $L$ most imminent tasks from $\mathbf{s}$ given the AMA's availability. From the definition of adjacent states in Definition \ref{Def:adjacent_states}, we have that $\bar{\mathbf{s}}_{1\textit{L}_{\sf a}^*}^{(\sf a)} \equiv \bar{\mathbf{s}}_{1\left(L_{\sf a}^*-1\right)}$. As a result,
    \begin{align}
        J_T^{\text{A}}\left(\mathbf{s}^{(\sf a)}\right) - J_T^{\text{A}}\left(\mathbf{s}\right) = C_{\sf o}.
    \end{align}
   Combining this with Ineq. (\ref{ineq:NOconds_rev}), we have the inequality:
   \begin{align}
       J_T^{\overline{\text{A}}}\left(\mathbf{s}^{(\sf a)}\right) - J_T^{\overline{\text{A}}}\left(\mathbf{s}\right)  < C_{\sf o}.
   \end{align}
   Using the fact that $\mathbf{s}$ is obtained by offloading the most imminent task from $\mathbf{s}^{(\sf a)}$, the above inequality becomes:
   \begin{align}
       J_T^{\overline{\text{A}}}\left(\bar{\mathbf{s}}_{10}^{(\sf a)}\right)  < J_T^{\overline{\text{A}}}\left(\bar{\mathbf{s}}_{11}^{(\sf a)}\right) + C_{\sf o},
   \end{align}
   in which state $\bar{\mathbf{s}}_{10}^{(\sf a)}$ is obtained by offloading 0 most imminent task from $\mathbf{s}^{(\sf a)}$, i.e., $\bar{\mathbf{s}}_{10}^{(\sf a)}=\mathbf{s}^{(\sf a)}$, and $\bar{\mathbf{s}}_{11}^{(\sf a)}$ is obtained by offloading the most imminent task from $\mathbf{s}^{(\sf a)}$, respectively. From the definition of function $F$ in Eq. (\ref{F_def_gen}), the above inequality is equivalent to
   \begin{align}
       F\left(T, \mathbf{s}^{(\sf a)}, 1, 0\right)  < F\left(T, \mathbf{s}^{(\sf a)}, 1, 1\right).
   \end{align}
   From this inequality, since function $F$ is convex as presented in Lemma \ref{Lem:Convexity}, we have the following:
	\begin{align}
	F\left(T, \mathbf{s}^{(\sf a)}, 1, 0\right)  < F\left(T, \mathbf{s}^{(\sf a)}, 1, L\right) + LC_{\sf o}, \forall L \in \mathbb{L}_1\left(\mathbf{s}^{(\sf a)}\right).
	\end{align}
	This suggests that 0 is the optimal offloading decision associated with state $\mathbf{s}^{(\sf a)}$ for the given time horizon $T$, which is contradict to our assumption that $\mathbf{s}^{(\sf a)}$ is an offloading state. Therefore, we can make a conclusion that: As Ineq. (\ref{optimcond_NO}) holds, $\mathbf{s}^{(\sf a)}$ is a non-offloading state.
	
	Hence, for two states $\mathbf{s}$ and $\mathbf{s}^{(\sf a)} \in \mathbb{S}_{\sf adj}\left(\mathbf{s}\right)$, state $\mathbf{s}^{(\sf a)}$ is a non-offloading state if and only if Ineq. (\ref{optimcond_NO}) holds. As a consequence, $\mathbf{s}^{(\sf a)}$ is an offloading state if and only if $J_T\left(\mathbf{s}^{(\sf a)}\right) - J_T\left(\mathbf{s}\right) \ge C_{\sf o}$.
	
	\section*{Appendix G}
	\section*{Proof of Theorem \ref{Theo:L*_thesmallest}}\label{proof_Theo:L*_thesmallest}
	We recall that given an current state $\mathbf{s}$, $\bar{\mathbf{s}}_{1L}$ denotes the resulting state by offloading $L$ most imminent task from $\mathbf{s}$. If $L^* = 0$ is the optimal offloading decision of state $\mathbf{s}$, then, $\mathbf{s}$ is a non-offloading state. It is trivially that $L^* = 0$ is the smallest offloading decision to reach a non-offloading state in this case.
    
    Let us consider the sequence of states $\bar{\mathbf{s}}_{10} = \mathbf{s}, \bar{\mathbf{s}}_{11}, \bar{\mathbf{s}}_{12}, \ldots, \bar{\mathbf{s}}_{1i}, \ldots$. By definition of the notation $\bar{\mathbf{s}}_{1i}$, state $\bar{\mathbf{s}}_{1\left(i+1\right)}$ is obtained by offloading the most imminent task from state $\bar{\mathbf{s}}_{1i}$ in the sequence. Therefore, state $\bar{\mathbf{s}}_{1i}$ is adjacent to $\bar{\mathbf{s}}_{1\left(i+1\right)}$. Assume $L^* > 0$ is the optimal offloading decision for $\mathbf{s}$. From the first point of Theorem \ref{Theo:adjacent_L*}, the optimal offloading decision of state $\bar{\mathbf{s}}_{11}$ would be $L^*_{1}=L^*-1$. By alternatively applying this property, the optimal offloading decisions $L^*_i$ of states $\bar{\mathbf{s}}_{1i}$ for $i=1, \ldots, L^*$ can be derived as
    \begin{align}
        L^*_i = L^* - i, \text{ for } i = 1, \ldots, L^*.
    \end{align}

    The above result suggests that the optimal offloading decision of the state $\bar{\mathbf{s}}_{1L^*}$ would be $L^*_{i} = 0$ for $i=L^*$. From the second point of Theorem \ref{Theo:adjacent_L*}, the optimal offloading decision of $\bar{\mathbf{s}}_{1\left(L^*+1\right)}$ would also be 0. Repeatedly applying this property allows us to derive the optimal decisions for state $\bar{\mathbf{s}}_{1i}, i > L^*$ as follows:
    \begin{align}
        L^*_i = 0, \text{ for } i > L^*.
    \end{align}
    
    In conclusion, states $\bar{\mathbf{s}}_{1i}$ for $i \le L^* - 1$ are offloading states, and states $\bar{\mathbf{s}}_{1i}$ for $i \ge L^*$ are non-offloading states. Therefore, $L^* > 0$ is the smallest offloading decision to reach a non-offloading state $\bar{\mathbf{s}}_{1L^*}$.

	\bibliographystyle{IEEEtran}
	\bibliography{IEEEabrv,DNK}

\begin{IEEEbiography}[{\includegraphics[width=1in,height=1.25in,clip,keepaspectratio]{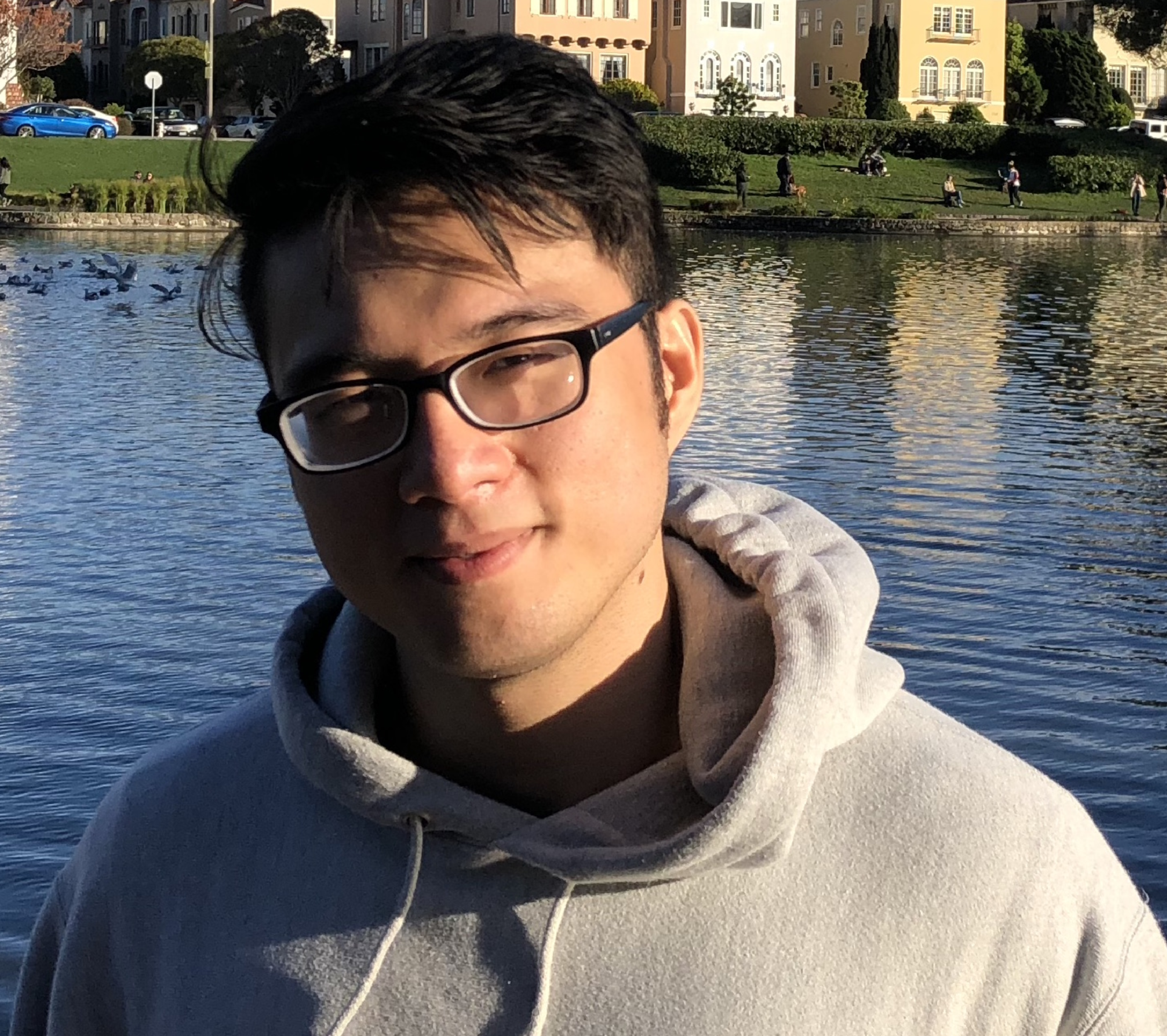}}]%
{Khai Doan}
    received his PhD in Information Systems Technology and Design from Singapore University of Technology and Design, Singapore, in 2020. He has worked as a Postdoctoral Fellow in the Systems and Computer Engineering Department at Carleton University, Canada, and as a Research Professor at the School of Electrical Engineering at Korea University, South Korea. His research interests include edge computing, machine learning, and satellite internet.
\end{IEEEbiography}

\begin{IEEEbiography}[{\includegraphics[width=1in,height=1.25in,clip,keepaspectratio]{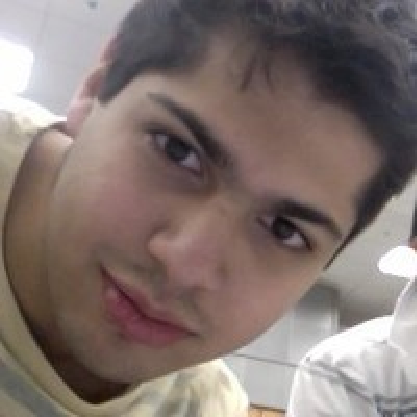}}]%
{Wesley Araujo}
    received his M.A.Sc. in Electrical and Computer Engineering from Carleton University in 2023. His research interests include computational offloading, mobile-edge computing, mobile cloud computing and the research area known as age of information.
\end{IEEEbiography}

	\begin{IEEEbiography}[{\includegraphics[width=1in,height=1.25in,clip,keepaspectratio]{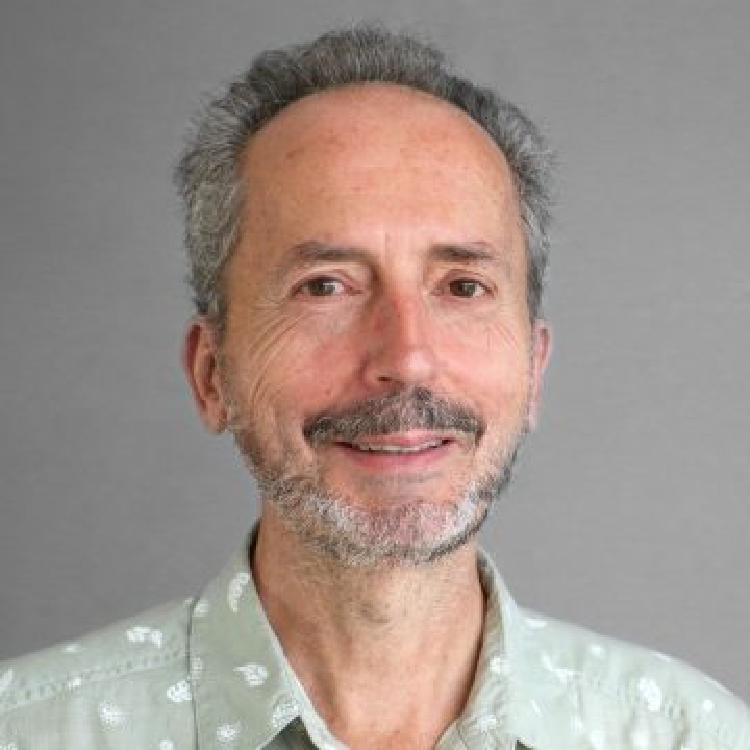}}]%
{Evangelos Kranakis}
    received a B.Sc. in Mathematics from the University of Athens, Greece, in 1973 and a Ph.D. also in Mathematics from the University of Minnesota, USA, in 1980. From 1980 to 1991 he held various faculty positions in Purdue University, USA, University of Heidelberg, Germany, Yale University, USA, Universiteit van Amsterdam, and Centrum voor Wiskunde en Informatica (CWI) in The Netherlands. He joined the faculty of the School of Computer Science of Carleton University, Ottawa, Canada, in the Fall of 1991. His current research interests include Algorithmics, Distributed and Computational Biology, Distributed and Mobile Agent Computing, Networks (Ad Hoc, Communication, Sensor, Social), and Cryptographic and Network Security.
\end{IEEEbiography}

\begin{IEEEbiography}[{\includegraphics[width=1in,height=1.25in,clip,keepaspectratio]{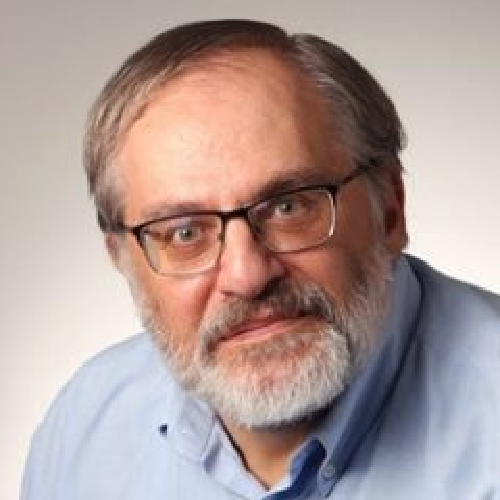}}]%
{Ioannis Lambadaris}
    was born in Thessaloniki, Greece. He received a diploma in Electrical Engineering from the Polytechnic School of the Aristotle University of Thessaloniki in 1984. He was a recipient at a Fulbright Fellowship (1984-1985) for graduate studies in USA. He received a  M.Sc. degree in Engineering from Brown University, Providence, RI, USA in 1985 and a Ph.D. degree in Electrical Engineering from the University of Maryland, College Park, MD, USA in 1991. 

He was employed as a research associate at Concordia University, Montreal, Quebec, Canada, in 1991-1992. He joined the Department of Systems and Computer Engineering in Carleton University in September 1992. Currently, he is a Chancellor's professor in the same department. While at Carleton he received the Premiere Research Excellence Award (2000), and the Carleton University Research Excellence Award (2000-2001) for his research achievements in the area of modeling and performance analysis of computer networks. In 2020 Prof. Lambadaris was awarded the Ericsson Chair in 5G Wireless Research (https://carleton.ca/ericsson/ericsson-chair-in-5g-wireless-research/)

Professor Lambadaris' interests lie in the area of applied stochastic processes, stochastic control, queueing theory and their application for modeling/simulation and performance analysis of computer communication networks. He has numerous contributions in the areas of quality of service (QoS) control for IP networks, resource allocation in optical networks, and optimal routing and flow control in ad-hoc wireless systems. His recent research focus is in the areas of hardware and software solutions for mobile applications and platforms with a focus on biomedical, remote monitoring and security applications. 
\end{IEEEbiography}

\begin{IEEEbiography}[{\includegraphics[width=1in,height=1.25in,clip,keepaspectratio]{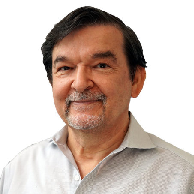}}]%
{Yannis Viniotis}
    received his Ph.D. from the University of Maryland, College Park, in 1988 and is currently a Professor with the Department of Electrical and Computer Engineering at North Carolina State University (http://www.ece.ncsu.edu/people/candice). Dr. Viniotis is the author of over one hundred technical publications, including two engineering textbooks. He has served as the cochair of two international conferences in computer networking. His research interests include virtualization, service engineering, IoT and design and analysis of stochastic algorithms as they apply to network management. Dr. Viniotis was the cofounder of Orologic, a successful startup networking company in Research Triangle Park, NC, that specialized in ASIC implementation of integrated traffic management solutions for high-speed networks.
\end{IEEEbiography}

\begin{IEEEbiography}[{\includegraphics[width=1in,height=1.25in,clip,keepaspectratio]{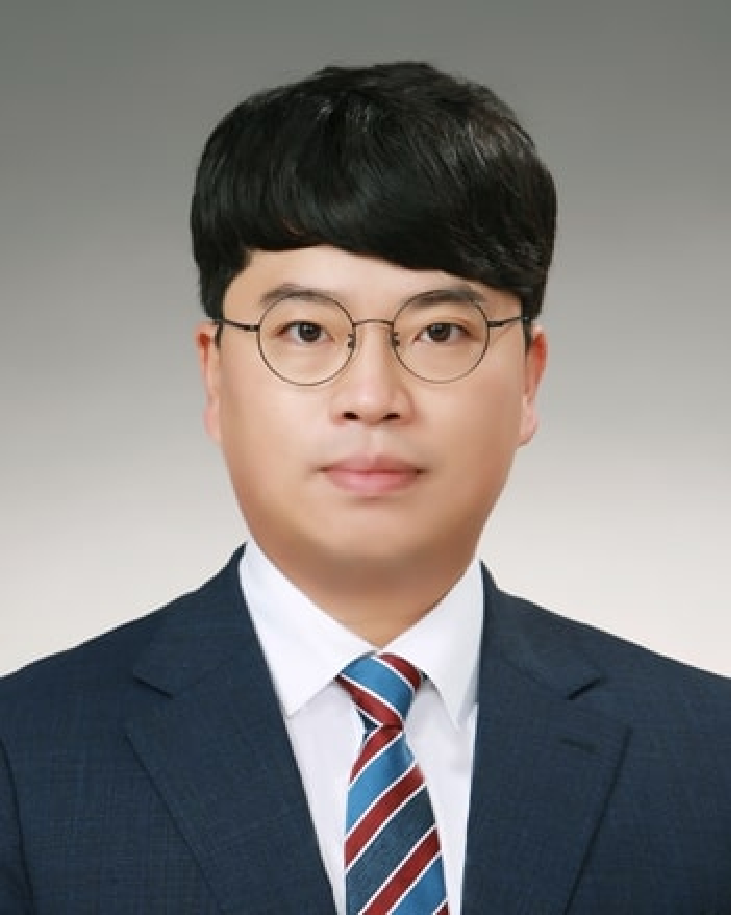}}]{Wonjae Shin} (Senior Member, IEEE) received the B.S. and M.S. degrees from the Korea Advanced Institute of Science and Technology (KAIST) in 2005 and 2007, respectively, and the Ph.D. degree from the Department of Electrical and Computer Engineering, Seoul National University, South Korea, in 2017. He has been a Visiting Scholar and a Postdoctoral Research Fellow at Princeton University, Princeton, NJ, USA, from 2016 to 2018. From 2007 to 2014, he was a Member of the Technical Staff at the Samsung Advanced Institute of Technology and Samsung Electronics Co., Ltd., South Korea, where he contributed to next-generation wireless communication networks, particularly in 3GPP LTE/LTE-advanced standardizations. Since 2023, he has been with the School of Electrical Engineering, Korea University, Seoul, South Korea, where he is currently an Associate Professor. Prior to joining Korea University, he was a Faculty Member at Pusan National University and Ajou University, South Korea. His research interests include the design and analysis of future wireless communication systems, such as interference-limited networks and machine learning for wireless networks.

Dr. Shin was the recipient of the Fred W. Ellersick Prize and the Asia-Pacific Outstanding Young Researcher Award from the IEEE Communications Society in 2020, the ICTC (International Conference on ICT Convergence) Best Workshop Paper Award in 2022, the Journal of Korean Institute of Communications and Information Sciences (J-KICS) Best Paper Award in 2021, the Best Ph.D. Dissertation Award from SNU in 2017, Gold Prize from the IEEE Student Paper Contest (Seoul Section) in 2014, and the Award of the Ministry of Science and ICT of Korea in the IDIS-Electronic News ICT Paper Contest in 2017. He was a co-recipient of the SAIT Patent Award (2010), the Samsung Journal of Innovative Technology Award (2010), the Samsung Human Tech Paper Contest (2010), and the Samsung CEO Award (2013). He was recognized as an Exemplary Reviewer by IEEE WIRELESS COMMUNICATIONS LETTERS in 2014 and IEEE TRANSACTIONS ON COMMUNICATIONS in 2019. He currently serves as an Associate Editor for the IEEE INTERNET of THINGS JOURNAL and the IEEE OPEN JOURNAL OF COMMUNICATIONS SOCIETY.
\end{IEEEbiography}


\end{document}